\documentclass[letterpaper,twocolumn]{aastex6}
\shorttitle{K2 Candidate Planetary Systems I: The Stars}
\shortauthors{Dressing et al.}
\usepackage{epsfig}
\usepackage{amsmath}
\usepackage{rotating}
\usepackage{natbib}
\usepackage{enumerate}
\usepackage{hyperref}
\usepackage{url}
\usepackage{longtable}
\usepackage{color}
\pdfoutput=1 

\def\msun{{\rm\,M_\odot}}                                                       
\def\rsun{{\rm\,R_\odot}}                                                       
\def\rearth{{\rm\,R_\oplus}} 

\def\lsun{{\rm\,L_\odot}}

\begin{document}
\title{Characterizing K2 Candidate Planetary Systems Orbiting Low-Mass Stars I: Classifying Low-mass Host Stars Observed During Campaigns 1-7}
\author{Courtney D. Dressing\altaffilmark{1,2,3}}
\author{Elisabeth R. Newton\altaffilmark{4,5}}
\author{Joshua E. Schlieder\altaffilmark{6}}
\author{David Charbonneau\altaffilmark{7}}
\author{Heather A. Knutson\altaffilmark{1}}
\author{Andrew Vanderburg\altaffilmark{7,8}}
\altaffiltext{1}{Division of Geological \& Planetary Sciences, California Institute of Technology, Pasadena, CA 91125}
\altaffiltext{2}{NASA Sagan Fellow}
\altaffiltext{3}{{\tt dressing@caltech.edu}}
\altaffiltext{4}{Department of Physics, Massachusetts Institute of Technology, Cambridge, MA 02139}
\altaffiltext{5}{National Science Foundation Astronomy \& Astrophysics Postdoctoral Fellow}
\altaffiltext{6}{NASA Exoplanet Science Institute, California Institute of Technology, Pasadena, CA 91125}
\altaffiltext{7}{Harvard-Smithsonian Center for Astrophysics, Cambridge, MA 02138}
\altaffiltext{8}{National Science Foundation Graduate Research Fellow}
\vspace{0.5\baselineskip}
\date{\today}
\slugcomment{Accepted to The Astrophysical Journal}

\begin{abstract}
We present near-infrared spectra for 144~candidate planetary systems identified during \mbox{Campaigns~1-7} of the NASA K2 Mission.  The goal of the survey was to characterize planets orbiting low-mass stars, but our IRTF/SpeX and Palomar/TripleSpec spectroscopic observations revealed that 49\% of our targets were actually giant stars or hotter dwarfs reddened by interstellar extinction. For the 72~stars with spectra consistent with classification as cool dwarfs (spectral types K3 - M4), we refined their stellar properties by applying empirical relations based on stars with interferometric radius measurements. Although our revised temperatures are generally consistent with those reported in the Ecliptic Plane Input Catalog (EPIC), our revised stellar radii are typically $0.13\rsun$ (39\%) larger than the EPIC values, which were based on model isochrones that have been shown to underestimate the radii of cool dwarfs. Our improved stellar characterizations will enable more efficient prioritization of K2 targets for follow-up studies. 
\end{abstract}


\maketitle

\section{Introduction}
Beginning in 2009, the NASA \emph{Kepler} mission revolutionized exoplanet science by searching for planets transiting roughly 190,000~stars and detecting thousands of planet candidates \citep{borucki_et_al2010, borucki_et_al2011a, borucki_et_al2011b, batalha_et_al2013, burke_et_al2014}. The main \emph{Kepler} mission ended in 2013 when the second of four reaction wheels failed, thereby destroying the ability of the spacecraft to point stably. Although the two-wheeled \emph{Kepler} was not able to continue observing the original targets, Ball Aerospace engineers and \emph{Kepler} team members realized that the torque from solar pressure could be mitigated by selecting fields along the ecliptic plane. In this new mode of operation (known as the K2 Mission), the spacecraft stares at 10,000 - 30,000 stars per field for roughly 80~days before switching to another field along the ecliptic \citep{howell_et_al2014, vancleve_et_al2016}. Unlike in the original \emph{Kepler} mission, all K2 targets are selected from community-driven Guest Observer (GO) \mbox{proposals.}

The K2 mission design is particularly well-matched for studies of planetary systems orbiting low-mass stars. Although M~dwarfs are intrinsically fainter than Sun-like stars, the prevalence of M~dwarfs within the galaxy \citep[e.g.,][]{henry_et_al2006, winters_et_al2015} ensures that there are several thousand reasonably bright low-mass stars per K2~field. Due to their smaller sizes and cooler temperatures, these stars are relatively easy targets for planet detection for two main reasons. First, the transit depth is deeper for a given planet radius. Second, the habitable zones are closer to the stars, thereby increasing both the geometric likelihood that planets within the habitable zone will appear to transit and the number of transits that could be observed during a single K2~campaign. For the coolest low-mass stars, the orbital periods of planets within the habitable zone are even short enough that potentially habitable planets would transit multiple times per campaign. 

The ``Small Star Advantage'' of deeper transit depths and higher transit probabilities within the habitable zone is partially offset by the challenge of identifying samples of low-mass stars for observation. When preparing for the original \emph{Kepler} mission, \citet{brown_et_al2011} conducted an extensive survey of the proposed field of view to identify advantageous targets and determine rough stellar properties. In contrast, the planning cycle for the K2 mission was too fast-paced to allow for such methodical preparation. During the early days of the K2 mission, the official Ecliptic Plane Input Catalog (EPIC) contained only coordinates, photometry, proper motions, and, when available, parallaxes. Proposers therefore had to use their own knowledge of stellar astrophysics to determine which stars were suitable for their investigations. 

More recently, \citet{huber_et_al2016} updated the EPIC to include stellar properties for 138,600 stars. After completing the messy tasks of matching sources from multiple catalogs, converting the photometry to standard systems, and enforcing quality cuts to discard low-quality photometry, \citet{huber_et_al2016} used the \emph{Galaxia} galactic model \citep{sharma_et_al2011} to generate synthetic realizations of different K2 fields. They then determined the most likely parameters for each K2 target star given the available photometric and kinematic information. When possible, the analysis also incorporated \emph{Hipparcos} parallaxes \citep{vanleeuwen2007} and spectroscopic estimates of $T_{\rm eff}$, $\log g$, and [Fe/H] from RAVE DR4 \citep{kordopatis_et_al2013}, LAMOST DR1 \citep{luo_et_al2015}, and APOGEE DR12 \citep{alam_et_al2015}. 

In all cases, \emph{Galaxia} used Padova isochrones \citep{girardi_et_al2000, marigo+girardi2007, marigo_et_al2008} to determine stellar properties. Aware that these isochrones tend to underpredict the radii of low-mass stars \citep{boyajian_et_al2012}, \citet{huber_et_al2016} therefore warned that the EPIC radii of low-mass stars may be up to roughly 20\% too small. Given that 41\% of selected K2 targets are low-mass M and K dwarfs \citep{huber_et_al2016}, improving the radius estimates of low-mass K2 targets is important for maximizing the scientific yield of the K2 mission. Both accurate characterization of individual planet candidates and ensemble studies of planetary occurrence demand reliable stellar properties.

Even during the more methodical \emph{Kepler} era, the properties of low-mass targets were frequently revised. Initially, \citet{brown_et_al2011} characterized all of the targets by comparing multi-band photometry to \citet{castelli+kurucz2004} stellar models. This approach worked well for characterizing Sun-like stars, but \citet{brown_et_al2011} cautioned that the Kepler Input Catalog (KIC) temperatures were untrustworthy for stars cooler than 3750~K. \citet{batalha_et_al2013} later improved the classifications for many \emph{Kepler} targets by replacing the original KIC values with parameters of the nearest model star selected from Yonsei-Yale isochrones \citep{demarque_et_al2004}, but those models noticeably underpredict the radii of low-mass stars \citep{boyajian_et_al2012}.

Considering the non-planet candidate host stars, \citet{mann_et_al2012} acquired medium-resolution ($1150 \lesssim R \lesssim 2300$) visible spectra of 382~putative low-mass dwarf targets. Using those stars as a ``training set,'' they found that the vast majority ($96\% \pm 1\%$) of cool, bright ($Kp < 14$) \emph{Kepler} target stars were actually giants. For fainter cool stars, giant contamination was much less pronounced ($7\% \pm 3\%$).  For stars that were correctly classified as dwarfs, \citet{mann_et_al2012} found that the KIC temperatures were systematically 110K hotter than the values determined by comparing their spectra to the BT-SETTL series of PHOENIX stellar models \citep{allard_et_al2011}.

In a following paper, \citet{mann_et_al2013c} obtained optical spectra of 123~putative low-mass stars hosting 188~planet candidates and NIR spectra for a smaller subset of host stars. Flux-calibrating their spectra and comparing them to BT-SETTL stellar models, they derived a set of empirically-based relations to determine stellar effective temperatures from spectral indices measured at visible and near-infrared wavelengths. \citet{mann_et_al2013c} also introduced a set of temperature--radius, temperature--mass, and temperature--luminosity relations based on the sample of stars with well-constrained radii, effective temperatures, and bolometric fluxes. 

Focusing specifically on the coolest \emph{Kepler} targets, \citet{muirhead_et_al2012b} re-characterized 84~cool Kepler Object of Interest (KOI) host stars by obtaining near-infrared spectra with TripleSpec at the Palomar Hale Telescope. As explained in \citet{rojas-ayala_et_al2012}, they estimated temperatures and metallicities using the H$_2$O-K2 index and the equivalent widths (EW) of the Na~I line at 2.210$\mu{\rm m}$ and the Ca~I line at 2.260$\mu{\rm m}$. Depending on stellar metallicity, the H$_2$O-K2 index saturates at approximately 3900K, so this approach cannot be used to characterize mid-K dwarfs.
\citet{muirhead_et_al2012b} then interpolated the temperatures and metallicities onto Dartmouth isochrones \citep{dotter_et_al2008, feiden_et_al2011} to estimate the radii and masses of their target stars. In a follow-up analysis, \citet{muirhead_et_al2014} expanded their sample to 103~cool KOI host stars and updated their mass and radius estimates using newer versions of the Dartmouth isochrones. 

Both KOIs and non-KOIs need to be accurately characterized in order to use the \emph{Kepler} data to investigate planet occurrence rates, which motivated \citet{dressing+charbonneau2013} to refit the KIC photometry using Dartmouth Stellar Evolutionary Models \citep{dotter_et_al2008, feiden_et_al2011} to determine revised properties for 3897 dwarfs cooler than 4000K. We then used the revised stellar properties to investigate the frequency of planetary systems orbiting low-mass stars.

Recognizing that the stellar parameters inferred in the previous studies were based on stellar models and therefore likely to underestimate stellar radii, \citet{newton_et_al2015} revised the properties of cool KOI host stars by employing empirical relations based on interferometrically characterized stars. Specifically, \citet{newton_et_al2015} established relationships between the EWs of Mg and Al features in $H$-band spectra from IRTF/SpeX and the temperatures, luminosities, and radii of low-mass stars. \citet{newton_et_al2015} found that the radii of M~dwarf planet candidates were typically 15\% larger than previously estimated in the \citet{huber_et_al2014} catalog, which contained a compilation of results from previous studies including \citet{dressing+charbonneau2013}, \citet{muirhead_et_al2012b, muirhead_et_al2014}, and \citet{mann_et_al2013c}.   

Accounting for the systematic effect of previously underestimated stellar radii, \citet{dressing+charbonneau2015} investigated low-mass star planet occurrence in more detail by employing their own pipeline to detect candidates and measure search completeness. Using the full four-year \emph{Kepler} dataset, we found that the mean number of small ($0.5-4\rearth$) planets per late K or early M~dwarf is $2.5 \pm 0.2$ planets per star for orbital periods shorter than 200~days. Within the habitable zone, we estimated occurrence rates of $0.24^{+0.18}_{-0.08}$ Earth-size planets and $0.21^{+0.11}_{-0.06}$ super-Earths ($1.5-2\rearth$) per star. Those estimates agree well with rates derived in independent studies \citep[e.g.,][]{gaidos2013, morton+swift2014, gaidos_et_al2014a, gaidos_et_al2016}.

In order to use the K2 data to conduct similar studies of planet occurrence rates and possibly investigate how the frequency of planetary systems orbiting low mass stars varies as a function of stellar mass, metallicity, or multiplicity, we first need to characterize the stellar sample. In this paper, we classify the subset of K2 target stars that appear to be low-mass stars harboring planetary systems. In the second paper in this series \citep{dressing_et_al2016b}, we use our new stellar classifications to revise the properties of the associated planet candidates and identify intriguing systems for follow-up analyses.  

In Section~\ref{sec:obs}, we describe our observation procedures and conditions. We then discuss the target sample in Section~\ref{sec:targets} and explain our data reduction and stellar characterization procedures in Section~\ref{sec:analysis}. Finally, we address the implications of our results and conclude in Section~\ref{sec:conc}.

\section{Observations}
\label{sec:obs}
We conducted our observations using the SpeX instrument on the NASA Infrared Telescope Facility (IRTF) over 15~partial nights during the 2015A, 2015B, 2016A, and 2016B semesters and the TripleSpec instrument on the Palomar 200'' over four~full nights during the 2016A semester. Eleven of our IRTF/SpeX nights were awarded to C.~Dressing via programs 2015B068, 2016A066, and 2016B057; the remaining SpeX time was provided by K.~Aller, W.~Best, A.~Howard, and E.~Sinukoff. All of our Palomar time was awarded to C.~Dressing for program P08. 

As detailed in Table~\ref{tab:obscon}, our observing conditions varied from photometric nights to nights with significant cloud cover through which only our brightest targets were observable. As recommended by \citet{vacca_et_al2003}, we removed telluric features from our science spectra using observations of A0V stars acquired under similar observing conditions. Accordingly, we interspersed our science observations with observations of nearby A0V stars. When possible, these A0V stars were within $15\degr$ of our target stars and observed within one hour at similar airmasses (difference $< 0.1$ airmasses). 

\begin{deluxetable*}{ccccccc}
\tablecolumns{7}
\scriptsize
\tablecaption{Observing Conditions \label{tab:obscon}
}
\tablehead{
\colhead{} &
\colhead{} &
\colhead{} &
\colhead{Date} & 
\colhead{Seeing} & 
\colhead{Weather}  &
\colhead{K2}  \\[0.1em]
\colhead{Semester} &
\colhead{Instru} &
\colhead{Program} &
\colhead{(UT)} &
\colhead{} &
\colhead{Conditions} &
\colhead{Targets\tablenotemark{a}} 
}
\startdata
2015A & SpeX & 989 & Apr 16, 2015 &  $0\farcs7 - 1\farcs0$ & Clear & 2\tablenotemark{b}\\
& SpeX & 989 & May 5, 2015 &  $0\farcs3 - 0\farcs8$ & Light wind, clear & 5\tablenotemark{c} \\
& SpeX & 981 & June 13, 2015 &  $0\farcs3 - 1\farcs0$ & Cirrus, patchy clouds & 2\tablenotemark{d} \\
\hline
2015B & SpeX & 057, 068 &  Aug 7, 2015 & $0\farcs5 - 1\farcs0$ & Clear at start; closed early due to high humidity & 3\tablenotemark{e} \\
& SpeX & 068 & Sep 24, 2015 & $0\farcs5 - 1\farcs0$ & Patchy clouds cleared slightly overnight & 20 \\
& SpeX &  072 & Oct 14, 2015 &  $0\farcs4 - 1\farcs0$ & Cirrus & 1\tablenotemark{f} \\
& SpeX &  068 & Nov 26, 2015 & $0\farcs5 - 2\farcs0$ & Patchy clouds; high humidity & 16 \\
& SpeX &  068 & Nov 27, 2015 & $0\farcs6$  & Cirrus & 16 \\
\hline
2016A & TSPEC & P08 & Feb 19, 2016 & $1\farcs2 - 2\farcs0$ & Cirrus clouds at start; moderately cloudy by morning & 3 \\
& SpeX & 066 & Mar 4, 2016  & $0\farcs5 - 1\farcs0$ & Clear & 10  \\
& SpeX & 066& Mar 8, 2016  &$0\farcs5 - 1\farcs0$  & Thick, patchy clouds at sunset; thinner clouds by morning & 12 \\
& SpeX & 986 & Mar 10, 2016  & $0\farcs9$ & Cirrus & 5\tablenotemark{g} \\
& TSPEC & P08& Mar 27, 2016  &  $0\farcs9$  & Clear & 15 \\
& TSPEC & P08& Mar 28, 2016  & $0\farcs9 - 2\farcs1$ & Patchy clouds; closed early due to high humidity \& fog & 9 \\
& TSPEC & P08& April 18, 2016  & $1\farcs1 - 1\farcs9$ & Clear & 11 \\
& SpeX & 066 & May 5, 2016  & $0\farcs5 - 1\farcs0$ & Patchy clouds & 11 \\
& SpeX & 066 & May 6, 2016  & $0\farcs3 - 0\farcs9$ & Clear & 6 \\
& SpeX & 066 & June 7, 2016  & $0\farcs4 - 1\farcs0$ & Clear & 8\\
\hline
2016B & SpeX & 057 & Oct 26, 2016 & $0\farcs5 - 1\farcs4$ & Clear & 5 \\
 \enddata
\tablenotetext{a}{We observed some stars twice on two different nights to assess the repeatability of our analysis.}
\tablenotetext{b}{Night awarded to Andrew Howard.}
\tablenotetext{c}{Night awarded to Andrew Howard, but observations obtained by Joshua Schlieder.}
\tablenotetext{d}{Observations obtained by Evan Sinukoff.}
\tablenotetext{e}{Includes one observation acquired by Will Best (Program 057) and two acquired by Courtney Dressing (Program 068).}
\tablenotetext{f}{Observations obtained by Kimberly Aller.}
\tablenotetext{g}{Night awarded to Andrew Howard, but observations obtained by Courtney Dressing.}
\end{deluxetable*}

\subsection{IRTF/SpeX}
For our SpeX observations, we selected the \mbox{$0\farcs3 \times 15"$}~slit and observed in SXD mode to obtain moderate resolution ($R \approx 2000$) spectra \citep{rayner_et_al2003, rayner_et_al2004}. Due the SpeX upgrade in 2014, our spectra include enhanced wavelength coverage from 0.7 - 2.55~$\mu$m.  

We carried out all of our observations using an ABBA nod pattern with the default settings of 7\farcs5 separation between positions A and B and 3\farcs75 separation between either pointing and the ends of the slit. For all targets except close binary stars, we aligned the slit with the parallactic angle to minimize systematic effects in our reduced spectra; for binary stars, we rotated the slit so that the sky spectra acquired in the B position would be free of contamination from the second star or so that spectra from both stars could be captured simultaneously. We scaled the exposure times for our targets and repeated the ABBA nod pattern as required so that the resulting spectra would have S/N of $100 - 200$ per resolution element.

We calibrated these spectra by running the standardized IRTF calibration sequence every few hours during our observations and ensuring that each region of the sky had a separate set of calibration frames. The calibration sequence includes flats taken using an internal quartz lamp and wavelength calibration spectra acquired using an internal thorium-argon lamp.

\subsection{Palomar/TripleSpec}
We acquired our TripleSpec observations using the fixed $1" \times 30"$ slit, which yields simultaneous coverage between 
1.0 and 2.4 $\mu$m at a spectral resolution of \mbox{$2500 - 2700$} \citep{herter_et_al2008}. In order to decrease the effect of bad pixels on the detector, we adopted the 4-position ABCD nod pattern used by \citet{muirhead_et_al2014} rather than the 2-position ABBA pattern we used for our SpeX observations. With the exception of double star systems for which we altered the slit rotation to place both stars in the slit when possible, we left the slit in a fixed East-West orientation. We calibrated our spectra using dome darks and dome flats acquired at both the beginning and end of the night. 

\begin{deluxetable*}{cccccccc}
\tablecolumns{8}
\tabletypesize{\normalsize}
\tablecaption{Targets Observed by K2 Campaign \label{tab:campaigns}}
\tablehead{
\multicolumn{4}{c}{Campaign} & 
\colhead{Total} & 
\multicolumn{3}{c}{Classification in This Paper}\\
\cline{1-4}
\cline{6-8}
\colhead{Field} &
\colhead{RA} & 
\colhead{Dec} &
\colhead{Galactic} &
\colhead{Targets} & 
\colhead{Cool} & 
\colhead{Hotter}&
\colhead{} \\
\colhead{Number} &
\colhead{(hh:mm:ss)} &
\colhead{(dd:mm:ss)} &
\colhead{Latitude ($^\circ$)} &
\colhead{Observed} &
\colhead{Dwarfs\tablenotemark{1}} &
\colhead{Dwarfs} &
\colhead{Giants}
}
\startdata
1 & 11:35:46 & $+01:25:02$ & $+59$ &  10 & 9 (90\%) & 1 (10\%) & 0 (0\%)\\
2 & 16:24:30 & $-22:26:50$ & $+19$ &  8 & 0 (0\%) & 4 (50\%) & 4 (50\%)\\
3 & 22:26:40 & $-11:05:48$ & $-52$ &  12 & 6 (50\%) & 5 (42\%) & 1 (8\%)\\
4 & 03:56:18 & $+18:39:38$ & $-26$ & 24 & 10 (42\%) & 10 (42\%) & 4 (17\%)\\
5 & 08:40:38 & $+16:49:47$ & $+32$ &  41 & 27 (66\%) & 13 (32\%) & 1 (2\%)\\
6 & 13:39:28 & $-11:17:43$ & $+50$ &  34 & 16 (47\%) & 12 (36\%) & 6 (18\%)\\
7 & 19:11:19 & $-23:21:36$ & $-15$ &  17 & 6 (35\%) & 4 (24\%) & 7 (41\%)\\
\hline
$1-7$ & $\cdots$ & $\cdots$ & $\cdots$ & 146 & 74 (51\%) & 49 (34\%) & 23 (16\%)\\
\enddata
\tablenotetext{1}{Two K2 targets (EPIC~211694226 and EPIC~212773309) have nearby companions that may or may not be physically associated. We classified 74 cool dwarfs in 72 systems.}
\end{deluxetable*}

\section{Target Sample}
\label{sec:targets}
The objective of our observing campaign was to determine the properties of K2 target stars and assess the planethood of associated planet candidates. Consequently, our targets were selected from lists of K2 planet candidates compiled by A.~Vanderburg and the K2 California Consortium (K2C2). These early target lists are preliminary versions of planet candidate catalogs such as those published in \citet{vanderburg_et_al2016} and \citet{crossfield_et_al2016}.

Of the 144~K2 targets observed, 99 (69\%) appear in unpublished lists provided by A. Vanderburg, 28 (19\%) were  published in the \citet{vanderburg_et_al2016} catalog, and 77 (53\%) were reported in previously unpublished lists generated by K2C2. (These totals sum to $>100\%$ due to partial overlap between the Vanderburg and K2C2 candidate lists.) The K2C2 planet candidates from K2 Campaigns $0-4$ were later published in \citet{crossfield_et_al2016}. Although we did not consult these catalogs for initial target selection, our target sample also contains 46~systems from \citet{barros_et_al2016}, 26~stars from \citet{pope_et_al2016}, 5~stars from \citet{foreman-mackey_et_al2015}, 5~stars from \citet{montet_et_al2015}, and 4~stars from \citet{adams_et_al2016}.  

The Vanderburg and K2C2 catalogs contain all of the planet candidates detected by the corresponding pipeline ({\tt K2SFF} and {\tt TERRA}, respectively) in the K2 light curves of stars proposed as individual Guest Observer targets. Neither pipeline considers stars observed as part of ``super-stamps.''  Due to the heterogenous nature of the K2 target lists and the limited information provided in the EPIC during early K2 campaigns, the selected target sample is heavily biased. As noted by \citet{huber_et_al2016}, the K2 target lists are biased toward cool dwarfs. Overall, the set of stars observed during Campaigns~1--8 consisted primarily of K and M dwarfs (41\%), F and G dwarfs (36\%), and K~giants (21\%), but the giant fraction was higher for fields close to the galactic plane (see Table~\ref{tab:campaigns}) than for fields at higher galactic latitude  \citep{huber_et_al2016}. Many Guest Observers used a magnitude cut when proposing targets, which may have increased the representation of multiple star systems within the selected sample.

\begin{figure}[tbp]
\centering
\includegraphics[width=0.45\textwidth]{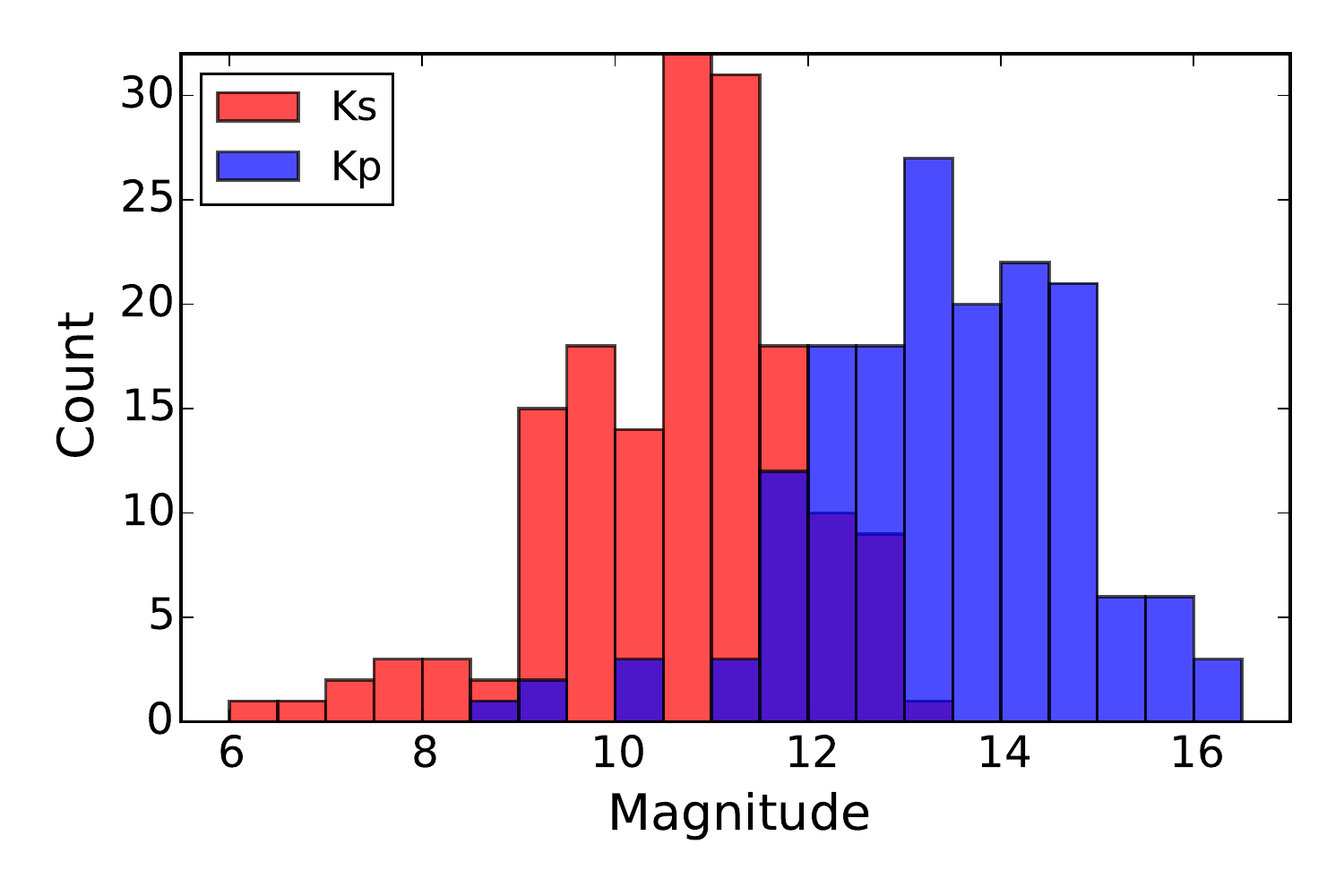}
\caption{Magnitude distribution of our full target sample in the $Kepler$ bandpass (Kp; blue) and $Ks$ (red). Our targets have median brightnesses of $Ks = 10.8$ and $Kp = 13.5$. The $Kepler$ bandpass extends from roughly 420~nm to 900~nm with maximum response at 575~nm \citep{kepler_instrument_handbook}; the $Ks$ bandpass is centered at $2.159~\mu{\rm m}$ \citep{cohen_et_al2003}.}
\label{fig:maghist}
\end{figure}

Due to the design of the K2 mission, our K2 targets were concentrated in distinct fields of the sky each spanning roughly 100 square degrees. We note the number of targets observed from each campaign in Table~\ref{tab:campaigns}. As shown in Figure~\ref{fig:maghist}, the magnitude distribution of our K2 targets ranged from $6.2 - 13.1$ in \emph{Ks}, with a median \emph{Ks} magnitude of 10.8. In the \emph{Kepler} bandpass (similar to $V$-band), our targets had brightnesses of  $Kp= 9.0 - 16.3$ and a median brightness of $Kp = 13.5$. 

With each K2 data release, we initially prioritized observations of stars harboring small planet candidates (estimated planet radius $< 4 \rearth$) and systems that could potentially be well-suited for high-precision radial velocity observations (host star brighter than $V = 12.5$ and estimated radial velocity semi-amplitude $K > 2$~m/s). Once we had exhausted the those targets, we worked down the target list and observed increasingly fainter host stars harboring larger planets. Our goal was to select late K dwarfs and M dwarfs, but the initial stellar classifications were uncertain, particularly for the first K2 fields when the \citet{huber_et_al2016} EPIC stellar catalog was not yet available. To ensure that few low-mass stars were excluded from our analysis, we adopted lenient criteria when selecting potential target stars. Our rough guidelines were $J-K > 0.5$ and, for stars with coarse initial temperature estimates, temperature cooler than 4900K. Concentrating on the brightest targets biased our sample towards giant stars and binary stars. Similarly, our selected $J-K$ color-cut also boosted the giant fraction by excluding hotter dwarfs with bluer $J-K$ colors without discarding giant stars with extremely red $J-K$ colors. The binary boost due to prioritizing bright targets may have been partially offset by our avoidance of stars with nearby companions detected in follow-up adaptive optics images.

\section{Data Analysis \& \\Stellar Characterization}
\label{sec:analysis}
We performed initial data reduction using the publicly-available {\tt Spextool} pipeline \citep{cushing_et_al2004} and a version customized for use with TripleSpec data (available upon request from M. Cushing). Both versions of the pipeline include the {\tt xtellcor} telluric correction package \citep{vacca_et_al2003}. As recommended in the {\tt Spextool} manual, we selected the Paschen $\delta$ line at $1.005 \mu$m when generating the convolution kernel used to apply the observed instrumental profile and rotational broadening to the Vega model spectrum. 

\subsection{Initial Classification}
\label{ssec:initial_class}
After completing the {\tt Spextool} reduction, we used an interactive Python-based plotting interface to compare our spectra to the spectra of standard stars from the IRTF Spectral Library \citep{rayner_et_al2009}. We allowed each model spectrum to shift slightly in wavelength space to accommodate differences in stellar radial velocities. Considering the $J$, $H$, and $K$ bandpasses independently, we assessed the $\chi^2$ of a fit of each model spectrum to our data and recorded the dwarf and giant models with the lowest $\chi^2$. 

We then considered the target spectrum holistically and assigned a single classification to the star. Although the focus of this analysis was to characterize planetary systems orbiting low-mass dwarfs, our target sample did include contamination from hotter and evolved stars. We list the 23~giants and 49~hotter dwarfs in Tables~\ref{tab:giants} and \ref{tab:hot_dwarfs}, respectively. We did not include either group in the more detailed analyses described in Section~\ref{ssec:starchar}. For the purposes of identifying contamination, we rejected all stars that we visually classified as giants or dwarfs with spectral types earlier than K3. Table~\ref{tab:hot_dwarfs} also includes all stars for which the \citet{newton_et_al2015} routines yielded estimated temperatures above 4800K or radii larger than $0.8\rsun$ (see Section~\ref{ssec:starchar}). We display the reduced spectra for all targets in Appendix~\ref{sec:appendix}. We have also posted our spectra and stellar classifications on the ExoFOP-K2 follow-up website.\footnote{\url{https://exofop.ipac.caltech.edu/k2/}}

\begin{figure}[tbhp]
\centering
\includegraphics[width=0.45\textwidth]{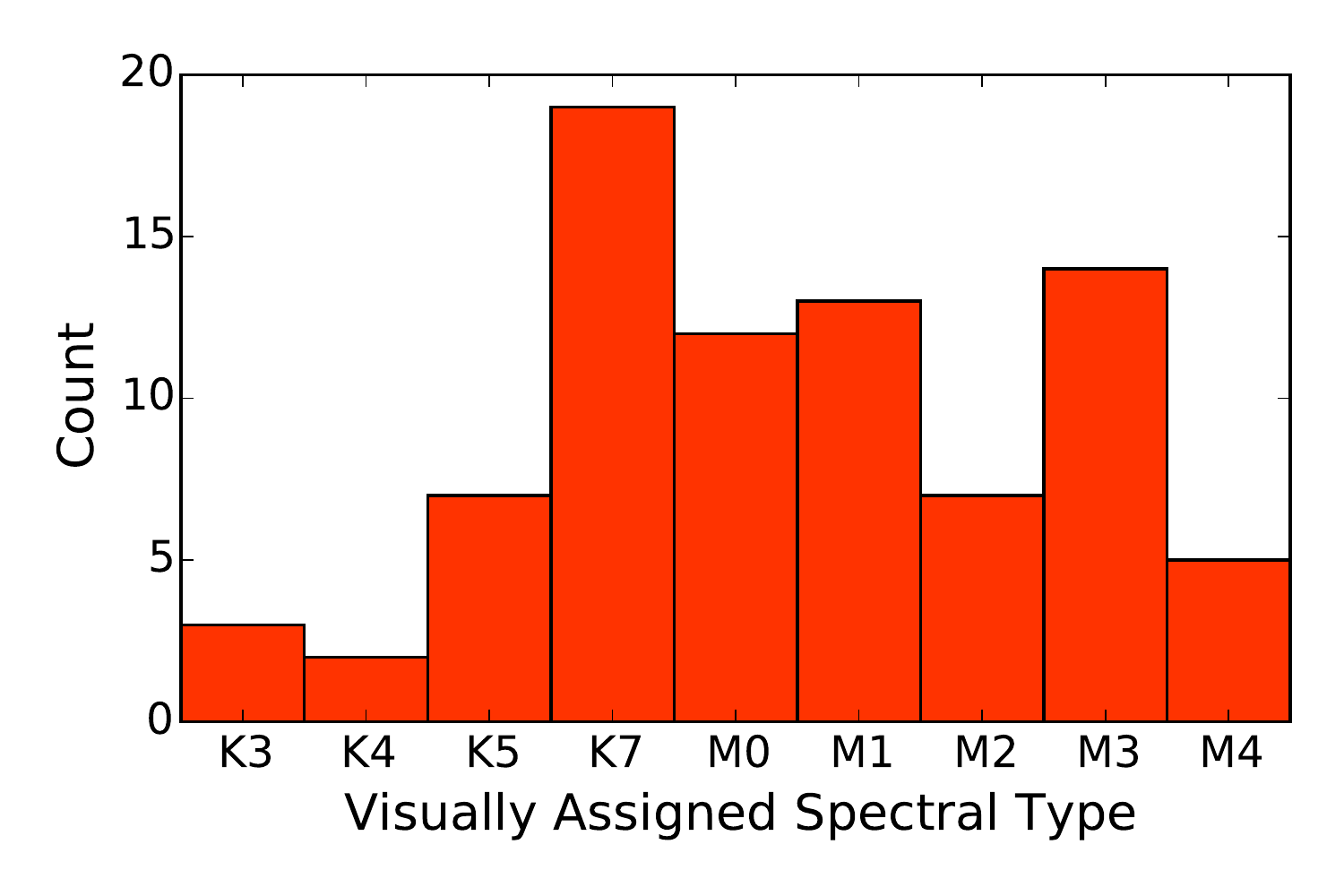}
\caption{Distribution of visually assigned spectral types for the 74~stars in our cool dwarf sample.}
\label{fig:spectype}
\end{figure}

Figure~\ref{fig:spectype} displays the spectral type distribution of the stars in the selected cool dwarf sample. The sample includes stars with spectral types between K3 and M4, with a median spectral type of M0. These spectral types are rather coarse visual assignments ($\pm 1$ subclass), so the spike at M3V may be a quirk of the particular template stars used for spectral type assignment rather than a true feature of the distribution. Due to the small sample size, the spike can also be explained by Poisson counting errors. 

\subsection{Detailed Stellar Characterization}
\label{ssec:starchar}
\begin{figure}[tbp]
\centering
\includegraphics[width=0.5\textwidth]{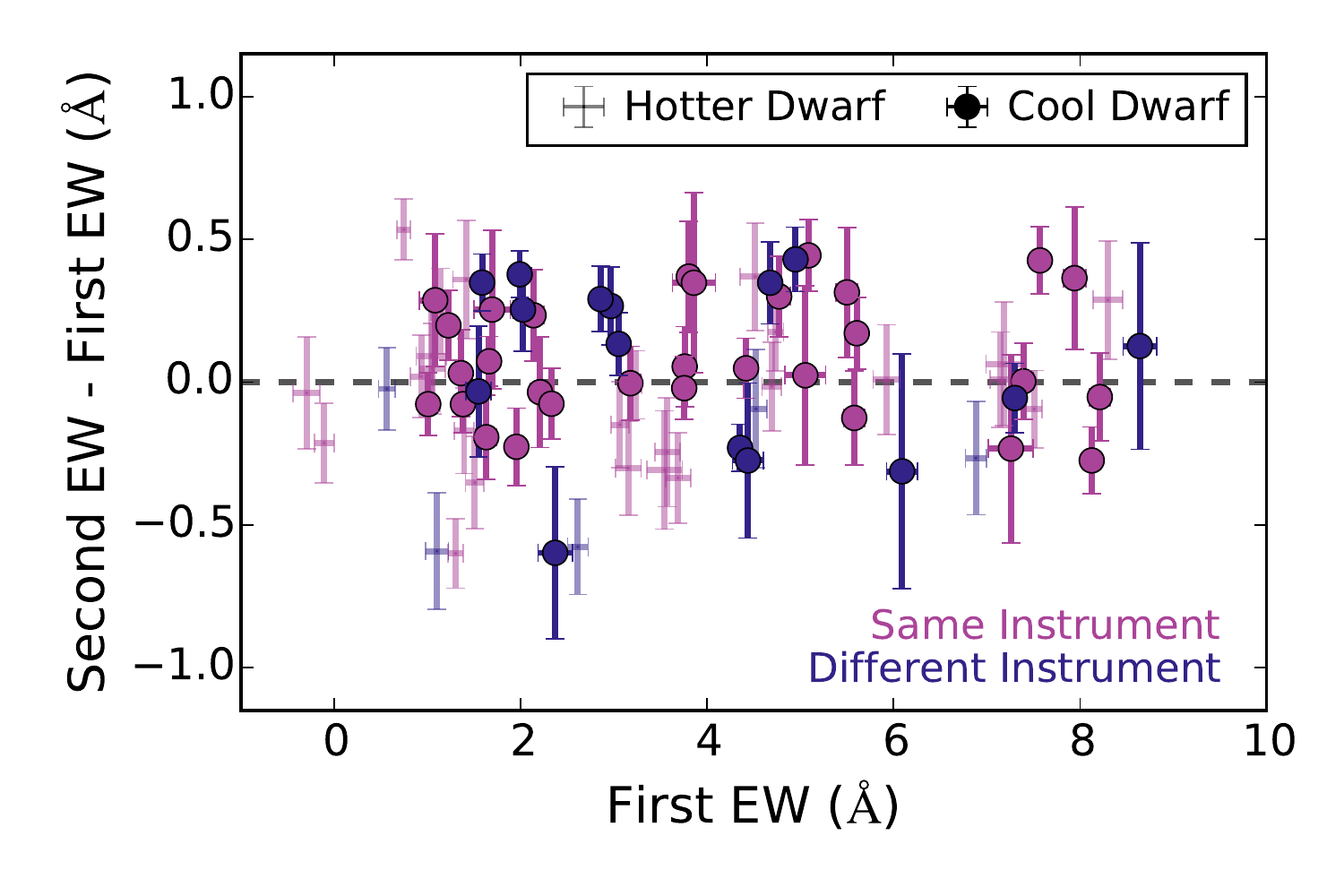}
\caption{Repeatability of our equivalent width measurements when using the same instrument for both observations (magenta points) or different instruments for each observation (navy points). The 75 data points plotted here are the EW measured for five Mg and Al features in thirty spectra of fifteen candidate low-mass dwarfs (two observations per star). Eight stars were later classified as cool dwarfs (large circles; spectral types K7, M0, and M1) and seven were classified as hotter dwarfs (small points). For reference, the gray dashed line marks zero difference between the two EW measurements.}
\label{fig:ew_repeats}
\end{figure}

For the stars that were visually identified as dwarfs with spectral types of K3 or later, we used a series of empirical relations to refine the stellar classification. We began by using the publicly-available, IDL-based {\tt tellrv}\footnote{\url{https://github.com/ernewton/tellrv}} and {\tt nirew}\footnote{\url{https://github.com/ernewton/nirew}} packages developed by \citet{newton_et_al2014, newton_et_al2015} to shift each spectrum to the stellar rest frame on an order-by-order basis, measure the equivalent widths of key spectral features, and estimate stellar properties. Specifically, the packages employ empirically-based relations linking the equivalent widths of H-band Al and Mg features to stellar temperatures, radii, and luminosities \citep{newton_et_al2015}. These relations are appropriate for stars with spectral types between mid-K and mid-M (i.e., temperatures of $3200 - 4800$K, radii of $0.18 < R_\star < 0.8 \rsun$, and luminosities of $-2.5 < \log L/\lsun < -0.5$). The relations were calibrated using IRTF/SpeX spectra \citep{newton_et_al2015} so we downgraded the Palomar/TSPEC spectra to match the lower resolution of IRTF/SpeX data before applying the relations. We note that neglecting the change in resolution can lead to systematic $0.1\rm{\AA}$ differences in the measured EW due to variations in the amount of contamination included in the designated wavelength interval \citep{newton_et_al2015}. As shown in Figure~\ref{fig:ew_repeats}, we find generally consistent equivalent widths in spectra acquired on different occasions even if the two observations used separate instruments under variable observing conditions. Specifically, the median absolute difference in equivalent widths for the five cool dwarfs with repeated measurements using the same instrument was 0.2\AA ($0.9\sigma$).  The median absolute difference for the three cool dwarfs with measurements from different instruments was 0.3\AA ($1.9\sigma$).

In the original formulation of the {\tt measure\_hband} stellar characterization routine, the errors on stellar parameters are determined via a Monte Carlo simulation in which multiple realizations of noise are added to the spectra and the equivalent widths of features are re-measured. The errors are then determined by combining the random errors in the resulting EWs with the intrinsic scatter in the relations. This approach yields useful errors, but the adopted stellar parameters are taken from a single realization of the noise. For high SNR spectra, variations in the simulated noise might not lead to large changes in stellar properties, but for lower SNR spectra the estimated properties can differ considerably from one realization to the next. Several of our spectra have SNR of less than 200, which was the threshold used in the \citet{newton_et_al2014} study. Accordingly, we altered {\tt measure\_hband} to calculate the temperatures, luminosities, and radii for each realization of the noise and report the 50th, 16th, and 84th percentiles as the best-fit values, lower error bars, and upper error bars, respectively. 

Our changes significantly improve the reproducibility of temperature, luminosity, and radius estimates for stars with lower SNR spectra. For example, we repeated the classification of the M2~dwarf EPIC~206209135 five times using both the original and modified versions of {\tt measure\_hband}. For each classification, we determined parameter errors by generating 1000~noise realizations. The original code yielded estimated temperatures ranging from $3267-3461$~K, radii of $0.32- 0.35\rsun$, and \mbox{$-1.94 \leq \log L/\lsun_\star \leq -1.85$}. The variations in the assigned temperatures and luminosities of 194~K and $0.09\log\lsun$ were significantly larger than the individual error estimates of $85$~K and $0.06\log \lsun$ and the spread in assigned radii of $0.03\rsun$ was equal to the individual radius errors. In comparison, our new method found $T_{\rm eff} = 3360 \pm 87$~K, $R_\star = 0.33 \pm 0.03 \rsun$, and $\log L_\star = -1.87 \pm 0.06 \log \lsun$ in all cases. Due to the asymmetry of the resulting temperature and radius distributions for some stars, we also report separate upper and lower error bounds instead of forcing the errors to be symmetric in all cases. (EPIC~2106209135 is an example of a star with naturally symmetric errors.)

\begin{figure}[tbhp]
\centering
\includegraphics[width=0.5\textwidth]{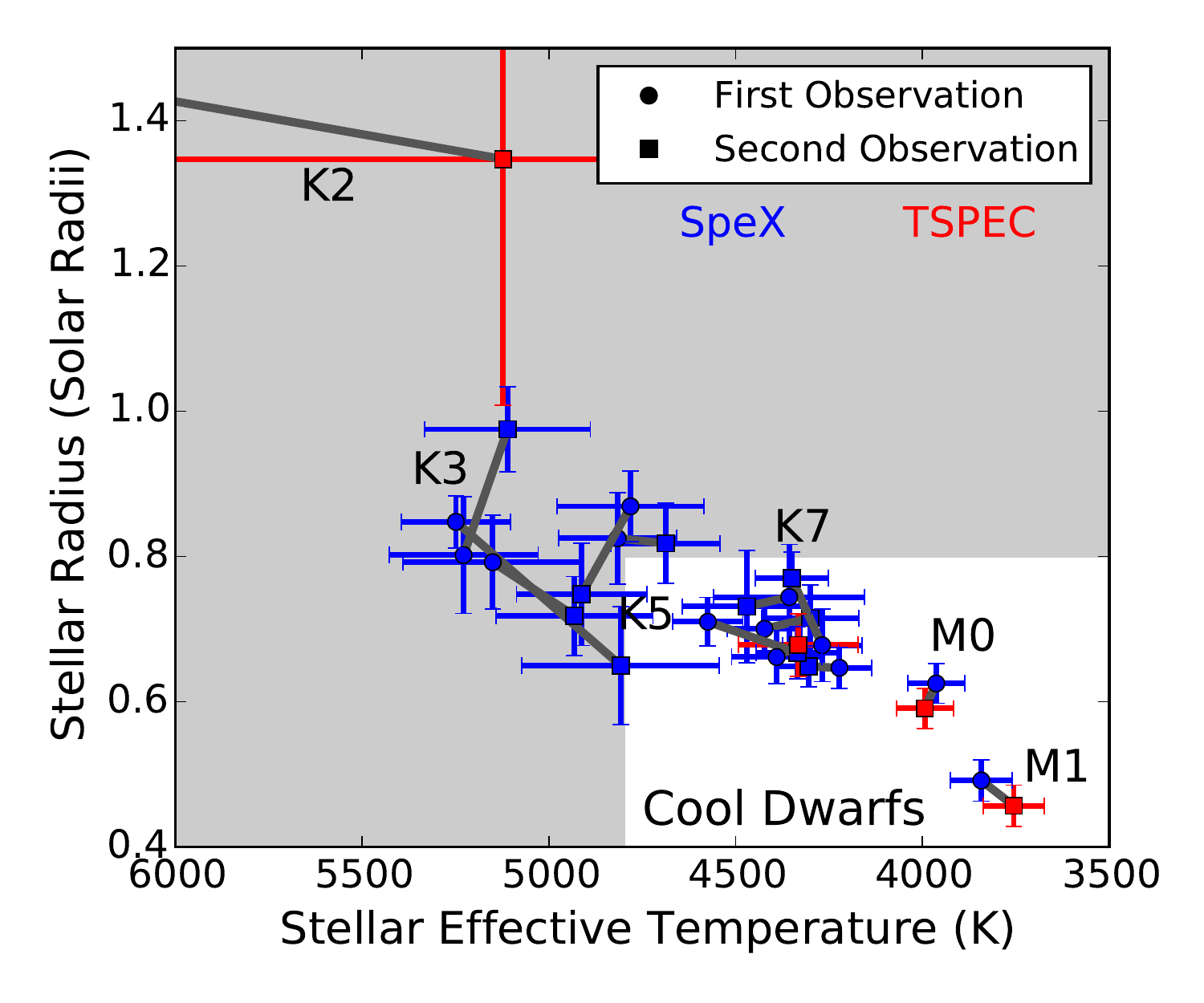}
\caption{Repeatability of parameter estimates for the subsample of fifteen stars with two observations.  The data points mark the estimated temperatures and radii found by applying the \citet{newton_et_al2015} EW relations to the first observations (circles) and second observations (squares) of each star. The colors differentiate between observations made using SpeX on the IRTF (blue) and TSPEC on the Palomar 200'' (red). The thick gray lines connect the two classifications for each star. The cluster of points near 4350K and $0.7\rsun$ contains five K7 dwarfs observed twice each. The white box indicates the boundaries of our cool dwarf sample: }$T_{\rm eff} < 4800$~K, $R_\star < 0.8 \rsun$.
\label{fig:repeat_classifications}
\end{figure}

We confirmed that our cool dwarf classifications were repeatable by comparing our parameter estimates for the fifteen stars observed on two different observing runs. Figure~\ref{fig:repeat_classifications} reveals satisfactory agreement in the temperature and radius estimates for the eight stars cooler than 4800K, the designated upper limit for our cool dwarf sample. Our results for the seven hotter stars are less consistent, but the relations from \citet{newton_et_al2015} are not valid at those temperatures. 

\subsubsection{Stellar Effective Temperature}
\label{sssec:teff}
For comparison, we also determined stellar effective temperatures using the $J$-, $H$-, and $K$-band temperature-sensitive indices and relations presented by \citet{mann_et_al2013c}. We then applied the temperature--metallicity--radius relation from \citet{mann_et_al2015} to assign stellar radii. Next, we determined luminosities and masses from the estimated stellar effective temperatures using relations 7 \& 8 from \citet{mann_et_al2013c}. These relations are based on stars with effective temperatures between 3238K and 4777K and radii between $0.19\rsun$ and $0.78\rsun$.  

\begin{figure*}[tbp]
\centering
\includegraphics[width=1\textwidth]{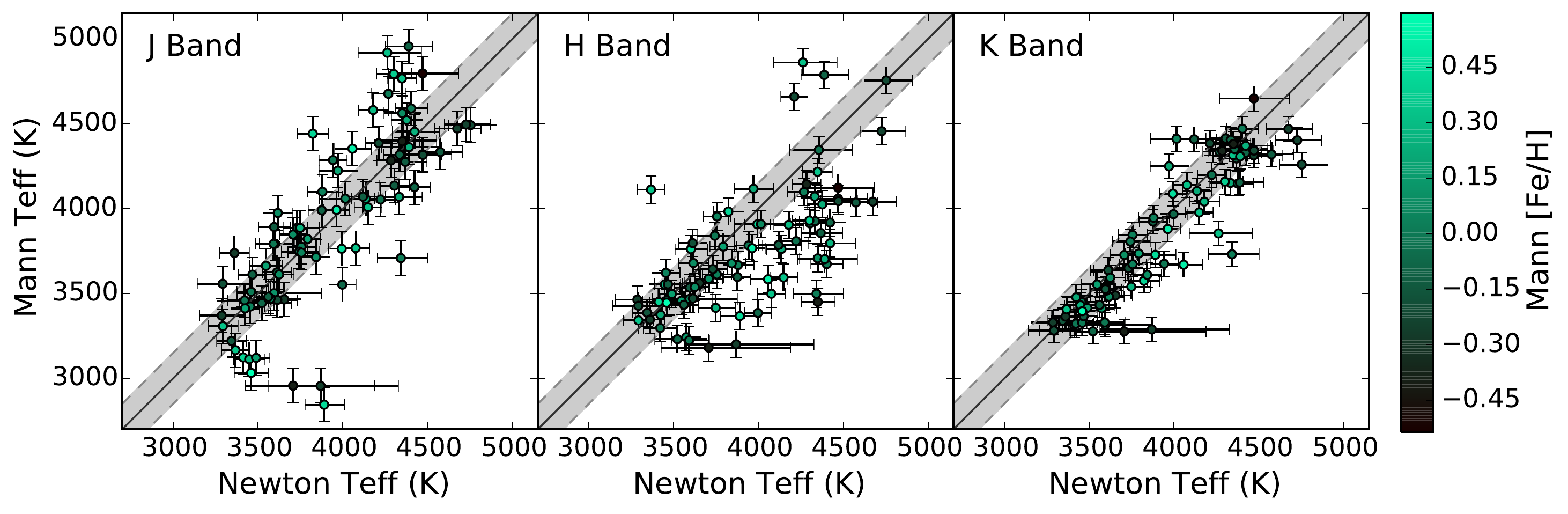}
\caption{Comparison of temperatures derived using EW-based estimates from \citet{newton_et_al2015} and spectral indices from \citet{mann_et_al2013c} in $J$-band (left), $H$-band (middle), and $K$-band (right). Points within the shaded region lie within $150$K of a one-to-one relation (solid line). All points are color-coded by [Fe/H] as indicated by the colorbar.}
\label{fig:newton_mann_teff}
\end{figure*}

In Figure~\ref{fig:newton_mann_teff}, we plot the temperature estimates generated using the \citet{newton_et_al2015} pipeline against those from the \citet{mann_et_al2013c} relations. The Mann $H$-band based temperatures display considerable scatter and are systematically lower than the three other estimates (the temperatures based on the \citet{newton_et_al2015} routines, the $J$-band temperatures, and the $K$-band temperatures). This discrepancy, which is most noticeable for stars hotter than 4000K, is likely caused by saturation of the index as the continuum flattens for hotter stars. The $J$-band temperatures also display large scatter, but they are more centered along a one-to-one relation than the $H$-band estimates. Due to the much tighter correlation observed between the $K$-band temperatures and the EW-based temperature estimates, we adopt the $K$-band temperatures as the ``Mann temperatures'' for our stars. We also see discrepancies for stars with $T_{\rm eff} < 3500$. There are three stars for which the temperature inferred using the \citet{newton_et_al2015} relations is larger than that inferred from the J, H and K band temperatures, The error bars in the temperature inferred from the \citet{newton_et_al2015} relations are also large. This is caused by the disappearance of the Mg and Al features in the coolest dwarf stars, which tends to result in an overestimate of Teff. Al is weaker at lower metallicity, consistent with this effect only being seen in metal-poor stars at the limits of the calibration.

\citet{newton_et_al2015} also compared temperature estimates derived using their empirical relations with those based on the \citet{mann_et_al2013c} temperature-sensitive indices. They found large standard deviations of \mbox{$\sigma_{\Delta T} = 140$~K} and \mbox{$\sigma_{\Delta T} = 170$~K} in $J$-band and \mbox{$H$-band,} respectively, between temperatures determined using each method, which they attributed to telluric contamination. In contrast, the standard deviation between the \citet{newton_et_al2015} estimates and the \citet{mann_et_al2013c} $K$-band estimates was only \mbox{$\sigma_{\Delta T}$ = 90~K}, suggesting that the $K$-band relation is less contaminated by telluric features. 
     
For our sample of stars, the agreement between the two methods is much worse: we measure standard deviations of 278~K, 311~K, and 162~K for the temperature differences between the EW-based estimates and the estimates based on the $J$-band, $H$-band, and $K$-band spectral indices, respectively. The median temperature differences are 13~K, 143~K, and 64~K for $J$-band, $H$-band, and $K$-band, respectively, with the EW-based estimates higher than the spectral index-based estimate for $H$- and $K$-band and lower for $J$-band.  The significantly poorer agreement is likely due to the differences between the \citet{newton_et_al2015} stellar sample and our stellar sample. The \citet{newton_et_al2015} sample was dominated by mid- and late-M dwarfs with effective temperatures between 3000~K and 3500~K. In contrast, our targets are primarily late K dwarfs and early M dwarfs. 

For an additional check on our stellar classifications, we applied the H$_{\rm 2}$O-K2 index - spectral type relation calibrated by \citet{newton_et_al2014} to estimate near-infrared spectral types. The H$_{\rm 2}$O-K2 index \citep{rojas-ayala_et_al2012} provides an estimate of the level of water absorption in an M dwarf spectrum by measuring the shape of the spectrum between 2.07$\mu$m and 2.38$\mu$m. Higher values indicate lower H$_{\rm 2}$O opacity and therefore hotter temperatures. The H$_{\rm 2}$O-K2 index is the second-generation version of the H$_{\rm 2}$O-K index introduced by \citet{covey_et_al2010} and uses slightly different portions of the spectrum to avoid contamination from atomic lines in early M~dwarfs.  The index is gravity-insensitive for stars with effective temperatures between 3000K and 3800K and metallicity-insensitive for stars cooler than 4000K. The H$_{\rm 2}$O-K2 index saturates near 4000K, so these index measurements and spectral types are not valid for the hotter stars in our sample. 

\begin{figure}[tbp]
\centering
\includegraphics[width=0.5\textwidth]{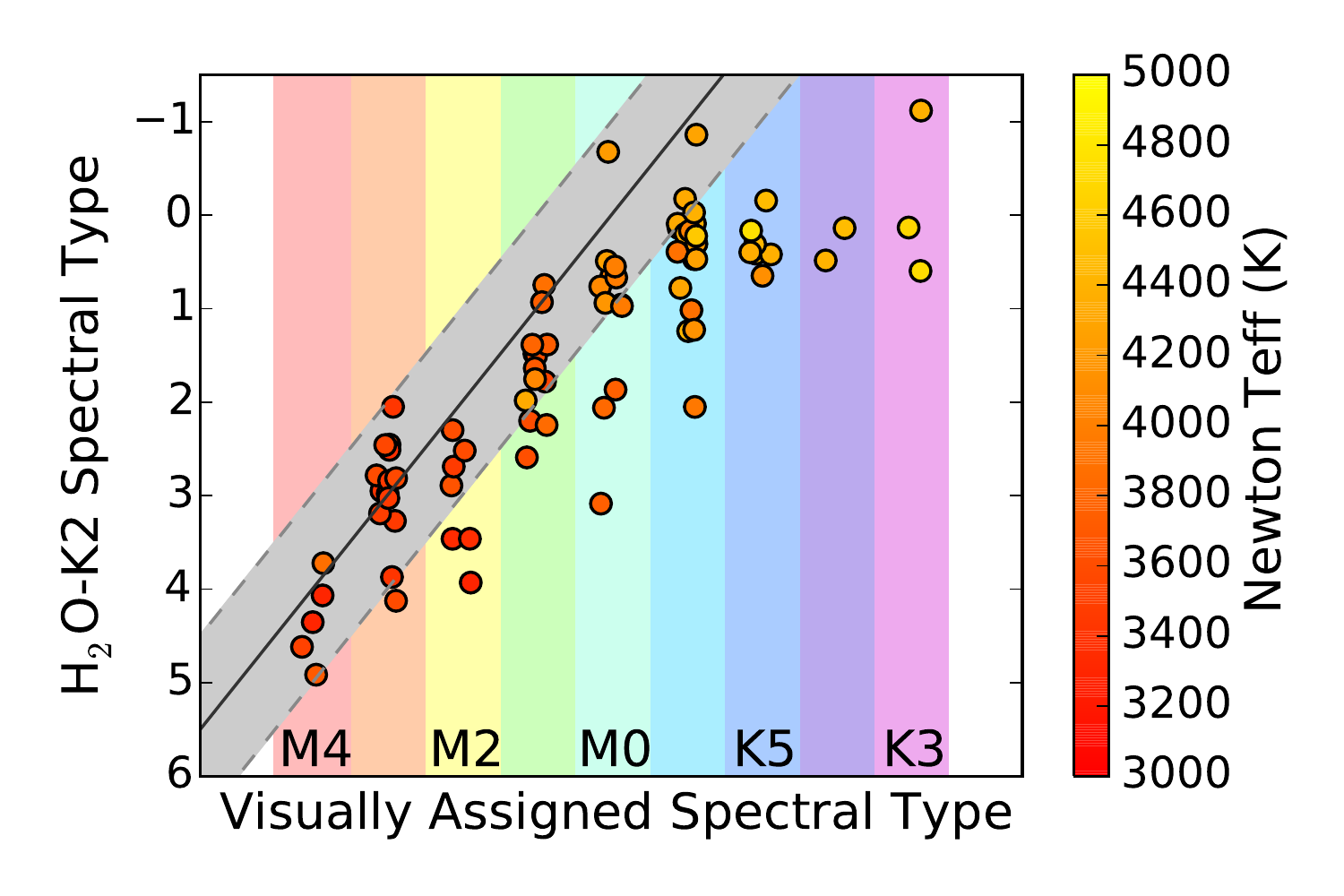}
\caption{Numerical spectral types automatically derived from the H$_{\rm 2}$O-K2 index versus our visually determined spectral types. The points are colorcoded based on the EW-based temperature estimate resulting from the \citet{newton_et_al2015} relations. The gray shaded region denotes spectral types that fall within one spectral type of a one-to-one relation. For reference, the rainbow shading also denotes the spectral type ranges. We assigned visual spectral types at integer values, but the points are horizontally offset for clarity.}
\label{fig:vis_index_sptype}
\end{figure}

\begin{figure*}[tbp]
\centering
\includegraphics[width=1\textwidth]{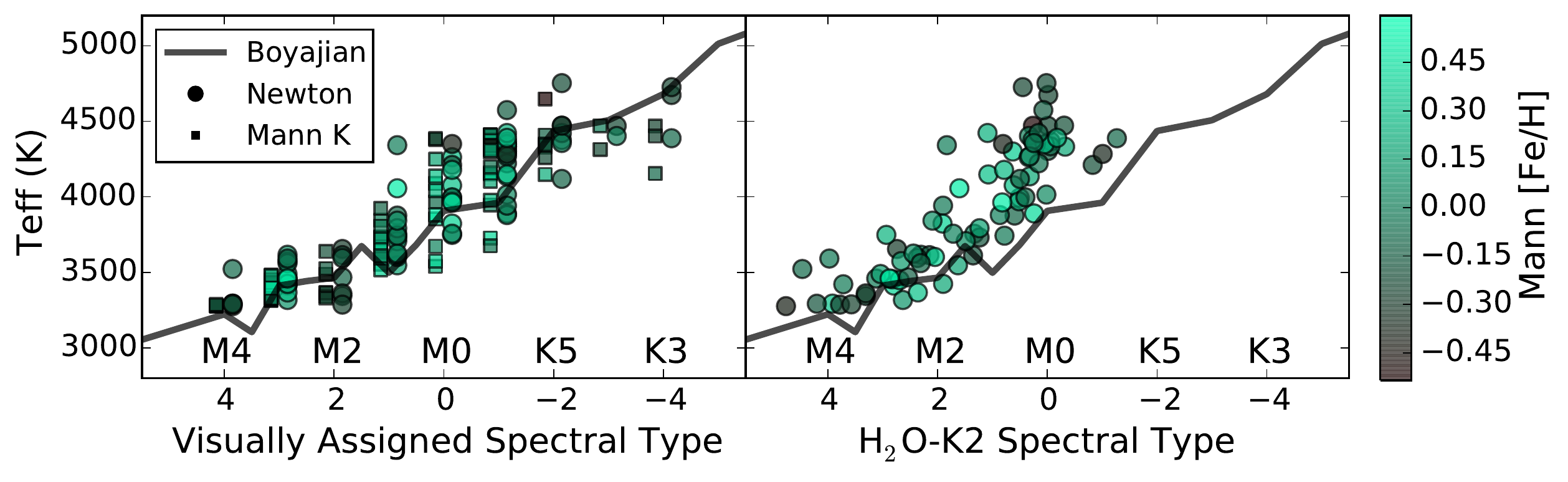}
\caption{Temperatures from \citet{newton_et_al2015} EW-based relation (circles) and \citet{mann_et_al2013c} $K$-band relation (squares) versus visually assigned spectral type (left) and automatically assigned H$_{\rm 2}$O-K2 index-based spectral types (right). For reference, the black line shows the spectral types and temperatures reported by \citet{boyajian_et_al2012} for interferometrically-characterized stars. Note that \citet{boyajian_et_al2012} report temperatures at half spectral types between M0 and M4. All points are color-coded by [Fe/H] as indicated by the colorbar.}
\label{fig:sptype_teff}
\end{figure*}

As shown in Figure~\ref{fig:vis_index_sptype}, our visually assigned spectral types and the index-based spectral types agree well for stars cooler than roughly 3800K. Above this temperature, the index-based spectral types plateau near M1 due to the inapplicability of the index for the earliest M~dwarfs.  The saturation of the H$_{\rm 2}$O-K2 index is highlighted in Figure~\ref{fig:sptype_teff}, which provides an alternative comparison of our spectral type assignments and temperature estimates. In the left panel, we show that our visually-assigned spectral types display the expected correlation with temperature throughout the spectral type range of our sample. In contrast, the index-based spectral types deviate from the expected correlation for stars earlier than M1V.  We list the visually-assigned and index-based spectral types for the cool dwarf sample in Table~\ref{tab:cdwarfs_st}. 

\subsubsection{Stellar Metallicities}
\label{sssec:met}
\begin{figure*}[tbp]
\centering
\includegraphics[width=0.45\textwidth]{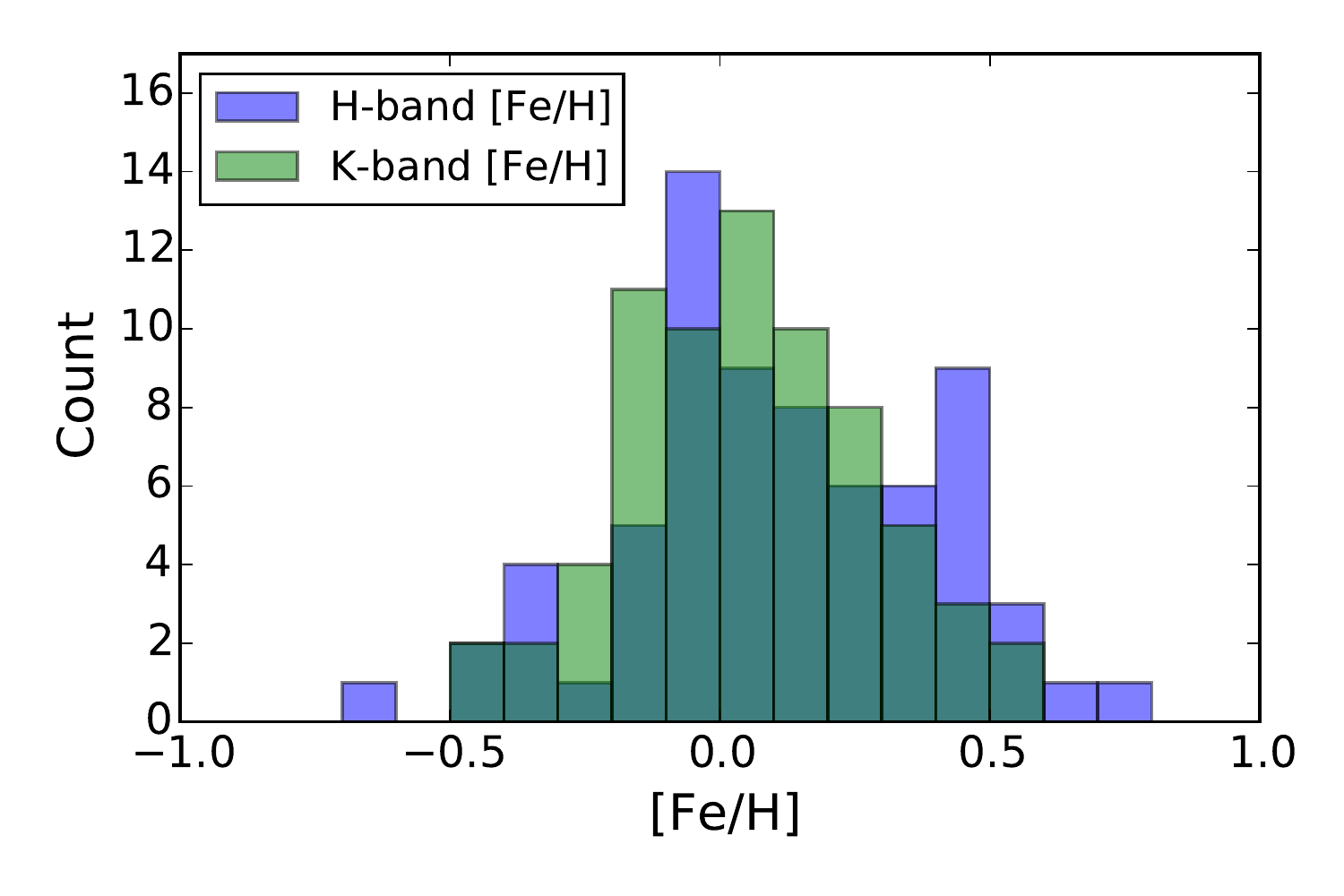}
\includegraphics[width=0.45\textwidth]{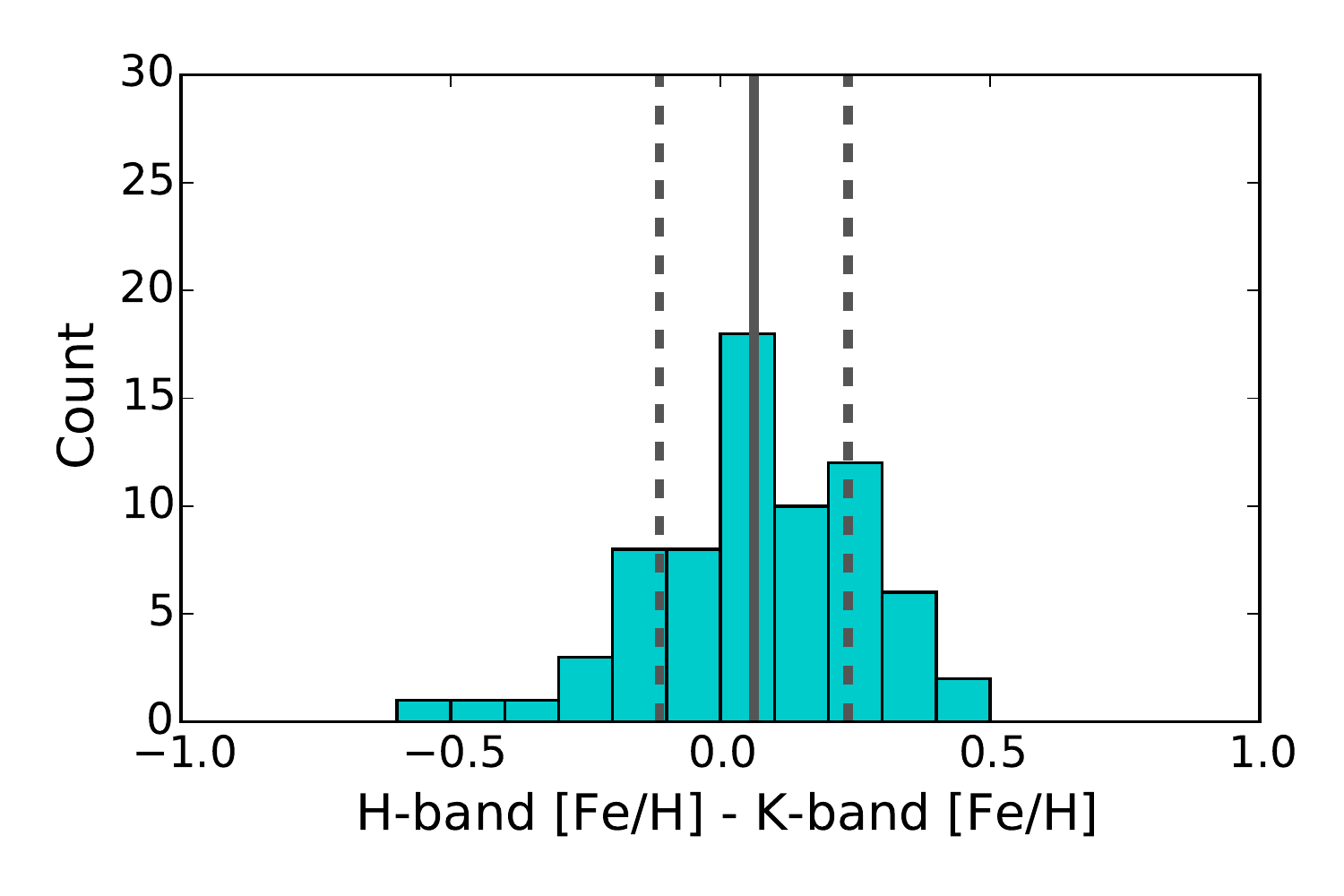}
\includegraphics[width=0.45\textwidth]{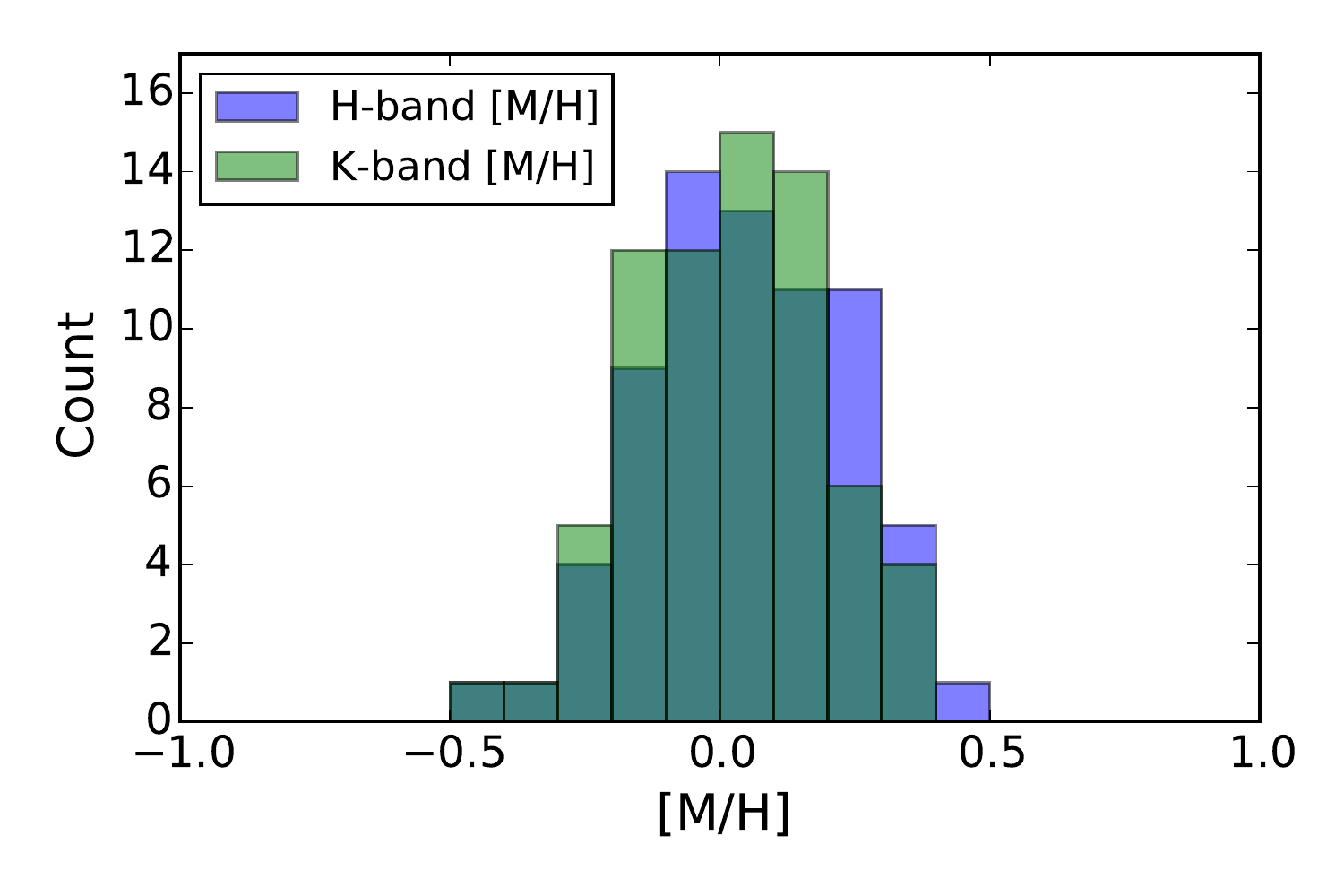}
\includegraphics[width=0.45\textwidth]{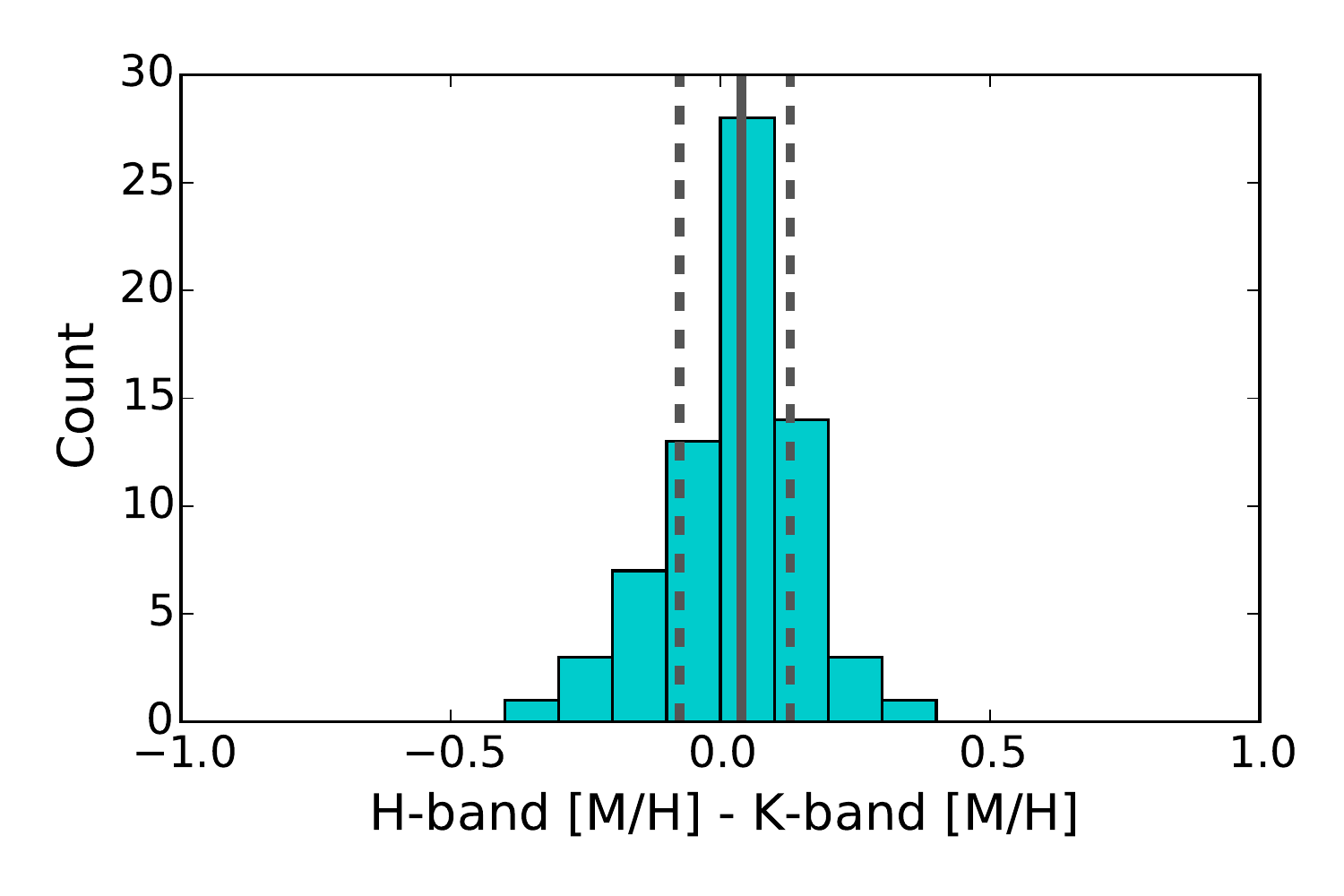}
\caption{Estimated metallicities for the 63 cool dwarfs with spectral types of K7 or later. The top two panels display the distribution of [Fe/H] (left) and [M/H] (right) calculated using separate relations from \citet{mann_et_al2013a} for $H$-band (blue) and $K$-band (green) spectra. The bottom two panels display the distributions of differences in the $H$-band and $K$-band estimates of [M/H] (left) and [Fe/H] (right). The green lines indicate the median values (solid lines) and the 16th and 84th percentile values (dashed lines).}
\label{fig:met}
\end{figure*}

We estimated [Fe/H] and [M/H] using the relations from \citet{mann_et_al2013a}. The latest stars in our sample are M4~dwarfs, so we did not need to transition from the metallicity relations for K7$-$M5 dwarfs provided by \citet{mann_et_al2013a} to the relations for M4.5$-$M9.5 dwarfs from \citet{mann_et_al2014}. We calculated metallicities using $H$-band and $K$-band spectra separately and compare the resulting distributions of [Fe/H] and [M/H] in Figure~\ref{fig:met}. On average, a typical star in our cool dwarf sample has near-solar metallicity. Averaging the $H$-band and $K$-band estimates for each star, we obtain median metallicities of [Fe/H]$ = 0.02$ and [M/H]$= 0.00$. Figure~\ref{fig:met} also displays distributions of the differences between the \mbox{$H$-band} and $K$-band metallicity estimates; they agree at the $1\sigma$ level. Although our cool dwarf sample includes 11 mid-K dwarfs, we restricted our metallicity analysis to the 63 cool dwarfs with spectral types of K7 or later. 

\subsubsection{Stellar Radii}
\label{sssec:rs}
We infer stellar radius using the methods from \citet{newton_et_al2015} and \citet{mann_et_al2015}. The former are derived directly from the EWs. The latter use $T_{\rm eff}$ and metallicity to estimate radii indirectly; for $T_{\rm eff}$ we use the $K$-band temperatures (which we refer to as ``Mann temperatures'', see Section \ref{sssec:teff}). The \citet{mann_et_al2015} temperature-metallicity-radius relation is valid for stars with temperatures between 2700K and 4100K, but many of the stars in our sample are hotter than this upper limit. For the stars for which the \citet{mann_et_al2015} relations yield temperatures hotter than 4100K, we instead compare the \citet{newton_et_al2015} radii to the radii estimated by applying the temperature-radius relation provided in Equation~8 of \citet{boyajian_et_al2012} using the Mann temperatures.

\begin{figure}[tbp]
\centering
\includegraphics[width=0.45\textwidth]{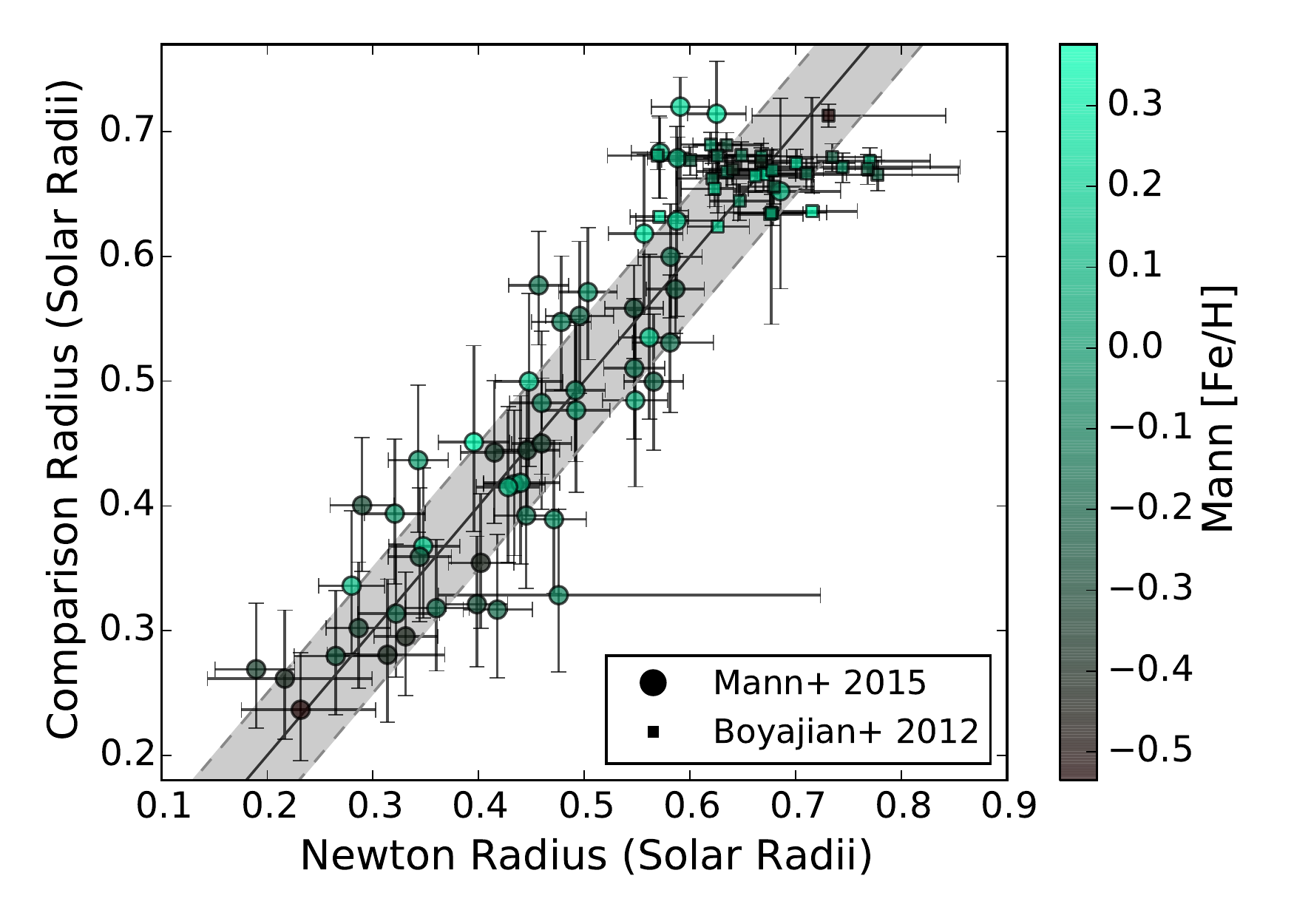}
\caption{Comparison of radii derived directly using the \citet{newton_et_al2015} relations and indirectly via the \citet[][circles]{mann_et_al2015} temperature--metallicity--radius relation or \citet[][squares]{boyajian_et_al2012} temperature--radius relation. Points within the shaded region lie within $0.05~\rsun$ of a one-to-one relation (solid line). The data points are color-coded by [M/H] as measured using relations from \citet{mann_et_al2013a}.}
\label{fig:radii}
\end{figure}

We display the resulting radius estimates in Figure~\ref{fig:radii}. The \citet{mann_et_al2015} methodology and the \citet{newton_et_al2015} routines yield similar radii: the median radius difference is $0.01\rsun$ (the Mann radii are larger) and the standard deviation of the differences is $0.06\rsun$. For comparison, the median reported radius errors are $0.03\rsun$ for the \citet{newton_et_al2015} values and $0.05\rsun$ for the \citet{mann_et_al2015} values.  Looking at the hotter stars, the median difference between the Newton radii and \citet{boyajian_et_al2012} radii is only $0.002\rsun$ and the standard deviation of the difference is $0.05\rsun$

\begin{figure}[tbp]
\centering
\includegraphics[width=0.45\textwidth]{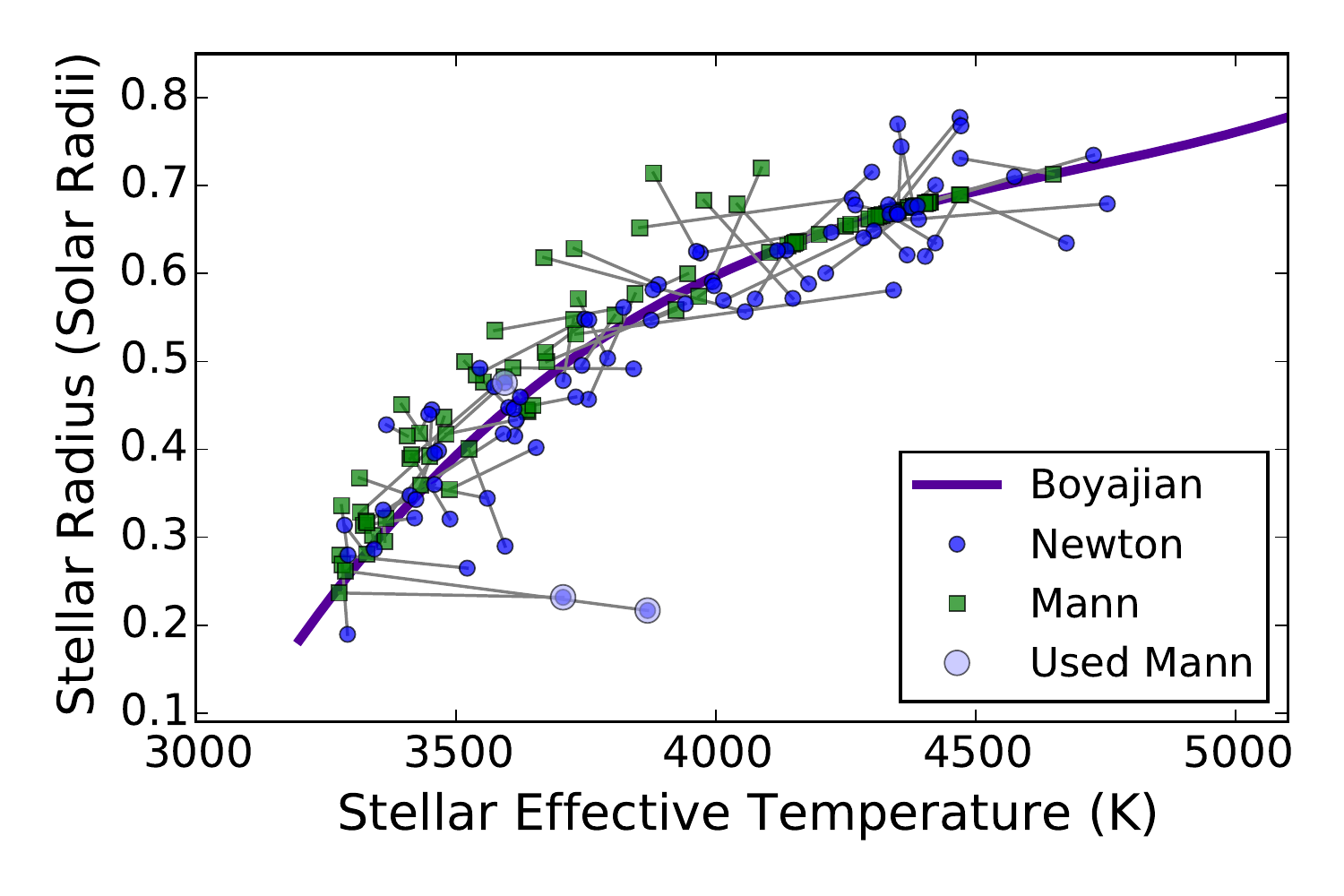}
\caption{Comparison of temperatures and radii derived using relations from \citet{newton_et_al2015} and \citet{mann_et_al2015}. The gray lines connect the values from the Newton relations (blue circles) and Mann relation (green squares) for each star. The three mid-M dwarfs highlighted with light blue circles have Al-a EW below the calibration range for the Newton temperature relations. For those three stars only, we use adopt the Mann parameters instead. For reference, the purple line displays the third-order temperature--radius polynomial presented in Equation~8 of \citet{boyajian_et_al2012}. }
\label{fig:teffrs_tracks}
\end{figure}

As shown in Figure~\ref{fig:teffrs_tracks}, the primary reason why the temperature agreement looks worse for the coolest stars is because three cool stars (EPIC~211817229, EPIC~211799258, and EPIC~211826814) have significantly different parameters using the two methods. Based on the sample of stars with interferometrically constrained properties, the expected temperatures and radii of M5.5$-$M3~dwarfs are \mbox{$3054-3412$}~K and $0.14-0.41\rsun$, respectively \citep{boyajian_et_al2012}. Although these stars were visually classified as M3 or M4~dwarfs, the \citet{newton_et_al2015} routines assigned them high temperatures of $3594 - 3869$~K because the Al-a EW measured in their spectra were below the lower limit of the calibration sample (see Table~\ref{tab:ew+met} for EW measurements). The Mann routines assigned the stars cooler temperatures of $3276 - 3317$~K. Due to the better agreement between the Mann temperatures and expected temperatures of mid-M dwarfs, we chose to adopt the Mann et al. classifications for those three stars.

\subsubsection{Stellar Luminosities}
\label{sssec:lum}
We compared the stellar luminosities estimated using the EW-based relation from \citet{newton_et_al2015} to those found using the temperature-luminosity relation from \citet{mann_et_al2013c}. Due to the functional nature of the \citet{mann_et_al2013c} relation, the Mann values followed a single track whereas the Newton values displayed scatter about that relation. Ignoring the three mid-M dwarfs that are too cool for the Newton relations, the luminosity differences (Newton - Mann) have a median value of $0.008\lsun$ and a standard deviation of $0.05\lsun$. The scatter increases as temperature increases. Dividing the sample into stars hotter and cooler than 4000~K, the luminosity differences for cooler sample have a median value of $0.005\lsun$ and a standard deviation of $0.03\lsun$ while the hotter sample has a median value of $0.034\lsun$ and a standard deviation of $0.07\lsun$. In the left panel of Figure~\ref{fig:masslum}, we display the adopted luminosities as a function of effective temperature.

\subsubsection{Stellar Masses}
\label{sssec:mass}
\begin{figure*}[tbp]
\centering
\includegraphics[width=0.45\textwidth]{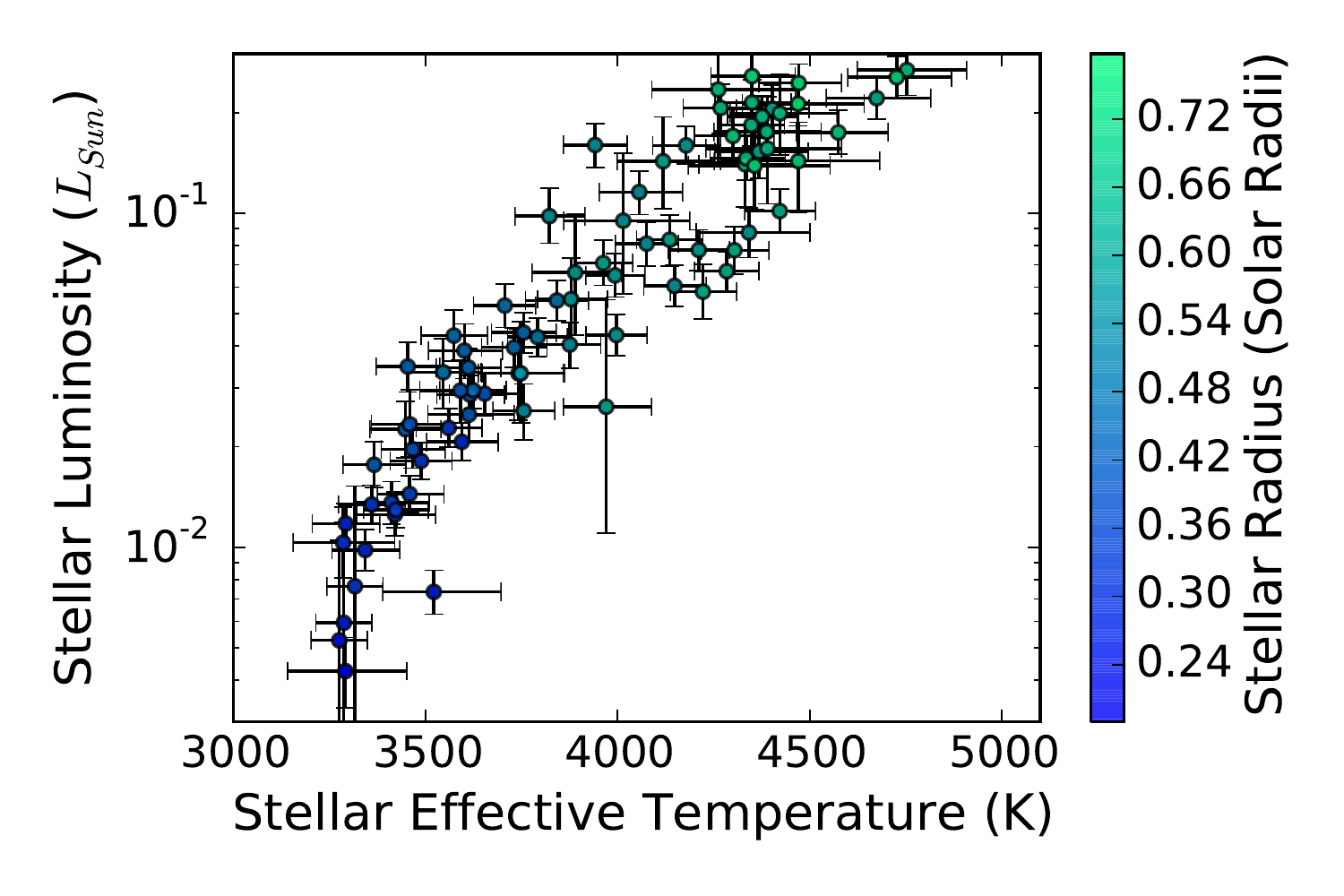}
\includegraphics[width=0.45\textwidth]{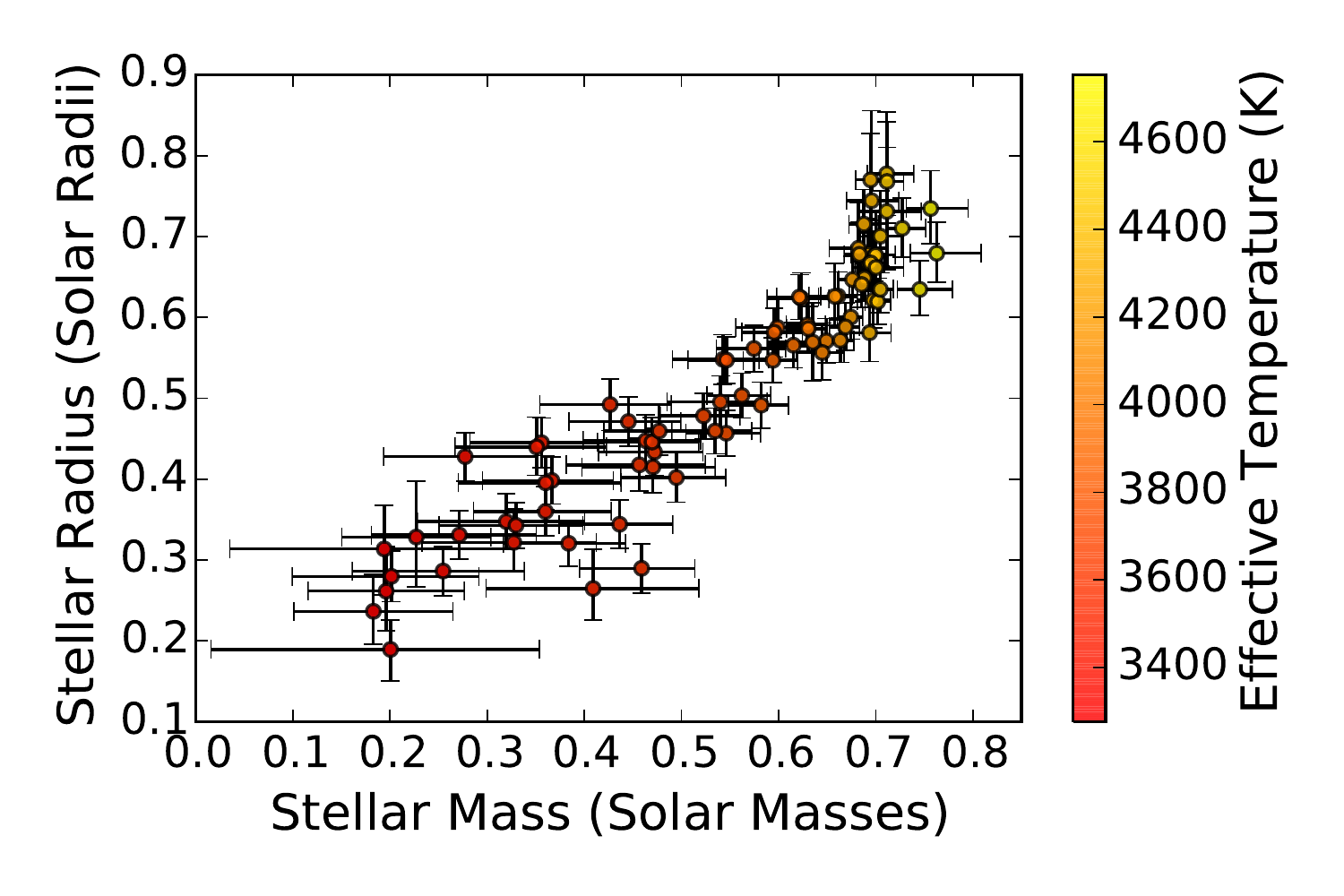}
\caption{Revised parameters for the cool dwarf sample. \emph{Left: } Revised stellar luminosity versus stellar effective temperature with points shaded according to revised stellar radii. \emph{Right: } Revised radii and masses with points shaded according to revised stellar effective temperatures.}
\label{fig:masslum}
\end{figure*}

The \citet{newton_et_al2015} relations do not include masses, so we computed the masses for all stars using the stellar effective temperature - mass relation from \citet{mann_et_al2013c}. The right panel of Figure~\ref{fig:masslum} displays the resulting mass estimates as a function of stellar radius. 

\subsection{Adopted Properties}
\label{ssec:starproperties}
After checking that the results from both classification schemes are generally consistent, we adopted parameters based on the \citet{newton_et_al2015} relations when possible because the calibrations are valid for hotter stars ($3100-4800$K versus $2700-4100$K), and because EWs are less susceptible to telluric contamination than the indices used by \citet{mann_et_al2013c}. Furthermore, the \citet{mann_et_al2013c} temperature calibrations have inflection points while the \citet{newton_et_al2015} relations do not.

Specifically, we report temperatures, radii, and luminosities estimated using the \citet{newton_et_al2015} relations, metallicities based on the \citet{mann_et_al2013a} relations,  masses generated by running the Newton temperatures through the temperature-mass relation from \citet{mann_et_al2013c}, and surface gravities computed from the radii and masses. (The exceptions are EPIC~211817229, EPIC~211799258, and~EPIC 211826814, for which we adopt the Mann parameters as explained in Section~\ref{sssec:rs}.) The \citet{newton_et_al2015} relations are not valid for early K dwarfs, so we rejected all of the stars with assigned temperatures hotter than 4800~K or radii larger than $0.8\rsun$. 

\begin{figure}[tbhp]
\centering
\includegraphics[width=0.5\textwidth]{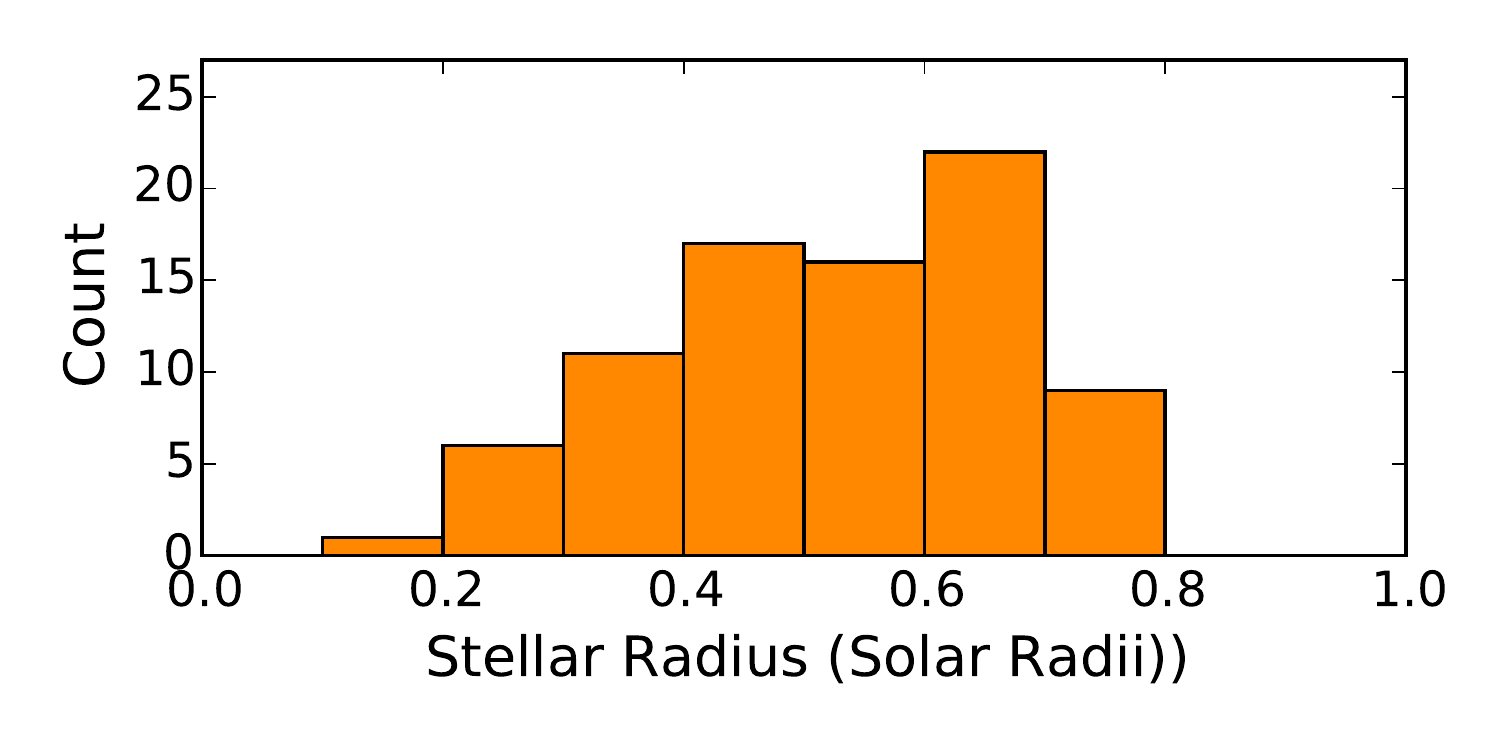}
\includegraphics[width=0.5\textwidth]{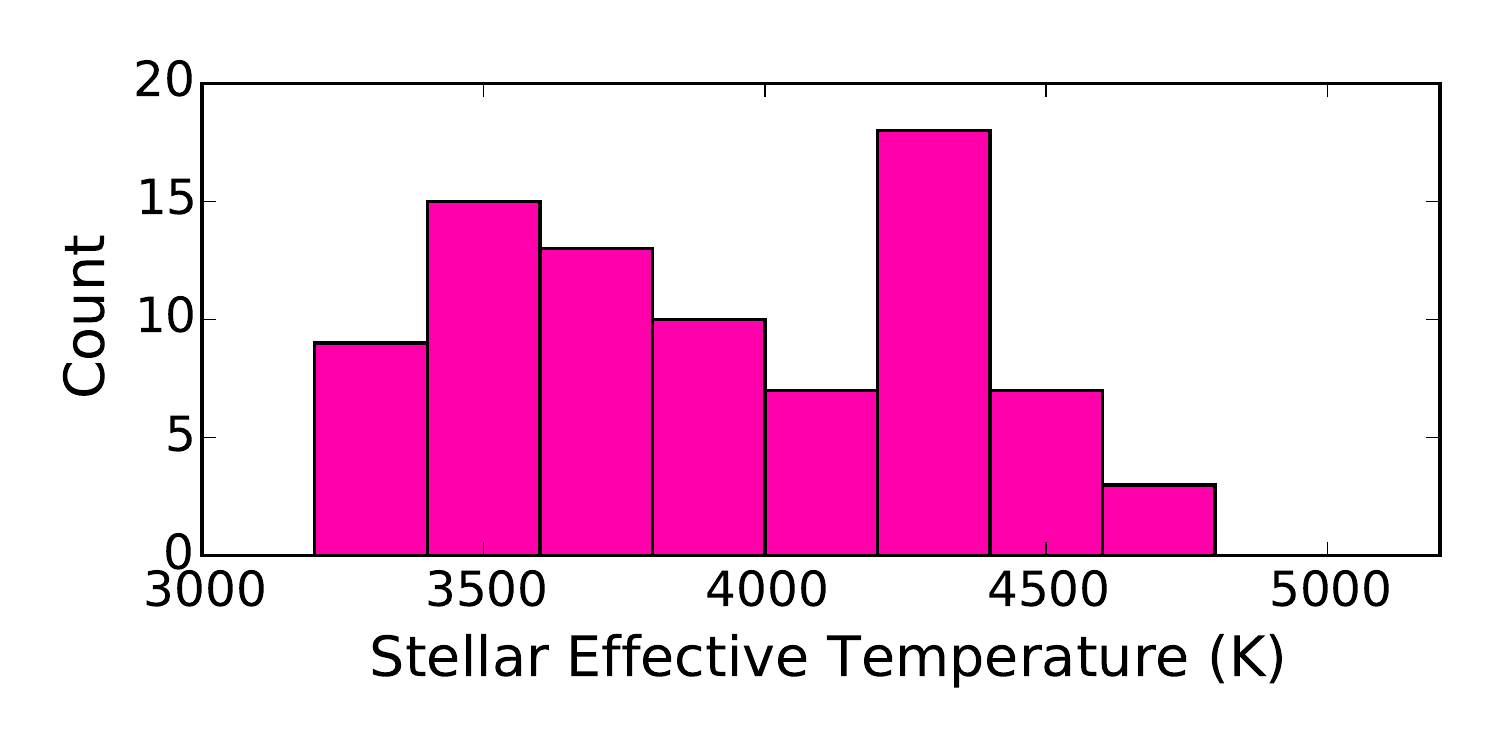}
\caption{Distribution of radii (top) and effective temperatures (bottom) for the stars in our cool dwarf sample.}
\label{fig:teff_rs_hist}
\end{figure}

As shown in the left panel of Figure~\ref{fig:teff_rs_hist}, our cool dwarf sample has a median radius of $0.56\rsun$. The temperature distribution in the right panel is bimodal, featuring a peak near 3500~K from the mid-M dwarfs in the sample and a second peak near 4350~K from late K~dwarfs. The median value of the distribution is 3884~K.

Our final cool dwarf sample consists of 74~stars in 72~systems; EPIC~211694226 and EPIC~212773309 are visual binaries. We obtained spectra of both components and consider all four stars as possible planetary host stars. As of 24~August 2016, there were no AO images of either system posted on the ExoFOP-K2 follow-up website. Using our data, we measured separations of roughly $1\farcs7$ and $11\farcs3$, respectively. The companion star to EPIC~212773309 is likely 2MASS~J13493168-0619267, which is listed on ExoFOP-K2 website\footnote{\mbox{\url{https://exofop.ipac.caltech.edu/k2/edit\_target.php?id=212773309}}} at a separation of $11\farcs4$. 2MASS~J13493168-0619267 is 2.6 $Kp$ magnitudes fainter than EPIC~212773309 and far enough away to lie outside the K2 target aperture. In contrast, both stars in the EPIC~211694226 system could fall within a single $3\farcs98$ K2 pixel.

The adopted parameters for the EPIC~211694226 and EPIC~212773309 visual binaries and all of the other stars in our cool dwarf sample are reported in Tables~\ref{tab:star_prop}. For reference, we also provide the intermediate measurements in Table~\ref{tab:ew+met} along with our metallicity estimates. 

\begin{figure*}[tbhp]
\centering
\includegraphics[width=0.32\textwidth]{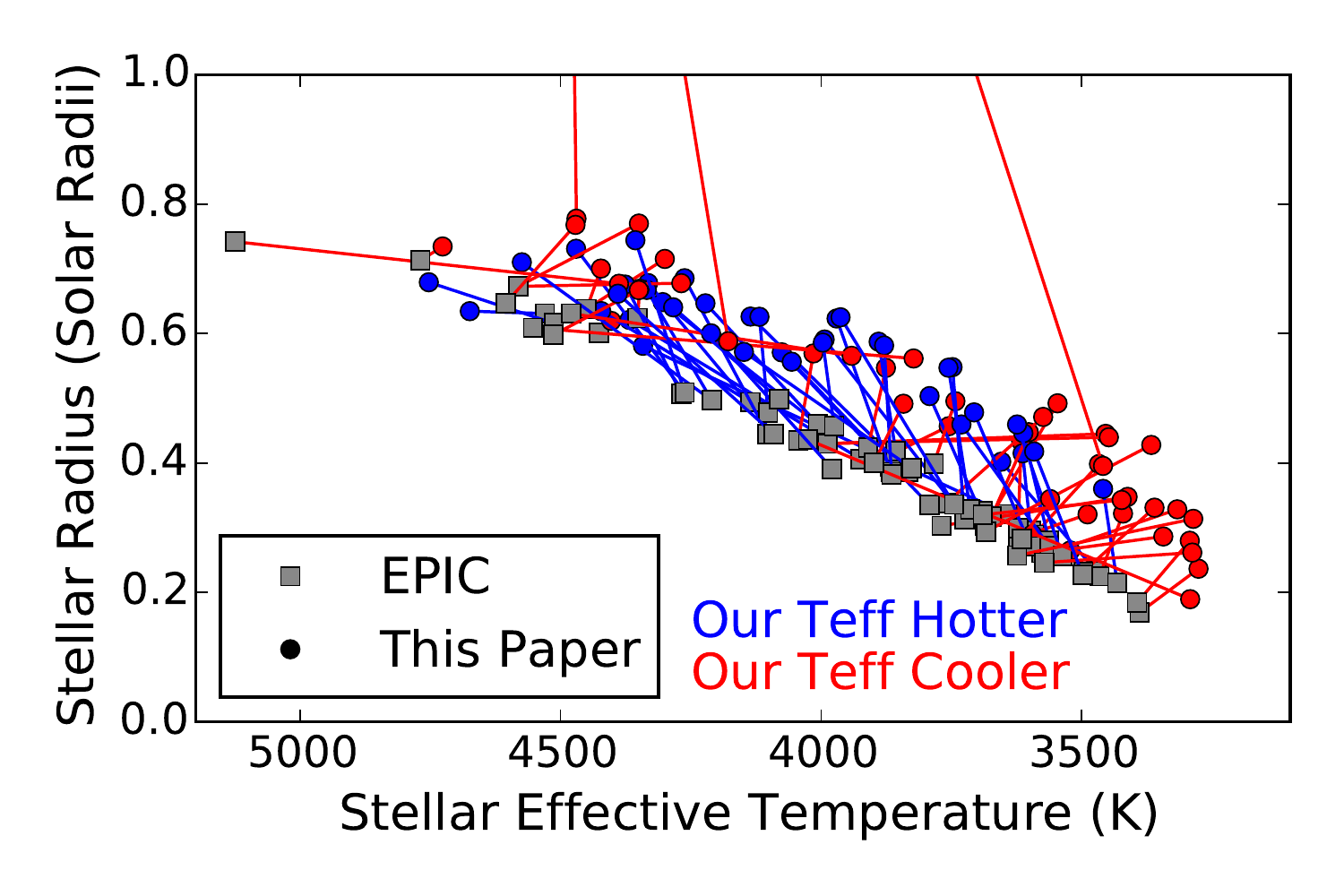}
\includegraphics[width=0.32\textwidth]{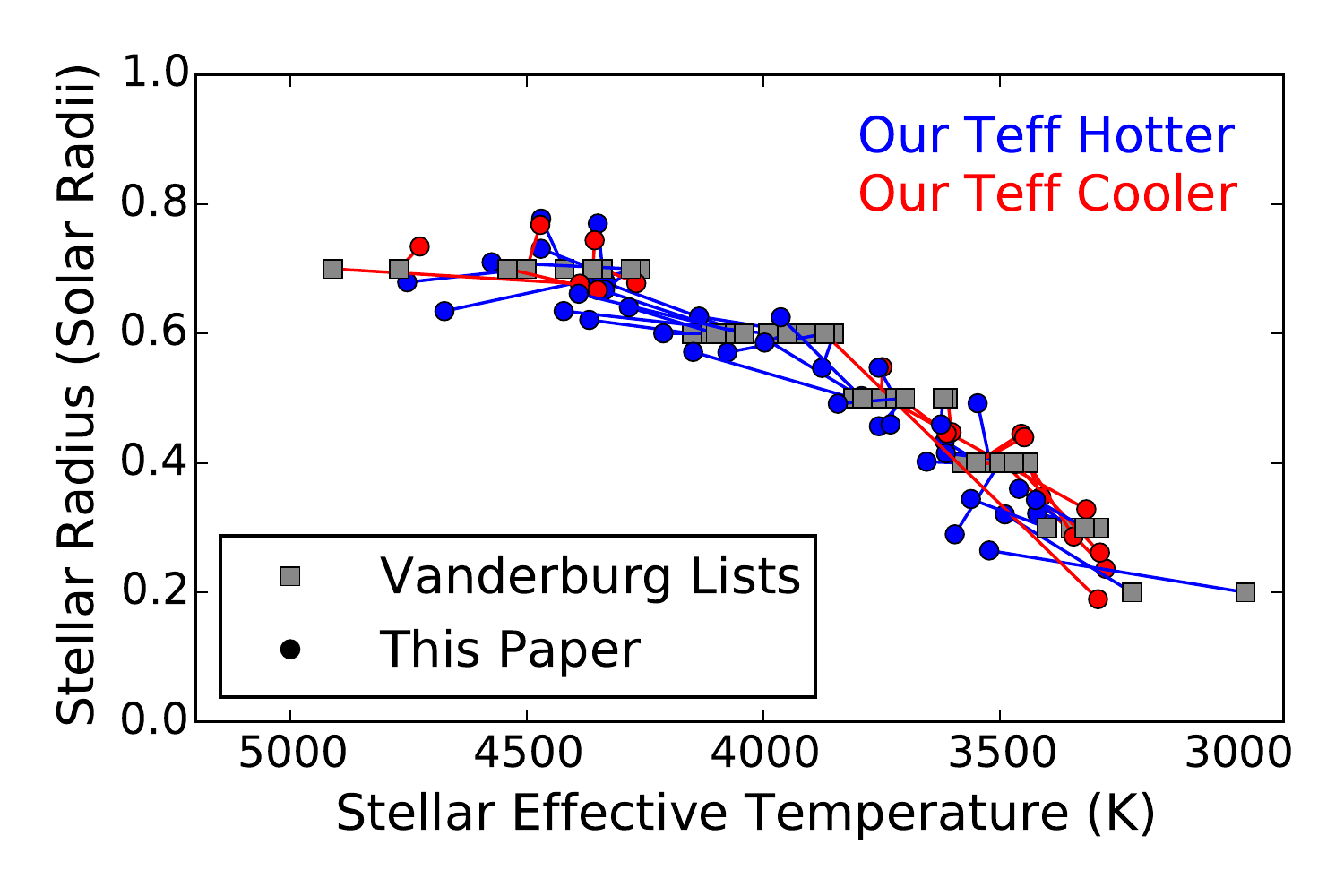}
\includegraphics[width=0.32\textwidth]{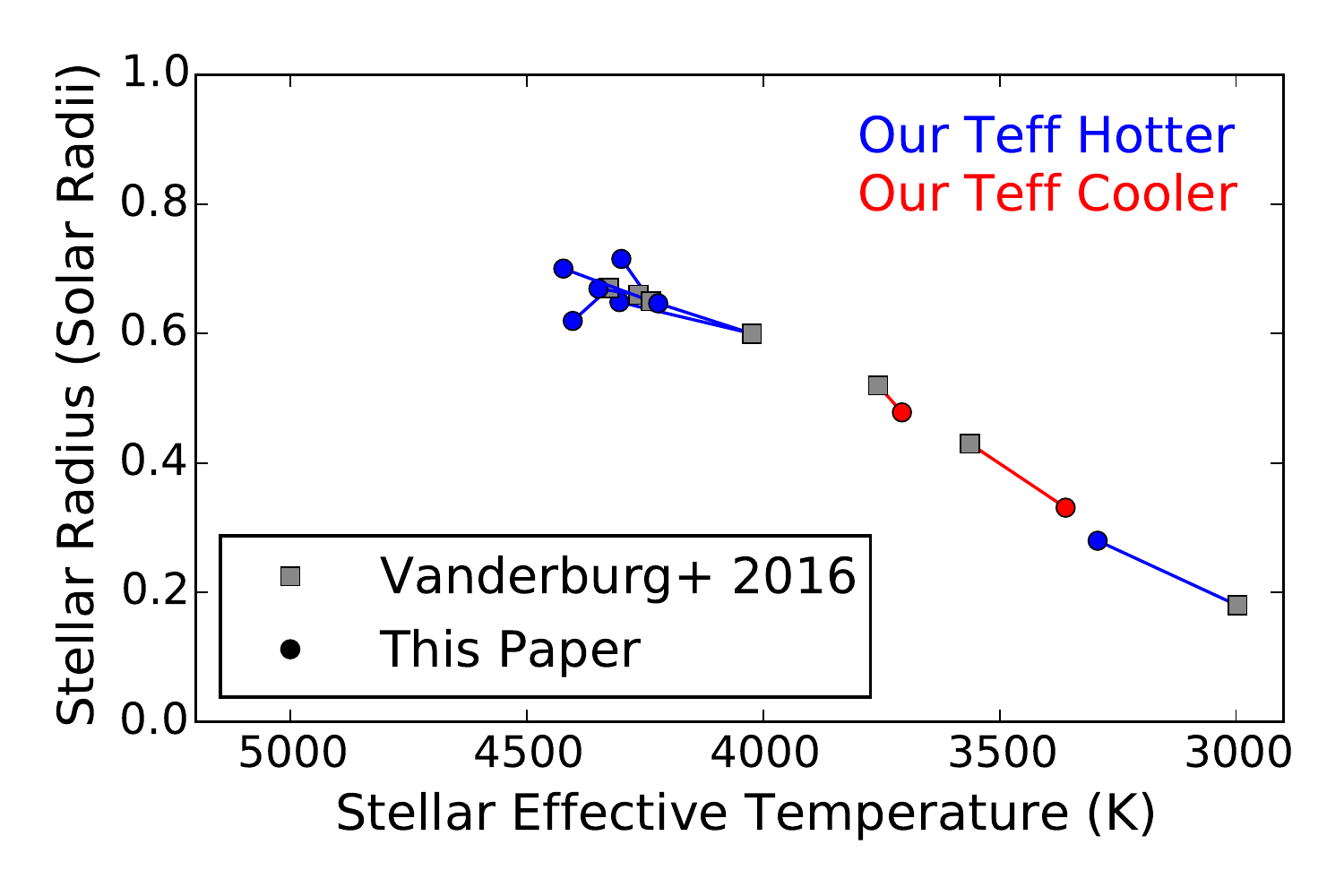}\\
\includegraphics[width=0.32\textwidth]{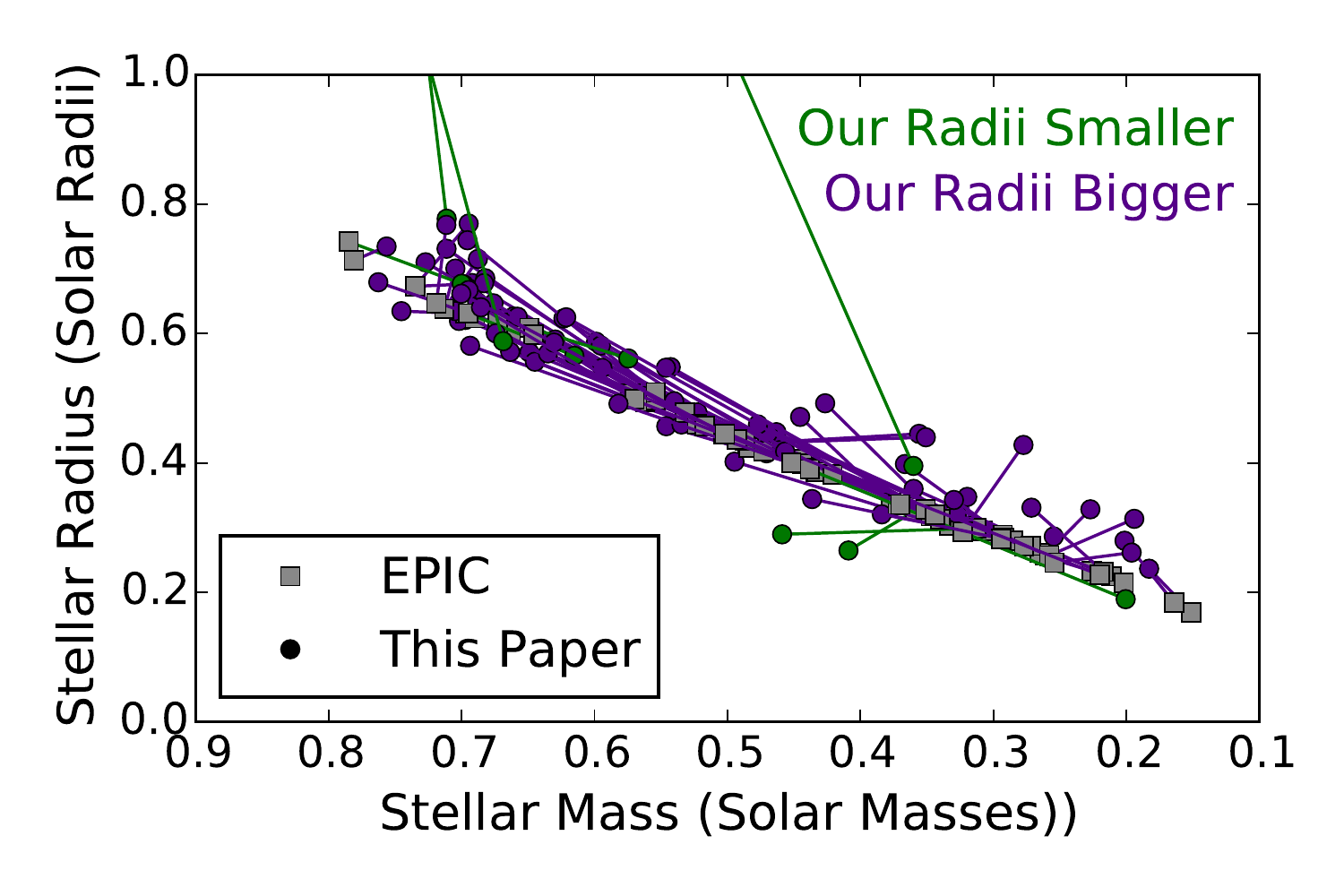}
\includegraphics[width=0.32\textwidth]{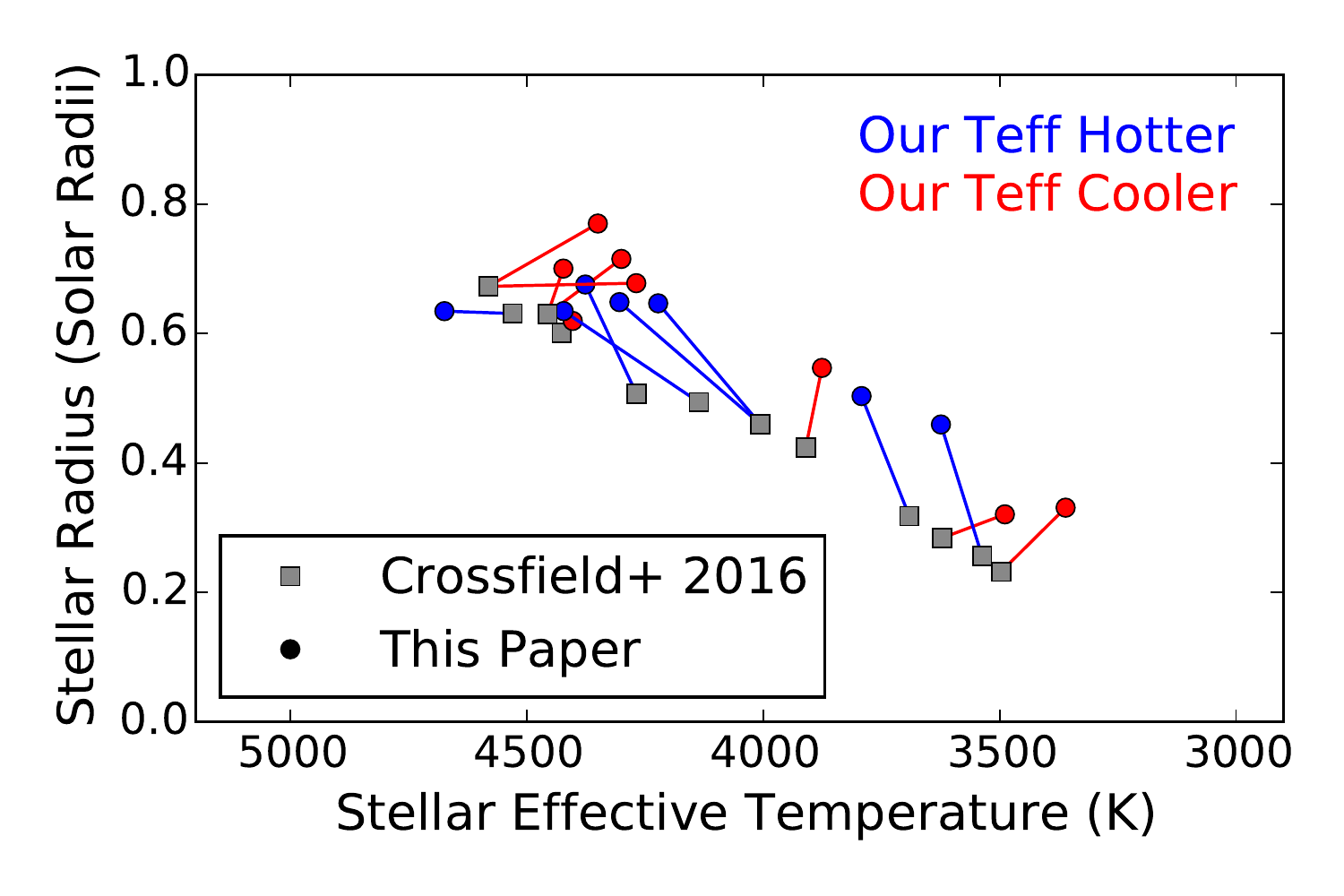}
\includegraphics[width=0.32\textwidth]{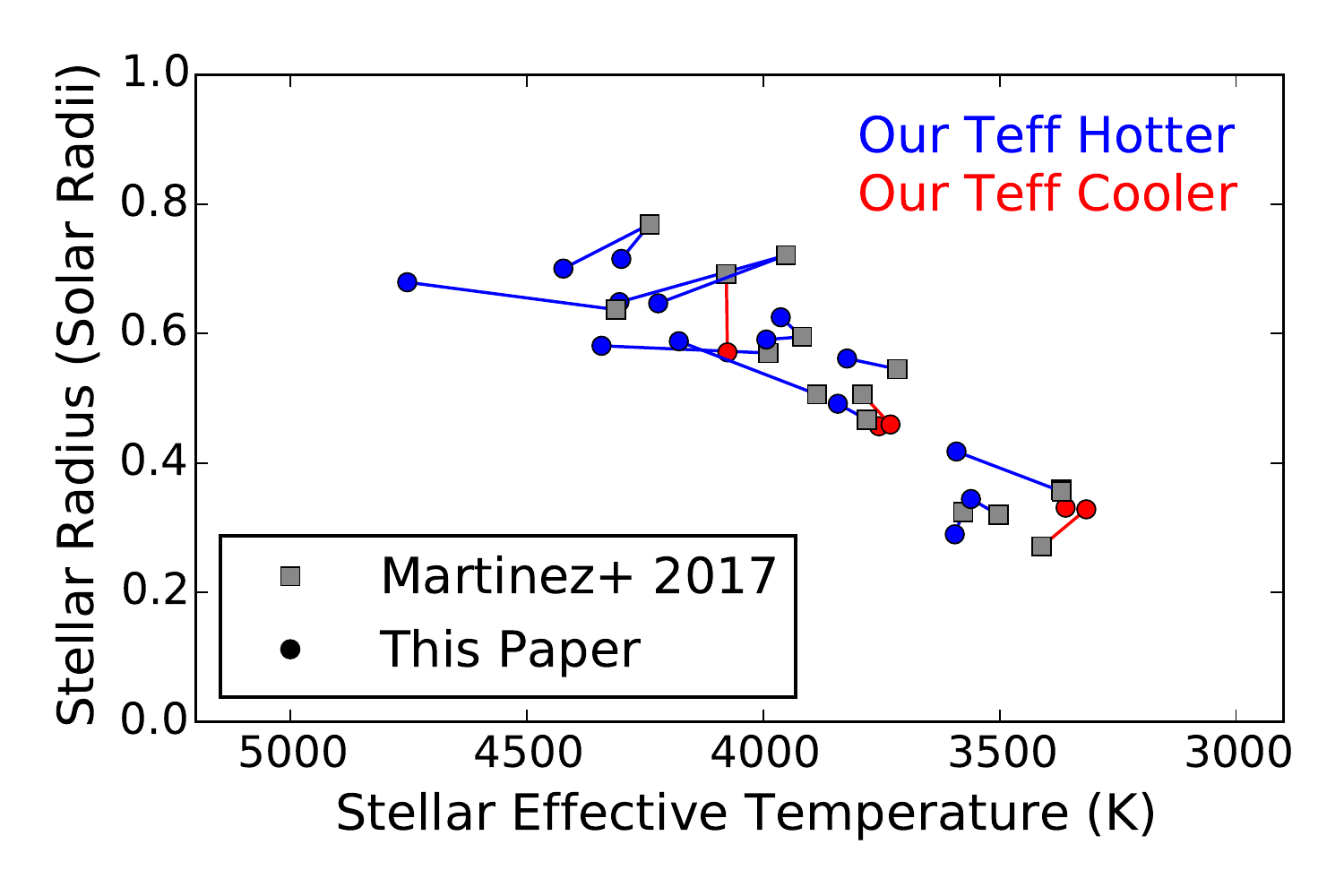}\\
\includegraphics[width=0.32\textwidth]{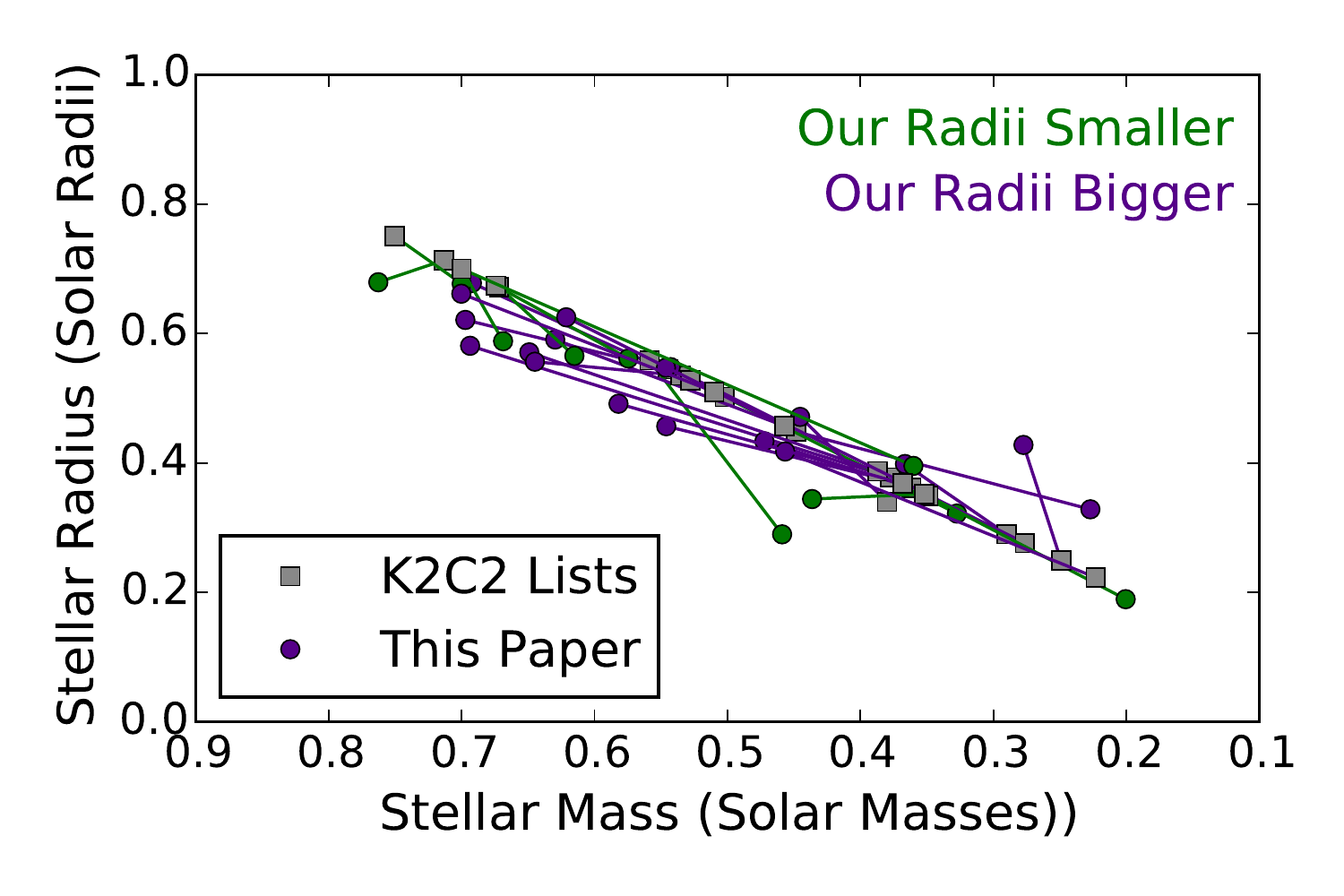}
\includegraphics[width=0.32\textwidth]{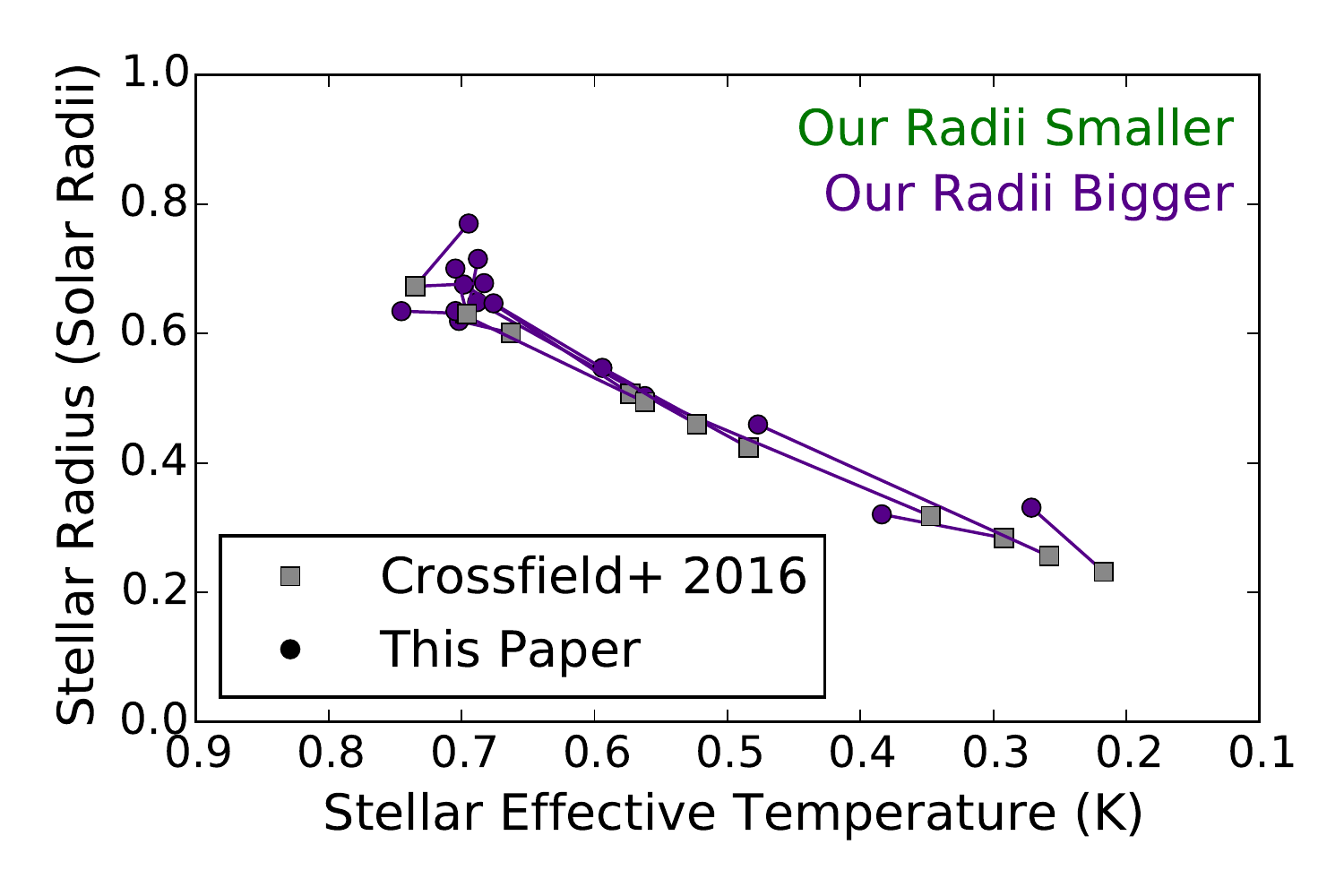}
\includegraphics[width=0.32\textwidth]{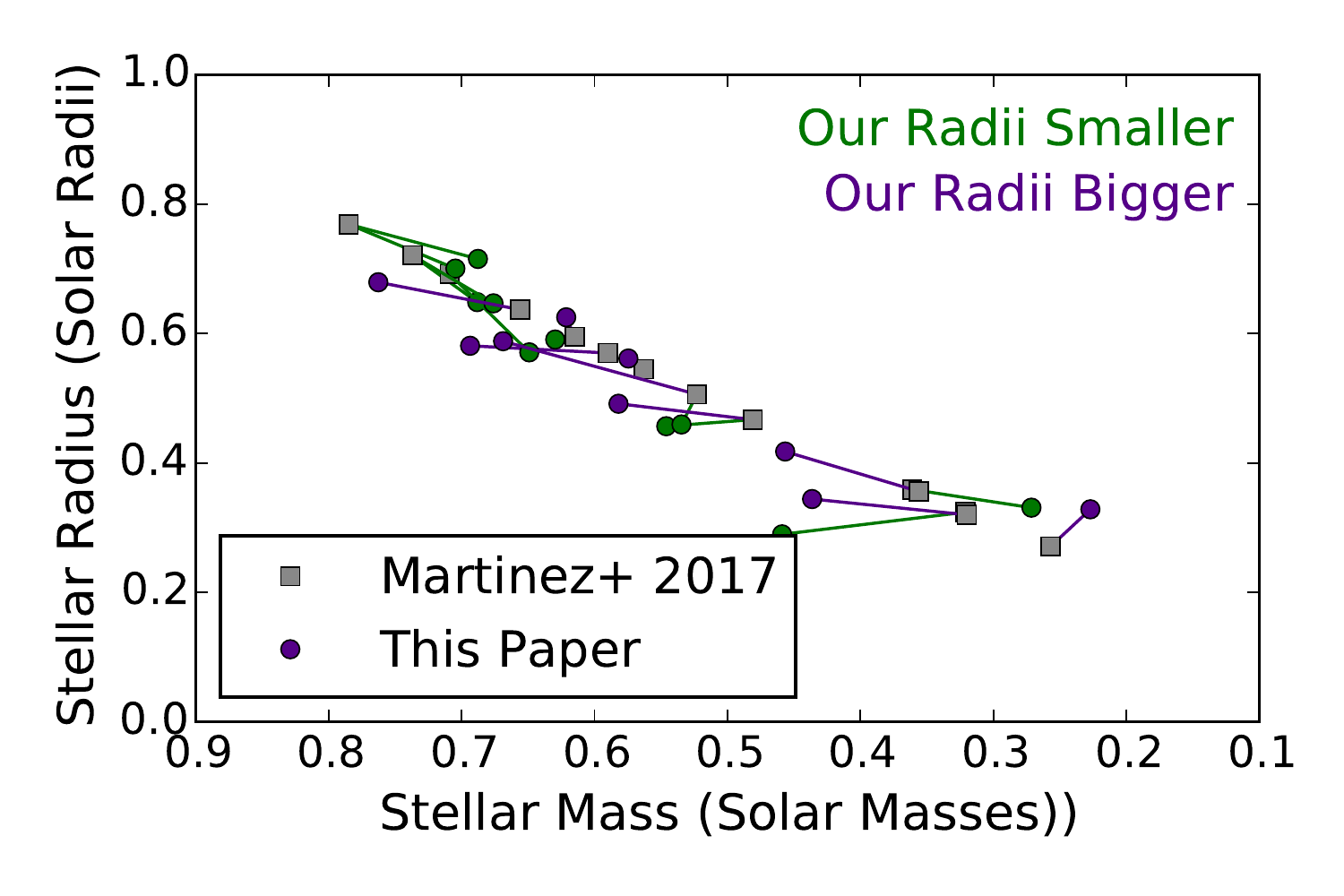}
\caption{Comparison of our revised stellar parameters (circles) to the earlier estimates from other studies (gray squares). Solid lines connect the before and after values for each star. \emph{Top Left:} Stellar radius versus effective temperature comparing values in the EPIC to our updated values. In this panel and in all other radius versus temperature panels, blue (red) lines connect the initial and revised values for stars for which our new effective temperature estimates are hotter (cooler). \emph{Top Center:}  Stellar radius versus stellar effective temperature comparing values in the unpublished planet candidate lists provided by A.~Vanderburg to our updated values. \emph{Top Right: } Stellar radius versus stellar effective temperature comparing values in \citet{vanderburg_et_al2016} to our updated values. \emph{Middle Center:} Stellar radius versus stellar effective temperature comparing values in \citet{crossfield_et_al2016} to our updated values. \emph{Middle Right: } Stellar radius versus stellar effective temperature comparing values in \citet{martinez_et_al2017} to our updated values. \emph{Bottom Left: } Stellar radius versus stellar mass comparing values in unpublished K2C2 planet candidate lists to our updated values. \emph{Bottom Center: } Stellar radius versus stellar mass comparing values in \citet{crossfield_et_al2016} to our updated values. \emph{Bottom Right: } Stellar radius versus stellar mass comparing values in \citet{martinez_et_al2017} to our updated values.}
\label{fig:star_changes}
\end{figure*}

As shown in Figure~\ref{fig:star_changes}, three of the stars in our cool dwarf sample were initially classified as giants in the EPIC. Considering only the stars originally classified as dwarfs, the median changes between our revised estimates and the EPIC values are \mbox{$+0.13\msun$ (+26\%)}, \mbox{$+0.13\rsun$ (+39\%)}, and \mbox{$-4$~K ($-0.1$\%).} For the 15~cool dwarfs with previous published estimates in \citet{crossfield_et_al2016}, we find median changes of  \mbox{$+0.09\msun$ (+23\%)}, \mbox{$+0.10\rsun$ (+28\%)}, and \mbox{$-23$~K ($-0.5$\%).} We find smaller radius changes \mbox{($+0.05\rsun$, +8\%)} but larger temperature changes \mbox{($+84$~K, +2\%)} for the nine~cool dwarfs with earlier estimates from \citet{vanderburg_et_al2016}. Consulting the unpublished planet candidate lists in which the stellar parameters are only coarsely estimated, we find median changes of \mbox{$+0.02\rsun$ (+4\%)} and \mbox{$+65$~K (+2\%)} for the 56 cool dwarfs in lists provided by A.~Vanderburg and \mbox{$+0.08\msun$ (+22\%)} and \mbox{$+0.07\rsun$ (+17\%)} for the 28~cool dwarfs in lists from the K2C2 Consortium. 

\citet{martinez_et_al2017} recently completed a parallel study in which they estimated the properties of low-mass K2 planet host stars using NTT/SOFI spectra covering the $0.95 - 2.52 \mu m$ wavelength range. Although their spectra are lower resolution than our data ($R \sim 1000$ rather than $R \sim 2000 - 2700$, they report consistent parameters for most of the 15~stars observed by both studies. Specifically, the median differences between our estimates (Dressing - Martinez) are 61K, $0.01\msun$, and $-0.004\rsun$.

\begin{figure}[tbp]
\centering
\includegraphics[width=0.45\textwidth]{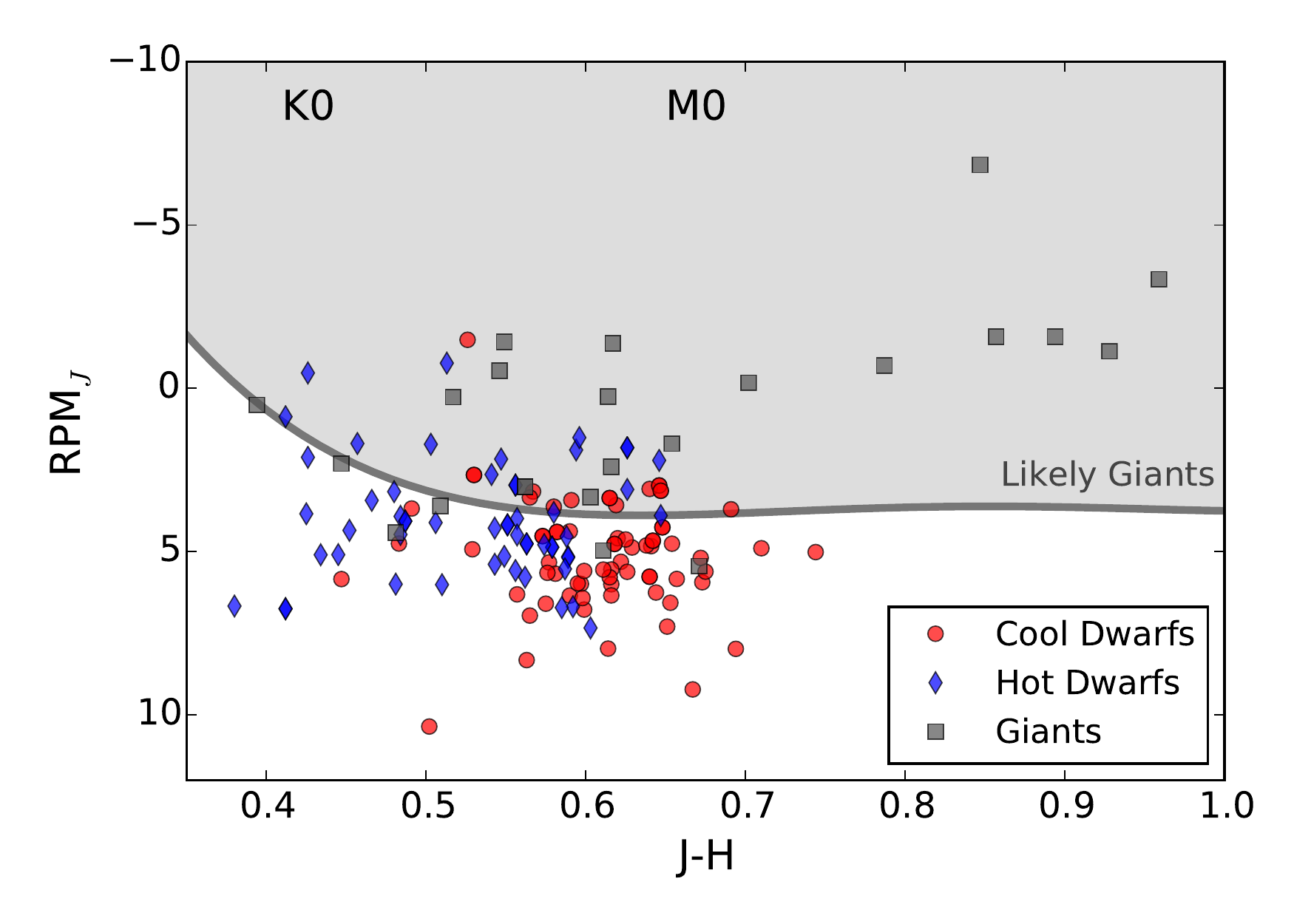}
\caption{Reduced proper motion in $J$-band versus $J-H$ for all of the stars we observed and later classified as giants (gray squares), hotter dwarfs (blue diamonds), or cool dwarfs (red circles). The gray line marks the dwarf/giant cut suggested by \citet{collier-cameron_et_al2007}; stars lying above this line (in the gray shaded region) are more likely to be giants while targets below the line are more likely to be dwarfs. For reference, we note the approximate $J-H$ colors of K0 and M0~stars.}
\label{fig:rpmj}
\end{figure}

\section{Discussion \& Conclusions}
\label{sec:conc}
In this paper, we presented revised system parameters for 144~targets observed by the NASA K2 mission. All of those stars were initially suspected to be cool dwarfs harbor transiting planets, but some of these systems have since been revealed to be false positives. Comparing our IRTF/SpeX and Palomar/TripleSpec spectra to standard spectra from the IRTF Spectral Library \citep{rayner_et_al2009}, we found that 49\% of our targets were contaminating giants or hotter dwarfs.

Intriguingly, one star (EPIC~211817229) has large proper motion \citep[380 mas/yr][]{roeser_et_al2010} and moderate radial velocity (28 km/s), indicating that the star likely does not belong to the thin disk population. Accordingly, we used the measured position, proper motion, and radial velocity of EPIC~211817229 along with an estimated photometric distance of $55 \pm 10$~pc \citep{pecaut+mamajek2013}, to calculate the star's UVW Galactic velocities corrected for the Sun's velocity \citep{coskunoglu_et_al2011}. We estimated $(UVW)_{\rm LSR} = (-11 \pm 9, -88 \pm 17, -15 \pm 8)$ km/s.
 
We then compared the Galactic velocities of EPIC~211817229 to distributions proposed in \citet{bensby_et_al2014} that approximately define the thin disk, thick disk, and halo populations. The total Galactic velocity of EPIC~211817229, $V_{\rm tot} = 90 \pm 21$~km/s, is consistent with the thick disk population ($V_{\rm tot}~70 - 180$ km/s). The placement of the star in a Toomre diagram and the estimated probability of membership in the three populations \citep[][Appendix A]{bensby_et_al2014} also point to a star in the  thick disk. This kinematic classification is consistent with EPIC~211817229 being metal poor and suggests an old age.

After classifying all of our targets, we revisited the initial selection of our sample to ask whether we could better identify low-mass stars in the future. As shown in the $J$-band reduced proper motion ($RPM_J$) versus $J-H$ color plot in Figure~\ref{fig:rpmj}, one possible avenue for improvement is to impose stricter cuts on the $J-H$ color and reduced proper motions of the target stars. For instance, confining our follow-up sample to stars with $0.45 < J-H < 0.8$ would have decreased the giant contamination by 30\% and hot dwarf contamination by 18\% while excluding only one cool dwarf from our sample. Imposing a further cut of $RPM_J > 1$ would decrease giant contamination by an additional 35\% and hot dwarf contamination by an additional 2\% at the cost of excluding two more cool dwarfs. Employing the more complicated polynomial cut suggested by \citet{collier-cameron_et_al2007} would remove 74\% of the giants and 29\% of the hot dwarfs along with 31\% of the cool dwarfs.

The main focus of this work was the sample of 74~cool dwarfs with spectral types between K3 and M4. For those stars, we estimated temperatures, radii, masses, luminosities, and metallicities using empirical relations \citep{mann_et_al2013a, mann_et_al2013c, mann_et_al2015, newton_et_al2015}. In most cases, we found that the original radius estimates were smaller than the actual radii of the stars: our revised estimates are typically $0.13\rsun$ (39\%) larger than the values reported in the EPIC \citep{huber_et_al2016}, $0.10\rsun$ (28\%) larger than the values in \citet{crossfield_et_al2016}, and $0.05\rsun$ (8\%) larger than those in \citet{vanderburg_et_al2016}. 

We defer a detailed discussion of the planetary implications of our revisions to the stellar parameters to the next paper in this series \citep{dressing_et_al2016b}, but assuming that the initial planet/star radius ratios are correct, we predict that the associated planet candidates are also $10 - 30\%$ larger than initially estimated. Accordingly, potentially habitable Earth-sized planets orbiting stars originally believed to be small, cool M~dwarfs may be larger and significantly less habitable than previously inferred. This result underscores the importance of characterizing TESS planet host stars before acquiring detailed atmospheric observations with JWST and the next-generation of extremely large ground-based telescopes.

\begin{acknowledgments}
Many of our targets were provided by the K2 California Consortium (K2C2). We thank K2C2 for sharing their candidate lists and vetting products. In particular, we thank K2C2 members Ian Crossfield and Arturo Martinez for their willingness to coordinate follow-up observations of low-mass stars. We are grateful to Michael Cushing for sharing a beta version of the Spextool pipeline designed for TripleSpec data. We thank Philip Muirhead and Juliette Becker for providing advice regarding TripleSpec data acquisition and reduction. We also acknowledge helpful conversations with Chas Beichman and Eric Gaidos. We thank Andrew Howard for donating SpeX time and Kimberly Aller, Will Best, and Evan Sinukoff for obtaining some of the SpeX observations described in this paper. Finally, we thank the anonymous referee for providing feedback that improved the quality of this paper.
 
This work was performed under contract with the Jet Propulsion Laboratory (JPL) funded by NASA through the Sagan Fellowship Program executed by the NASA Exoplanet Science Institute. This publication was made possible through the support of a grant from the John Templeton Foundation. The opinions expressed here are those of the authors and do not necessarily reflect the views of the John Templeton Foundation. This paper includes data collected by the K2 mission, which is funded by the NASA Science Mission directorate. Our follow-up observations were obtained at the Infrared Telescope Facility, which is operated by the University of Hawaii under contract NNH14CK55B with the National Aeronautics and Space Administration and at Palomar Observatory. We thank the staff at both observatories and the Caltech Remote Observing Facilities staff for supporting us during our many observing runs. We are grateful to the IRTF and Caltech TACs for awarding us telescope time. This research has made use of the NASA Exoplanet Archive, which is operated by the California Institute of Technology, under contract with the National Aeronautics and Space Administration under the Exoplanet Exploration Program.

The authors wish to recognize and acknowledge the very significant cultural role and reverence that the summit of Mauna Kea has always had within the indigenous Hawaiian community.  We are most fortunate to have the opportunity to conduct observations from this mountain. 
\end{acknowledgments}

\facilities{IRTF (SpeX), Palomar (TripleSpec)}

\bibliography{../../mdwarf_biblio.bib}
\clearpage

\begin{deluxetable*}{lcccccccccc}
\tablecolumns{11}
\tabletypesize{\small}
\tablecaption{Observations of K2 Targets Classified as Giant Stars \label{tab:giants}}
\tablehead{
\colhead{} &
\multicolumn{2}{c}{Observation } &
\colhead{Spectral} &
\colhead{} &
\multicolumn{6}{c}{EPIC Classification}\\
\cline{2-3}
\cline{6-8}
\cline{9-11}
\colhead{EPIC} &
\colhead{Date} &
\colhead{Instru}  &
\colhead{Type\tablenotemark{1}} &
\colhead{Campaign} &
\colhead{$T_{\rm eff}$ (K)} &
\colhead{ep\_$T_{\rm eff}$} &
\colhead{em\_$T_{\rm eff}$} &
\colhead{logg (cgs)} &
\colhead{ep\_logg} &
\colhead{em\_logg} 
}
\startdata
 202710713 &  Aug-07-2015 &   SpeX &   K4III &    2 &      3817 &            92 &            92 &     0.523 &         0.168 &         0.168 \\
 203485624 &   Jun-7-2016 &   SpeX &   F2III &    2 &      6237 &           449 &           187 &     3.848 &         0.228 &         0.020 \\
 203776696 &  Mar-27-2016 &  TSPEC &  F8III &    2 &      6113 &          1219 &           508 &     4.143 &         0.270 &         0.315 \\
 205064326 &   Jun-7-2016 &   SpeX &   K0III &    2 &      4734 &            75 &            75 &     2.946 &         0.144 &         0.144 \\
 206049452 &  Sep-24-2015 &   SpeX &   M2III &    3 &      4553 &           191 &           109 &     4.671 &         0.035 &         0.042 \\
 210769880 &  Sep-24-2015 &   SpeX &   K2III &    4 &      4018 &           118 &           802 &     4.809 &         2.400 &         0.060 \\
 210843708 &  Sep-24-2015 &   SpeX &   K3III &    4 &      4823 &           120 &            90 &     2.456 &         0.075 &         0.450 \\
 211098117 &  Sep-24-2015 &   SpeX &   K0III &    4 &      3858 &           186 &           186 &     4.870 &         0.070 &         0.084 \\
 211106187 &  Nov-27-2015 &   SpeX &   G5III &    4 &      5321 &            96 &           192 &     4.561 &         0.164 &         0.020 \\
 211351816 &  Nov-27-2015 &   SpeX &   K2III &    5 &      4742 &            96 &            76 &     2.984 &         0.483 &         0.345 \\
 212311834 &  Apr-18-2016 &  TSPEC &   M1III &    6 &      5199 &           156 &           188 &     3.631 &         0.890 &         0.890 \\
 212443457 &   Mar-8-2016 &   SpeX &   K0III &    6 &      4804 &           144 &           173 &     4.598 &         0.025 &         0.030 \\
 212443457 &   Jun-7-2016 &   SpeX &   K0III &    6 &      4804 &           144 &           173 &     4.598 &         0.025 &         0.030 \\
 212473154 &   Jun-7-2016 &   SpeX &   K0III &    6 &      4570 &           136 &           136 &     2.365 &         0.682 &         0.186 \\
 212586030 &   Mar-8-2016 &   SpeX &   K1III &    6 &      4814 &            76 &            76 &     3.328 &         0.144 &         0.144 \\
 212644491 &  Apr-18-2016 &  TSPEC &   K1III &    6 &      4940 &            96 &            96 &     2.505 &         0.306 &         0.663 \\
 212786391 &  Mar-27-2016 &  TSPEC &   G5III &    6 &      4688 &           109 &            73 &     2.164 &         0.912 &         0.570 \\
 214629283 &   May-5-2016 &   SpeX &   M3III &    7 &      3508 &           150 &           150 &     0.241 &         0.310 &         0.558 \\
 214799621 &   May-5-2016 &   SpeX &   K4III &    7 &      4375 &           132 &           132 &     2.184 &         0.360 &         0.216 \\
 215030652 &   Jun-7-2016 &   SpeX &   M0III &    7 &      3935 &            79 &            79 &     0.778 &         0.250 &         0.300 \\
 215090200 &   May-5-2016 &   SpeX &   K0III &    7 &      4596 &           115 &           172 &     2.422 &         0.145 &         0.203 \\
 215174656 &   May-6-2016 &   SpeX &   K7III &    7 &      3814 &            92 &           115 &     0.538 &         0.150 &         0.150 \\
 215346008 &   Jun-7-2016 &   SpeX &   K4III &    7 &      4038 &           165 &           132 &     1.357 &         1.216 &         0.228 \\
 218006248 &   May-5-2016 &   SpeX &   M2III &    7 &      3330 &            33 &            33 &     0.088 &         0.070 &         0.182 \\
 \enddata
 \tablenotetext{1}{Spectral types are coarse assignments based on visual inspection of the near-infrared spectra collected in this paper. The assigned spectral types have errors of roughly $\pm1$ subtype. (See Section~\ref{ssec:initial_class} for details.)}
\end{deluxetable*}

\begin{deluxetable*}{lcccccccccc}
\tablecolumns{11}
\tabletypesize{\small}
\tablecaption{Observations of K2 Targets Classified as Hotter Dwarfs \label{tab:hot_dwarfs}}
\tablehead{
\colhead{} &
\multicolumn{2}{c}{Observation } &
\colhead{Spectral} &
\colhead{} &
\multicolumn{6}{c}{EPIC Classification}\\
\cline{2-3}
\cline{6-8}
\cline{9-11}
\colhead{EPIC} &
\colhead{Date} &
\colhead{Instru}  &
\colhead{Type\tablenotemark{1}} &
\colhead{Campaign} &
\colhead{$T_{\rm eff}$ (K)} &
\colhead{ep\_$T_{\rm eff}$} &
\colhead{em\_$T_{\rm eff}$} &
\colhead{logg (cgs)} &
\colhead{ep\_logg} &
\colhead{em\_logg} 
}
 \startdata
 201754305 &  Jun-13-2015 &   SpeX &    K3V\tablenotemark{2} &    1 &      4755 &           113 &           113 &     4.642 &         0.045 &         0.045 \\
 204890128 &  Mar-27-2016 &  TSPEC &    K2V &    2 &      5213 &           188 &           707 &     3.848 &         0.535 &         0.535 \\
 205084841 &  Mar-27-2016 &  TSPEC &    K0V &    2 &      4793 &           207 &           207 &     2.369 &         0.205 &         0.656 \\
 205145448 &   Jun-7-2016 &   SpeX &    G5V &    2 &      5700 &           390 &            57 &     3.841 &         1.362 &         0.020 \\
 205145448 &   May-5-2016 &   SpeX &    G5V &    2 &      5700 &           390 &            57 &     3.841 &         1.362 &         0.020 \\
 205686202 &   May-5-2016 &   SpeX &    K1V &    2 &      3809 &            68 &          1432 &     4.889 &         0.399 &         0.084 \\
 206055981 &  Oct-26-2016 &   SpeX &    K3V\tablenotemark{2} &    3 &      4522 &            45 &            73 &     4.668 &         0.028 &         0.024 \\
 206055981 &  Nov-26-2015 &   SpeX &    K3V\tablenotemark{2} &    3 &      4522 &            45 &            73 &     4.668 &         0.028 &         0.024 \\
 206056433 &  Oct-26-2016 &   SpeX &    K4V\tablenotemark{2} &    3 &      4506 &           109 &            54 &     4.666 &         0.025 &         0.045 \\
 206056433 &  Nov-26-2015 &   SpeX &    K4V\tablenotemark{2} &    3 &      4506 &           109 &            54 &     4.666 &         0.025 &         0.045 \\
 206096602 &  Aug-07-2015 &   SpeX &    K3V\tablenotemark{2} &    3 &      4617 &           138 &           138 &     4.649 &         0.030 &         0.036 \\
 206096602 &  Sep-24-2015 &   SpeX &    K3V\tablenotemark{2} &    3 &      4617 &           138 &           138 &     4.649 &         0.030 &         0.036 \\
 206135267 &  Sep-24-2015 &   SpeX &    K2V &    3 &      5165 &           123 &           215 &     3.678 &         0.286 &         0.130 \\
 206144956 &  Sep-24-2015 &   SpeX &    K2V &    3 &      4848 &            78 &            97 &     4.611 &         0.025 &         0.025 \\
210414957\tablenotemark{3} &  Nov-26-2015 &   SpeX &    G2V &    4 &      5404 &           107 &            86 &     3.779 &         0.196 &         0.020 \\
 210423938 &  Nov-27-2015 &   SpeX &    K3V\tablenotemark{2} &    4 &      4856 &           114 &           171 &     2.876 &         0.582 &         0.485 \\
 210577548 &  Nov-26-2015 &   SpeX &    K2V &    4 &       $\cdots$ &   $\cdots$ &       $\cdots$ &     $\cdots$ &     $\cdots$ &      $\cdots$ \\
 210609658 &  Sep-24-2015 &   SpeX &    K2V &    4 &      4963 &            97 &            97 &     3.268 &         0.416 &         0.260 \\
 210731500 &  Nov-27-2015 &   SpeX &    K1V &    4 &      5406 &           168 &           168 &     4.472 &         0.476 &         0.068 \\
 210754505 &  Nov-26-2015 &   SpeX &    G5V &    4 &      6041 &           120 &           120 &     4.224 &         0.168 &         0.140 \\
 210793570 &  Nov-26-2015 &   SpeX &    K3V\tablenotemark{2} &    4 &      4896 &           118 &           118 &     3.242 &         0.609 &         0.435 \\
 210852232 &  Nov-27-2015 &   SpeX &    K0V &    4 &      5437 &           167 &           301 &     4.527 &         0.384 &         0.040 \\
 211058748 &  Nov-27-2015 &   SpeX &    K2V &    4 &      5070 &            81 &           243 &     4.615 &         0.060 &         0.110 \\
 211133138 &  Nov-26-2015 &   SpeX &    K2V &    4 &      5742 &           367 &           275 &     3.965 &         0.150 &         0.500 \\
 211418290 &  Nov-27-2015 &   SpeX &    G5V &    5 &      5182 &           126 &           126 &     2.461 &         0.055 &         1.111 \\
 211529065 &  Mar-28-2016 &  TSPEC &    K4V\tablenotemark{2} &    5 &      4742 &           167 &           167 &     4.621 &         0.036 &         0.030 \\
 211579683 &  Mar-28-2016 &  TSPEC &    K3V\tablenotemark{2} &    5 &      4829 &            57 &            76 &     3.432 &         1.045 &         1.254 \\
 211619879 &   Mar-4-2016 &   SpeX &    K3V\tablenotemark{2} &    5 &      4403 &           303 &           216 &     4.706 &         0.045 &         0.081 \\
 211779390 &  Nov-26-2015 &   SpeX &    K3V\tablenotemark{2} &    5 &      4472 &           122 &            87 &     4.705 &         0.065 &         0.195 \\
 211783206 &  Mar-28-2016 &  TSPEC &    K5V\tablenotemark{2} &    5 &      4855 &            94 &            94 &     3.324 &         0.655 &         1.310 \\
 211796070 &   Mar-4-2016 &   SpeX &    K3V\tablenotemark{2} &    5 &      4564 &            91 &            91 &     4.665 &         0.025 &         0.035 \\
 211797637 &  Mar-27-2016 &  TSPEC &    K5V\tablenotemark{2} &    5 &      4521 &           108 &           135 &     4.696 &         0.055 &         0.121 \\
 211913977 &  Nov-27-2015 &   SpeX &    K3V\tablenotemark{2} &    5 &      4825 &            58 &            77 &     4.607 &         0.025 &         0.040 \\
 211970147 &   Mar-8-2016 &   SpeX &    K3V\tablenotemark{2} &    5 &      4576 &            54 &            72 &     4.667 &         0.035 &         0.025 \\
 212012119 &  Nov-27-2015 &   SpeX &    K3V\tablenotemark{2} &    5 &      4837 &            78 &            58 &     3.178 &         0.715 &         0.325 \\
 212132195 &  Nov-27-2015 &   SpeX &    K3V\tablenotemark{2} &    5 &      4631 &            75 &           112 &     4.656 &         0.036 &         0.020 \\
 212138198 &  Nov-27-2015 &   SpeX &    K3V\tablenotemark{2} &    5 &      4975 &            99 &           139 &     4.577 &         1.218 &         0.030 \\
 212315941 &  Mar-28-2016 &  TSPEC &    K3V\tablenotemark{2} &    6 &      4909 &            78 &           118 &     4.628 &         0.025 &         0.040 \\
 212470904 &   Mar-8-2016 &   SpeX &    K5V\tablenotemark{2} &    6 &      4761 &            97 &            97 &     4.617 &         0.042 &         0.030 \\
 212521166\tablenotemark{4} &  Mar-10-2016 &   SpeX &    K2V &    6 &      4841 &           145 &           174 &     4.628 &         0.030 &         0.025 \\
 212525174 &  Mar-27-2016 &  TSPEC &    K4V\tablenotemark{2} &    6 &      4163 &            41 &           100 &     4.876 &         0.084 &         0.020 \\
 212530118 &   Mar-4-2016 &   SpeX &    K5V\tablenotemark{2} &    6 &      4175 &            41 &            49 &     4.824 &         0.045 &         0.108 \\
 212532636 &  Mar-28-2016 &  TSPEC &    K3V\tablenotemark{2} &    6 &      4519 &           109 &            73 &     4.698 &         0.030 &         0.042 \\
 212572439 &  Mar-10-2016 &   SpeX &    K2V &    6 &      4972 &            59 &            49 &     4.593 &         0.020 &         0.039 \\
 212572439 &  Mar-27-2016 &  TSPEC &    K2V &    6 &      4972 &            59 &            49 &     4.593 &         0.020 &         0.039 \\
 212730483 &   Mar-4-2016 &   SpeX &    K3V\tablenotemark{2} &    6 &      4612 &            55 &            55 &     4.657 &         0.040 &         0.020 \\
 212737443 &  Mar-28-2016 &  TSPEC &    K3V\tablenotemark{2} &    6 &      4542 &           298 &           149 &     4.708 &         0.040 &         0.088 \\
 212756297 &  Mar-10-2016 &   SpeX &    K5V\tablenotemark{2} &    6 &      4429 &            78 &           131 &     4.729 &         0.078 &         0.104 \\
 212757039 &  Apr-18-2016 &  TSPEC &    K1V &    6 &      5510 &           223 &           223 &     4.574 &         0.088 &         0.066 \\
 212779596 &   Jun-7-2016 &   SpeX &    K5V\tablenotemark{2} &    6 &      4731 &            77 &            77 &     4.623 &         0.036 &         0.036 \\
 212779596 &   Mar-8-2016 &   SpeX &    K5V\tablenotemark{2} &    6 &      4731 &            77 &            77 &     4.623 &         0.036 &         0.036 \\
 214173069 &  Oct-26-2016 &   SpeX &    K3V\tablenotemark{2} &    7 &      4659 &           150 &            75 &     4.633 &         0.035 &         0.025 \\
 214173069 &   May-6-2016 &   SpeX &    K3V\tablenotemark{2} &    7 &      4659 &           150 &            75 &     4.633 &         0.035 &         0.025 \\
 216111905 &   May-6-2016 &   SpeX &    G8V &    7 &      5221 &           126 &            84 &     4.543 &         0.760 &         0.040 \\
 217192839 &   May-6-2016 &   SpeX &    K2V &    7 &      4563 &            89 &           107 &     4.682 &         0.042 &         0.133 \\
 219114906 &   May-6-2016 &   SpeX &    K2V &    7 &      4523 &           108 &            90 &     4.662 &         0.030 &         0.042 \\
\enddata
 \tablenotetext{1}{Spectral types are coarse assignments based on visual inspection of the near-infrared spectra collected in this paper. The assigned spectral types have errors of roughly $\pm1$ subtype. (See Section~\ref{ssec:initial_class} for details.)}
 \tablenotetext{2}{In general, we list stars with spectral types of K3V or later in the cool dwarf sample rather than the hotter dwarf sample. However, these stars had an estimated temperatures $>4800$~K or estimated radii $>0.8\rsun$, which are beyond the validity range of the \citet{newton_et_al2015} relations.}
 \tablenotetext{3}{Possible fainter nearby star identified in Gemini AO image acquired by D.~Ciardi.\footnote{\url{https://exofop.ipac.caltech.edu/k2/edit\_target.php?id=210414957}}} 
\tablenotetext{4}{Characterized by \citet{osborn_et_al2016} as a K3 dwarf with $M_\star = 0.739 \pm 0.017\msun$, $R_\star = 0.713 \pm 0.020 \rsun$, $T_{\rm eff} = 5010 \pm 48$~K, and [Fe/H]~$=-0.343 \pm 0.032$.} 
\end{deluxetable*}

\begin{deluxetable*}{lccccccc}
\tablecolumns{8}
\tabletypesize{\normalsize}
\tablecaption{Observation Dates, Spectral Types, \& Radial Velocities for Stars Classified as Cool Dwarfs \label{tab:cdwarfs_st}}
\tablehead{
\colhead{} &
\colhead{} & 
\multicolumn{2}{c}{Observation} &
\colhead{Spectral} &
\multicolumn{2}{c}{H$_2$O-K2 } &
\colhead{RV\tablenotemark{4} } \\
\cline{3-4}
\cline{6-7}
\colhead{EPIC} &
\colhead{Campaign} &
\colhead{Date} & 
\colhead{Instru} &
\colhead{Type\tablenotemark{1}} & 
\colhead{Index\tablenotemark{2}} &
\colhead{SpType\tablenotemark{3}} & 
\colhead{(km/s)}
}
\startdata
201205469 &     1 &  Jun-13-2015 &   SpeX &    K7V &   1.03 &   0.39 &    -4.0 \\
  201208431 &     1 &  May-05-2015 &   SpeX &    K7V &   1.04 &   0.17 &    16.4 \\
  201345483 &     1 &  May-05-2015 &   SpeX &    M0V &   1.03 &   0.49 &    4.5 \\
  201549860 &     1 &  Nov-26-2015 &   SpeX &    K4V &   1.03 &   0.49 &    54.7 \\
  201617985 &     1 &  Apr-16-2015 &   SpeX &    M1V &   1.01 &   0.93 &    4.4 \\
  201635569 &     1 &  May-05-2015 &   SpeX &    M0V &   1.02 &   0.67 &    6.6 \\
  201637175 &     1 &  May-05-2015 &   SpeX &    K7V &   1.01 &   1.02 &    -8.4 \\
  201717274 &     1 &  May-05-2015 &   SpeX &    M2V &   0.89 &   3.93 &    43.1 \\
  201855371 &     1 &  Apr-16-2015 &   SpeX &    K5V &   1.02 &   0.65 &    -11.9 \\
  205924614 &     3 &  Sep-24-2015 &   SpeX &    K7V &   1.00 &   1.24 &    0.9 \\
  205924614 &     3 &  Nov-26-2015 &   SpeX &    K7V &   1.02 &   0.78 &    4.4 \\
  206011691 &     3 &  Aug-07-2015 &   SpeX &    K7V &   1.04 &   0.14 &    9.5 \\
  206011691 &     3 &  Sep-24-2015 &   SpeX &    K7V &   1.04 &   0.31 &     4.2 \\
  206119924 &     3 &  Sep-24-2015 &   SpeX &    K7V &   1.04 &   0.20 &   -16.8 \\
  206209135 &     3 &  Sep-24-2015 &   SpeX &    M2V &   0.91 &   3.46 &   -38.1 \\
  206312951 &     3 &  Sep-24-2015 &   SpeX &    M1V &   0.98 &   1.64 &   -14.0\\
  206318379 &     3 &  Sep-24-2015 &   SpeX &    M4V &   0.88 &   4.07 &    11.7 \\
  210448987 &     4 &  Nov-27-2015 &   SpeX &    K3V &   1.04 &   0.13 &    -15.9 \\
  210489231 &     4 &  Sep-24-2015 &   SpeX &    M1V &   0.98 &   1.75 &  -56.6\\
  210508766 &     4 &  Sep-24-2015 &   SpeX &    M1V &   1.02 &   0.75 &   -0.4 \\
  210558622 &     4 &  Oct-14-2015 &   SpeX &    K7V &   1.03 &   0.47 &    -0.1 \\
  210558622 &     4 &  Nov-26-2015 &   SpeX &    K7V &   1.03 &   0.36 &    -2.6 \\
  210564155 &     4 &  Nov-27-2015 &   SpeX &    M2V &   0.91 &   3.46 &    36.5 \\
  210707130 &     4 &  Sep-24-2015 &   SpeX &    K5V &   1.03 &   0.42 &   -2.4 \\
  210750726 &     4 &  Sep-24-2015 &   SpeX &    M1V &   0.94 &   2.59 &    2.5 \\
  210838726 &     4 &  Sep-24-2015 &   SpeX &    M1V &   0.99 &   1.39 &    18.6 \\
  210968143 &     4 &  Sep-24-2015 &   SpeX &    K5V &   1.04 &   0.31 &   20.9 \\
  211077024 &     4 &  Nov-26-2015 &   SpeX &    M3V &   0.92 &   3.19 &    23.2 \\
  211305568 &     5 &  Nov-27-2015 &   SpeX &    M1V &   0.99 &   1.50 &    29.7 \\
  211331236 &     5 &  Nov-26-2015 &   SpeX &    M1V &   0.99 &   1.48 &    2.0 \\
  211331236 &     5 &  Apr-18-2016 &  TSPEC &    M1V &   0.96 &   2.24 &    -5.3\\
  211336288 &     5 &  Mar-27-2016 &  TSPEC &    M0V &   1.03 &   0.55 &    19.4 \\
  211357309 &     5 &  Nov-27-2015 &   SpeX &    M1V &   0.99 &   1.38 &    18.5 \\
  211428897\tablenotemark{5} &     5 &  Nov-26-2015 &   SpeX &    M2V &   0.95 &   2.52 &    25.6 \\
  211509553 &     5 &  Mar-27-2016 &  TSPEC &    M0V &   0.97 &   1.87 &    -14.7 \\
  211680698 &     5 &  Mar-28-2016 &  TSPEC &    K3V &   1.02 &   0.60 &     -29.4 \\
 211694226A &     5 &   Mar-8-2016 &   SpeX &    M3V &   0.93 &   2.98 &    21.2 \\
 211694226B &     5 &   Mar-8-2016 &   SpeX &    M3V &   0.93 &   2.84 &    24.0 \\
  211762841 &     5 &   Mar-4-2016 &   SpeX &    K7V &   1.03 &   0.47 &   24.6 \\
  211770795 &     5 &  Apr-18-2016 &  TSPEC &    K5V &   1.04 &   0.17 &   -44.3 \\
  211791178 &     5 &  Mar-27-2016 &  TSPEC &    M0V &   1.01 &   0.96 &    61.5 \\
  211799258 &     5 &   Mar-8-2016 &   SpeX &    M3V &   0.93 &   2.78 &   44.6 \\
  211817229 &     5 &   Mar-4-2016 &   SpeX &    M4V &   0.85 &   4.91 &   28.2 \\
  211818569 &     5 &  Feb-19-2016 &  TSPEC &    K5V &   1.06 &  -0.16 &    24.9 \\
  211822797 &     5 &  Mar-27-2016 &  TSPEC &    K7V &   1.00 &   1.23 &    28.3 \\
  211826814 &     5 &  Feb-19-2016 &  TSPEC &    M4V &   0.90 &   3.72 &    24.1 \\
  211831378 &     5 &  Apr-18-2016 &  TSPEC &    M0V &   0.92 &   3.08 &    3.7 \\
  211839798 &     5 &   Mar-4-2016 &   SpeX &    M4V &   0.86 &   4.62 &    30.5 \\
  211924657 &     5 &   Mar-8-2016 &   SpeX &    M3V &   0.89 &   3.87 &    40.0 \\
  211965883 &     5 &  Mar-27-2016 &  TSPEC &    M0V &   1.08 &  -0.68 &    37.3 \\
  211969807 &     5 &   Mar-8-2016 &   SpeX &    M1V &   0.98 &   1.78 &    33.5 \\
  211970234 &     5 &  Apr-18-2016 &  TSPEC &    M4V &   0.87 &   4.35 &    -8.5 \\
  211988320 &     5 &  Mar-27-2016 &  TSPEC &    K7V &   1.09 &  -0.86 &   79.1 \\
  212006344 &     5 &  Nov-26-2015 &   SpeX &    M0V &   1.02 &   0.65 &   -13.3 \\
  212006344 &     5 &  Feb-19-2016 &  TSPEC &    M0V &   1.01 &   0.97 &   -15.5 \\
  212069861 &     5 &  Nov-26-2015 &   SpeX &    M0V &   1.02 &   0.76 &    25.3\\
  212154564 &     5 &  Mar-27-2016 &  TSPEC &    M3V &   0.95 &   2.46 &    20.9 \\
  212354731 &     6 &  Mar-28-2016 &  TSPEC &    M3V &   0.88 &   4.12 &   -24.4 \\
  212398486 &     6 &   Mar-4-2016 &   SpeX &    M2V &   0.93 &   2.89 &   -19.0 \\
  212443973 &     6 &  Mar-27-2016 &  TSPEC &    M3V &   0.96 &   2.05 &   0.7 \\
  212460519 &     6 &   Mar-8-2016 &   SpeX &    K7V &   1.05 &   0.09 &   -1.6 \\
  212554013 &     6 &  Apr-18-2016 &  TSPEC &    K3V &   1.10 &  -1.12 &   -60.0 \\
  212565386 &     6 &  Mar-10-2016 &   SpeX &    M1V &   0.97 &   1.98 &   -38.7 \\
  212572452 &     6 &  Mar-10-2016 &   SpeX &    K7V &   1.06 &  -0.17 &    5.7 \\
  212572452 &     6 &  Mar-27-2016 &  TSPEC &    K7V &   1.05 &  -0.03 &   6.0 \\
  212628098 &     6 &  Apr-18-2016 &  TSPEC &    K7V &   0.96 &   2.05 &    -2.2 \\
  212634172 &     6 &   Mar-4-2016 &   SpeX &    M3V &   0.93 &   2.95 &   23.2 \\
  212679181 &     6 &   Mar-4-2016 &   SpeX &    M3V &   0.95 &   2.45 &   13.3 \\
  212679798 &     6 &  Apr-18-2016 &  TSPEC &    M0V &   0.96 &   2.06 &   4.0\\
  212686205 &     6 &   Mar-8-2016 &   SpeX &    K4V &   1.04 &   0.14 &  -9.6 \\
  212690867 &     6 &   Mar-8-2016 &   SpeX &    M2V &   0.95 &   2.30 &    6.5 \\
  212773272 &     6 &  Apr-18-2016 &  TSPEC &    M3V &   0.95 &   2.51 &   -7.2 \\
  212773309 &     6 &  Mar-28-2016 &  TSPEC &    M0V &   1.01 &   0.94 &  -13.6 \\
 212773309B &     6 &  Mar-28-2016 &  TSPEC &    M3V &   0.92 &   3.03 &   -4.1 \\
  213951550 &     7 &   May-6-2016 &   SpeX &    M3V &   0.93 &   2.81 &   -77.2 \\
  214254518 &     7 &   May-5-2016 &   SpeX &    K7V &   1.05 &   0.09 &    17.6 \\
  214254518 &     7 &  Oct-26-2016 &   SpeX &    K7V &   1.04 &   0.22 &    17.3 \\
  214522613 &     7 &   May-5-2016 &   SpeX &    M1V &   0.96 &   2.20 &    35.9 \\
  214787262 &     7 &   May-5-2016 &   SpeX &    M3V &   0.91 &   3.27 &    -24.1 \\
  216892056 &     7 &   May-5-2016 &   SpeX &    M2V &   0.94 &   2.69 &   -82.8 \\
  217941732 &     7 &   May-5-2016 &   SpeX &    K5V &   1.03 &   0.41 &     -49.8 \\
  217941732 &     7 &  Oct-26-2016 &   SpeX &    K5V &   1.03 &   0.40 &     -50.9 \\
\enddata
 \tablenotetext{1}{Spectral types are coarse assignments based on visual inspection of the near-infrared spectra collected in this paper. The assigned spectral types have errors of roughly $\pm1$ subtype. (See Section~\ref{ssec:initial_class} for details.)}
 \tablenotetext{2}{H$_{\rm 2}$O-K2 index \citep{rojas-ayala_et_al2012}. Although we report H$_{\rm 2}$O-K2 indices and index-based spectral types for the full cool dwarf sample, these values are meaningless for the hotter stars.}
 \tablenotetext{3}{Spectral type estimated using the H$_{\rm 2}$O-K2 - spectral type relation introduced by \citet{newton_et_al2014}. On this scale, a spectral type of 0 corresponds to MV0 and positive values indicate correspondingly later M dwarf spectral types (e..g, $2=$~M2V). Negative values indicate K subtypes (i.e., $-1=$~K7V, $-2=$~K5V). }
  \tablenotetext{4}{Reported absolute radial velocities are the median of the values estimated by cross-correlating the telluric lines in our $J$-, $H$-, and $K$-band spectra with a theoretical atmospheric transmission spectrum using the {\tt tellrv} framework developed by \citet{newton_et_al2014}.}
 \tablenotetext{5}{Keck AO imaging by D.~Ciardi and Gemini speckle imaging by M.~Everett revealed that the star is actually a visual binary with a separation of roughly $1\farcs1$.\footnote{\url{https://exofop.ipac.caltech.edu/k2/edit\_target.php?id=211428897}}}  
\end{deluxetable*}

\begin{deluxetable*}{ccccccccccccccc}
\tablecolumns{15}
\tabletypesize{\scriptsize}
\tablecaption{Inferred Stellar Parameters for Low-Mass Dwarfs \label{tab:star_prop}}
\tablehead{
\colhead{} & 
\colhead{} & 
\colhead{} & 
\multicolumn{3}{c}{Teff (K)} &
\multicolumn{3}{c}{Radius ($\rsun$)} &
\multicolumn{3}{c}{Mass ($\msun$)} &
\multicolumn{3}{c}{Luminosity ($\log L_*/\lsun$)} \\
\cline{4-6}
\cline{10-12}
\colhead{EPIC} &
\colhead{Date} &
\colhead{SpType} &
\colhead{Val} &
\colhead{-Err} &
\colhead{+Err} &
\colhead{Val} &
\colhead{-Err} &
\colhead{+Err} &
\colhead{Val} &
\colhead{-Err} &
\colhead{+Err} &
\colhead{Val} &
\colhead{-Err} &
\colhead{+Err}
}
\startdata
201205469 &  Jun-13-2015 &    K7V &       3890 &           121 &           113 &    0.587 &       0.039 &       0.039 &    0.599 &       0.043 &       0.035 &    -1.178 &        0.188 &        0.175 \\
  201208431 &  May-05-2015 &    K7V &       4015 &           173 &           155 &    0.569 &       0.047 &       0.049 &    0.635 &       0.046 &       0.035 &    -1.023 &        0.219 &        0.202 \\
  201345483 &  May-05-2015 &    M0V &       4262 &           201 &           173 &    0.686 &       0.045 &       0.057 &    0.682 &       0.030 &       0.028 &    -0.630 &        0.218 &        0.198 \\
  201549860 &  Nov-26-2015 &    K4V &       4403 &            96 &            93 &    0.620 &       0.028 &       0.029 &    0.702 &       0.013 &       0.013 &    -0.688 &        0.073 &        0.071 \\
  201617985 &  Apr-16-2015 &    M1V &       3742 &           116 &           105 &    0.496 &       0.032 &       0.032 &    0.540 &       0.055 &       0.048 &    -1.480 &        0.141 &        0.134 \\
  201635569 &  May-05-2015 &    M0V &       3970 &           118 &           112 &    0.623 &       0.032 &       0.032 &    0.623 &       0.035 &       0.028 &    -1.580 &        0.378 &        0.321 \\
  201637175 &  May-05-2015 &    K7V &       3879 &            95 &            87 &    0.582 &       0.031 &       0.030 &    0.595 &       0.033 &       0.029 &    -1.258 &        0.135 &        0.124 \\
  201717274 &  May-05-2015 &    M2V &       3286 &           134 &           130 &    0.314 &       0.057 &       0.054 &    0.194 &       0.159 &       0.133 &    -1.986 &        0.106 &        0.106 \\
  201855371 &  Apr-16-2015 &    K5V &       4118 &           133 &           119 &    0.626 &       0.036 &       0.041 &    0.658 &       0.027 &       0.023 &    -0.845 &        0.142 &        0.133 \\
  205924614 &  Sep-24-2015 &    K7V &       4423 &           149 &           130 &    0.700 &       0.045 &       0.056 &    0.705 &       0.018 &       0.022 &    -0.701 &        0.125 &        0.116 \\
  205924614\tablenotemark{2}  &  Nov-26-2015 &    K7V &       4300 &           107 &           100 &    0.715 &       0.040 &       0.043 &    0.688 &       0.015 &       0.015 &    -0.769 &        0.079 &        0.081 \\
  206011691 &  Aug-07-2015 &    K7V &       4304 &            90 &            86 &    0.649 &       0.029 &       0.029 &    0.688 &       0.013 &       0.012 &    -1.111 &        0.072 &        0.071 \\
  206011691\tablenotemark{2}  &  Sep-24-2015 &    K7V &       4222 &            88 &            84 &    0.647 &       0.028 &       0.029 &    0.676 &       0.015 &       0.013 &    -1.235 &        0.082 &        0.083 \\
  206119924 &  Sep-24-2015 &    K7V &       4348 &            86 &            88 &    0.669 &       0.030 &       0.030 &    0.695 &       0.013 &       0.012 &    -0.736 &        0.063 &        0.063 \\
  206209135 &  Sep-24-2015 &    M2V &       3360 &            87 &            86 &    0.331 &       0.030 &       0.030 &    0.271 &       0.091 &       0.079 &    -1.872 &        0.059 &        0.058 \\
  206312951 &  Sep-24-2015 &    M1V &       3707 &            80 &            81 &    0.478 &       0.028 &       0.028 &    0.523 &       0.045 &       0.037 &    -1.277 &        0.066 &        0.064 \\
  206318379 &  Sep-24-2015 &    M4V &       3293 &            89 &            87 &    0.280 &       0.031 &       0.031 &    0.201 &       0.102 &       0.090 &    -1.929 &        0.059 &        0.061 \\
  210448987 &  Nov-27-2015 &    K3V &       4674 &           141 &           131 &    0.635 &       0.032 &       0.035 &    0.745 &       0.023 &       0.034 &    -0.656 &        0.062 &        0.059 \\
  210489231 &  Sep-24-2015 &    M1V &       4056 &           113 &           104 &    0.557 &       0.034 &       0.037 &    0.645 &       0.027 &       0.022 &    -0.937 &        0.067 &        0.063 \\
  210508766 &  Sep-24-2015 &    M1V &       3876 &            81 &            80 &    0.547 &       0.028 &       0.028 &    0.594 &       0.031 &       0.025 &    -1.393 &        0.071 &        0.066 \\
  210558622\tablenotemark{2} &  Oct-14-2015 &    K7V &       4268 &           105 &            98 &    0.678 &       0.036 &       0.040 &    0.683 &       0.016 &       0.015 &    -0.685 &        0.076 &        0.070 \\
  210558622 &  Nov-26-2015 &    K7V &       4350 &           112 &           106 &    0.770 &       0.050 &       0.057 &    0.695 &       0.015 &       0.016 &    -0.590 &        0.076 &        0.070 \\
  210564155 &  Nov-27-2015 &    M2V &       3344 &            90 &            87 &    0.286 &       0.031 &       0.030 &    0.255 &       0.093 &       0.084 &    -2.008 &        0.062 &        0.061 \\
  210707130 &  Sep-24-2015 &    K5V &       4376 &            95 &            90 &    0.676 &       0.031 &       0.031 &    0.698 &       0.013 &       0.013 &    -0.711 &        0.063 &        0.062 \\
  210750726 &  Sep-24-2015 &    M1V &       3624 &            88 &            87 &    0.460 &       0.030 &       0.032 &    0.477 &       0.057 &       0.048 &    -1.530 &        0.055 &        0.054 \\
  210838726 &  Sep-24-2015 &    M1V &       3792 &            78 &            78 &    0.503 &       0.028 &       0.028 &    0.562 &       0.036 &       0.030 &    -1.371 &        0.058 &        0.057 \\
  210968143 &  Sep-24-2015 &    K5V &       4422 &            93 &            91 &    0.635 &       0.029 &       0.029 &    0.705 &       0.013 &       0.013 &    -0.994 &        0.064 &        0.066 \\
  211077024 &  Nov-26-2015 &    M3V &       3489 &            81 &            80 &    0.321 &       0.029 &       0.029 &    0.384 &       0.067 &       0.058 &    -1.742 &        0.054 &        0.054 \\
  211305568 &  Nov-27-2015 &    M1V &       3612 &            85 &            84 &    0.446 &       0.030 &       0.031 &    0.470 &       0.056 &       0.048 &    -1.462 &        0.057 &        0.056 \\
  211331236 &  Nov-26-2015 &    M1V &       3755 &            85 &            83 &    0.457 &       0.028 &       0.028 &    0.546 &       0.042 &       0.035 &    -1.358 &        0.061 &        0.059 \\
  211331236\tablenotemark{2} & Apr-18-2016 &    M1V &       3842 &            82 &            82 &    0.492 &       0.028 &       0.028 &    0.582 &       0.034 &       0.028 &    -1.262 &        0.060 &        0.060 \\
  211336288 &  Mar-27-2016 &    M0V &       3997 &            80 &            79 &    0.586 &       0.027 &       0.027 &    0.630 &       0.022 &       0.019 &    -1.365 &        0.062 &        0.061 \\
  211357309 &  Nov-27-2015 &    M1V &       3731 &            86 &            85 &    0.460 &       0.028 &       0.028 &    0.535 &       0.045 &       0.038 &    -1.402 &        0.060 &        0.059 \\
  211428897 &  Nov-26-2015 &    M2V &       3595 &            95 &            91 &    0.290 &       0.030 &       0.030 &    0.459 &       0.064 &       0.055 &    -1.685 &        0.056 &        0.058 \\
  211509553 &  Mar-27-2016 &    M0V &       3756 &            81 &            80 &    0.547 &       0.029 &       0.029 &    0.546 &       0.040 &       0.034 &    -1.592 &        0.087 &        0.081 \\
  211680698 &  Mar-28-2016 &    K3V &       4726 &           143 &           127 &    0.735 &       0.043 &       0.047 &    0.756 &       0.025 &       0.039 &    -0.593 &        0.063 &        0.061 \\
 211694226a &   Mar-8-2016 &    M3V &       3454 &            83 &            82 &    0.445 &       0.031 &       0.031 &    0.356 &       0.074 &       0.064 &    -1.459 &        0.076 &        0.073 \\
 211694226b &   Mar-8-2016 &    M3V &       3448 &            93 &            92 &    0.440 &       0.035 &       0.037 &    0.351 &       0.084 &       0.072 &    -1.647 &        0.086 &        0.084 \\
  211762841 &   Mar-4-2016 &    K7V &       4136 &            87 &            86 &    0.626 &       0.029 &       0.030 &    0.661 &       0.018 &       0.015 &    -1.080 &        0.078 &        0.075 \\
  211770795 &  Apr-18-2016 &    K5V &       4753 &           155 &           129 &    0.679 &       0.036 &       0.038 &    0.763 &       0.027 &       0.046 &    -0.572 &        0.076 &        0.070 \\
  211791178 &  Mar-27-2016 &    M0V &       4350 &           102 &            96 &    0.667 &       0.034 &       0.038 &    0.695 &       0.014 &       0.014 &    -0.669 &        0.068 &        0.068 \\
  211799258\tablenotemark{3} &   Mar-8-2016 &    M3V &       3317 &            73 &            73 &    0.328 &       0.062 &       0.069 &    0.227 &       0.077 &       0.077 &    -2.117 &       0.373 &       0.373 \\
  211817229\tablenotemark{3} &   Mar-4-2016 &    M4V &       3276 &            73 &            73 &    0.237 &       0.041 &       0.046 &    0.183 &       0.082 &       0.082 &    -2.279 &       0.676 &       0.676 \\
  211818569 &  Feb-19-2016 &    K5V &       4471 &           112 &           104 &    0.768 &       0.042 &       0.042 &    0.712 &       0.014 &       0.017 &    -0.611 &        0.058 &        0.057 \\
  211822797 &  Mar-27-2016 &    K7V &       4148 &            82 &            80 &    0.572 &       0.027 &       0.027 &    0.663 &       0.016 &       0.014 &    -1.218 &        0.061 &        0.061 \\
  211826814\tablenotemark{3} &  Feb-19-2016 &    M4V &       3288 &            73 &            73 &    0.262 &       0.049 &       0.055 &    0.196 &       0.080 &       0.080 &    -2.226 &       0.539 &       0.539 \\
  211831378 &  Apr-18-2016 &    M0V &       3748 &           115 &           101 &    0.548 &       0.031 &       0.031 &    0.543 &       0.052 &       0.047 &    -1.480 &        0.148 &        0.154 \\
  211839798 &   Mar-4-2016 &    M4V &       3522 &           175 &           133 &    0.265 &       0.039 &       0.049 &    0.409 &       0.110 &       0.109 &    -2.134 &        0.067 &        0.065 \\
  211924657 &   Mar-8-2016 &    M3V &       3421 &           106 &            98 &    0.322 &       0.036 &       0.041 &    0.327 &       0.095 &       0.085 &    -1.902 &        0.064 &        0.063 \\
  211965883 &  Mar-27-2016 &    M0V &       4211 &            80 &            79 &    0.600 &       0.027 &       0.027 &    0.674 &       0.014 &       0.012 &    -1.110 &        0.061 &        0.060 \\
  211969807 &   Mar-8-2016 &    M1V &       3546 &            99 &            95 &    0.492 &       0.032 &       0.032 &    0.427 &       0.072 &       0.063 &    -1.476 &        0.109 &        0.100 \\
  211970234 &  Apr-18-2016 &    M4V &       3292 &           159 &           150 &    0.190 &       0.039 &       0.036 &    0.200 &       0.185 &       0.153 &    -2.371 &        0.111 &        0.101 \\
  211988320 &  Mar-27-2016 &    K7V &       4284 &            84 &            84 &    0.641 &       0.028 &       0.029 &    0.685 &       0.013 &       0.012 &    -1.174 &        0.059 &        0.058 \\
  212006344\tablenotemark{2} & Nov-26-2015 &   M0V &       3993 &            78 &            76 &    0.591 &       0.027 &       0.027 &    0.630 &       0.022 &       0.018 &    -1.186 &        0.065 &        0.066 \\
  212006344 &  Feb-19-2016 &    M0V &       3963 &            77 &            76 &    0.625 &       0.028 &       0.028 &    0.621 &       0.024 &       0.020 &    -1.150 &        0.066 &        0.069 \\
  212069861 &  Nov-26-2015 &    M0V &       4076 &            83 &            81 &    0.571 &       0.028 &       0.028 &    0.649 &       0.019 &       0.016 &    -1.091 &        0.068 &        0.063 \\
  212154564 &  Mar-27-2016 &    M3V &       3561 &            87 &            84 &    0.344 &       0.030 &       0.030 &    0.436 &       0.062 &       0.054 &    -1.643 &        0.058 &        0.058 \\
  212354731 &  Mar-28-2016 &    M3V &       3591 &           119 &           106 &    0.418 &       0.032 &       0.033 &    0.457 &       0.075 &       0.068 &    -1.531 &        0.096 &        0.091 \\
  212398486 &   Mar-4-2016 &    M2V &       3654 &           100 &            92 &    0.402 &       0.031 &       0.031 &    0.495 &       0.057 &       0.051 &    -1.540 &        0.067 &        0.064 \\
  212443973 &  Mar-27-2016 &    M3V &       3423 &            84 &            84 &    0.343 &       0.028 &       0.028 &    0.330 &       0.079 &       0.069 &    -1.888 &        0.054 &        0.054 \\
  212460519 &   Mar-8-2016 &    K7V &       4368 &           128 &           115 &    0.621 &       0.034 &       0.036 &    0.697 &       0.016 &       0.018 &    -0.816 &        0.080 &        0.075 \\
  212554013 &  Apr-18-2016 &    K3V &       4388 &           142 &           137 &    0.677 &       0.045 &       0.052 &    0.700 &       0.019 &       0.020 &    -0.757 &        0.080 &        0.078 \\
  212565386 &  Mar-10-2016 &    M1V &       4342 &           159 &           137 &    0.581 &       0.036 &       0.041 &    0.694 &       0.020 &       0.022 &    -1.058 &        0.075 &        0.074 \\
  212572452 &  Mar-27-2016 &    K7V &       4390 &           193 &           160 &    0.662 &       0.043 &       0.053 &    0.700 &       0.023 &       0.028 &    -0.807 &        0.165 &        0.155 \\
212572452\tablenotemark{2} &  Mar-10-2016 &    K7V &       4332 &           135 &           121 &    0.678 &       0.037 &       0.044 &    0.692 &       0.018 &       0.019 &    -0.854 &        0.128 &        0.120 \\
  212628098 &  Apr-18-2016 &    K7V &       3942 &            84 &            82 &    0.566 &       0.028 &       0.028 &    0.615 &       0.027 &       0.022 &    -0.796 &        0.067 &        0.065 \\
  212634172 &   Mar-4-2016 &    M3V &       3412 &            98 &            94 &    0.348 &       0.033 &       0.034 &    0.320 &       0.092 &       0.081 &    -1.866 &        0.064 &        0.062 \\
  212679181 &   Mar-4-2016 &    M3V &       3616 &            89 &            87 &    0.434 &       0.029 &       0.029 &    0.472 &       0.058 &       0.050 &    -1.544 &        0.056 &        0.058 \\
  212679798 &  Apr-18-2016 &    M0V &       3823 &            92 &            89 &    0.562 &       0.029 &       0.029 &    0.575 &       0.039 &       0.032 &    -1.009 &        0.081 &        0.084 \\
  212686205 &   Mar-8-2016 &    K4V &       4470 &           172 &           145 &    0.778 &       0.061 &       0.076 &    0.711 &       0.020 &       0.028 &    -0.673 &        0.066 &        0.065 \\
  212690867 &   Mar-8-2016 &    M2V &       3614 &           118 &           107 &    0.415 &       0.032 &       0.033 &    0.471 &       0.073 &       0.064 &    -1.603 &        0.078 &        0.077 \\
  212773272 &  Apr-18-2016 &    M3V &       3367 &            82 &            81 &    0.428 &       0.030 &       0.030 &    0.277 &       0.084 &       0.074 &    -1.753 &        0.067 &        0.069 \\
  212773309 &  Mar-28-2016 &    M0V &       4178 &            90 &            87 &    0.588 &       0.029 &       0.029 &    0.669 &       0.016 &       0.014 &    -0.797 &        0.056 &        0.057 \\
 212773309B &  Mar-28-2016 &    M3V &       3459 &           103 &           100 &    0.396 &       0.034 &       0.034 &    0.360 &       0.090 &       0.078 &    -1.632 &        0.097 &        0.104 \\
  213951550 &   May-6-2016 &    M3V &       3574 &            88 &            85 &    0.471 &       0.030 &       0.030 &    0.445 &       0.061 &       0.054 &    -1.367 &        0.075 &        0.076 \\
  214254518\tablenotemark{2} &   May-5-2016 &    K7V &       4335 &           102 &            94 &    0.668 &       0.033 &       0.037 &    0.693 &       0.014 &       0.014 &    -0.836 &        0.066 &        0.066 \\
  214254518 &  Oct-26-2016 &    K7V &       4574 &           130 &           110 &    0.710 &       0.036 &       0.038 &    0.727 &       0.017 &       0.024 &    -0.758 &        0.065 &        0.067 \\
  214522613 &   May-5-2016 &    M1V &       3602 &            99 &            94 &    0.448 &       0.032 &       0.032 &    0.463 &       0.065 &       0.056 &    -1.412 &        0.084 &        0.080 \\
  214787262 &   May-5-2016 &    M3V &       3459 &            89 &            84 &    0.360 &       0.030 &       0.031 &    0.360 &       0.074 &       0.068 &    -1.841 &        0.056 &        0.055 \\
  216892056 &   May-5-2016 &    M2V &       3467 &            84 &            82 &    0.398 &       0.029 &       0.029 &    0.367 &       0.071 &       0.063 &    -1.707 &        0.057 &        0.056 \\
  217941732 &   May-5-2016 &    K5V &       4470 &           211 &           202 &    0.731 &       0.072 &       0.111 &    0.711 &       0.028 &       0.035 &    -0.844 &        0.153 &        0.116 \\
  217941732\tablenotemark{2} &  Oct-26-2016 &    K5V &       4356 &           197 &           172 &    0.744 &       0.078 &       0.111 &    0.696 &       0.026 &       0.028 &    -0.858 &        0.132 &        0.126 \\
  \enddata
 \tablenotetext{1}{Spectral types are coarse assignments based on visual inspection of the near-infrared spectra collected in this paper. The assigned spectral types have errors of roughly $\pm1$ subtype. (See Section~\ref{ssec:initial_class} for details.)}
  \tablenotetext{2}{Star observed twice to check the repeatability of our analysis. These are the higher precision estimates.}
 \tablenotetext{3}{The Al-a EW for these stars are below the calibration range for the \citet{newton_et_al2015} relations. Adopted parameters are based on the \citet{mann_et_al2013a, mann_et_al2013c, mann_et_al2015} relations.} 
\end{deluxetable*}

\begin{deluxetable*}{cccccccccccccccc}
\tablecolumns{16}
\tabletypesize{\footnotesize}
\tablecaption{Equivalent Widths \& Metallicities for Cool Dwarfs \label{tab:ew+met}}
\tablehead{
\colhead{} &
\colhead{} &
\multicolumn{6}{c}{EW of Mg Features (A)} &
\multicolumn{4}{c}{EW of Al Features (A)} &
\multicolumn{4}{c}{Metallicity\tablenotemark{1}} \\
\cline{3-8}
\cline{13-16}
\colhead{} &
\colhead{} &
\multicolumn{2}{c}{(1.50 $\mu$m)} &
\multicolumn{2}{c}{(1.57 $\mu$m)} &
\multicolumn{2}{c}{(1.71 $\mu$m)} &
\multicolumn{2}{c}{a (1.67 $\mu$m)} &
\multicolumn{2}{c}{b (1.67 $\mu$m)} &
\multicolumn{2}{c}{[Fe/H]} &
\multicolumn{2}{c}{[M/H]} \\
\colhead{EPIC} &
\colhead{Date} &
\colhead{Val} & 
\colhead{Err} & 
\colhead{Val} & 
\colhead{Err} & 
\colhead{Val} & 
\colhead{Err} & 
\colhead{Val} & 
\colhead{Err} & 
\colhead{Val} & 
\colhead{Err} &
\colhead{Val} &
\colhead{Err} &
\colhead{Val} &
\colhead{Err}
}
\startdata
201205469 &  Jun-13-2015 &     5.84 &      0.37 &     3.82 &      0.30 &     3.59 &      0.33 &   2.43 &    0.21 &   3.01 &    0.23 &   0.433 &    0.166 &   0.307 &    0.146 \\
  201208431 &  May-05-2015 &     7.76 &      0.33 &     2.87 &      0.59 &     3.52 &      0.32 &   1.43 &    0.27 &   2.74 &    0.35 &   0.066 &    0.191 &  -0.024 &    0.170 \\
  201345483 &  May-05-2015 &     8.23 &      0.41 &     6.14 &      0.51 &     3.79 &      0.39 &   1.94 &    0.23 &   2.36 &    0.31 &   0.316 &    0.202 &   0.130 &    0.164 \\
  201549860 &  Nov-26-2015 &     8.13 &      0.10 &     5.08 &      0.10 &     3.86 &      0.09 &   1.72 &    0.07 &   2.15 &    0.09 &  $\cdots$ & $\cdots$ &  $\cdots$ & $\cdots$\\
  201617985 &  Apr-16-2015 &     5.26 &      0.26 &     3.35 &      0.22 &     4.29 &      0.20 &   1.56 &    0.15 &   2.60 &    0.20 &  -0.010 &    0.143 &  -0.022 &    0.116 \\
  201635569 &  May-05-2015 &     7.44 &      0.39 &     5.13 &      0.30 &     5.02 &      0.36 &   2.08 &    0.20 &   2.92 &    0.25 &   0.196 &    0.180 &   0.138 &    0.147 \\
  201637175 &  May-05-2015 &     7.00 &      0.21 &     4.53 &      0.22 &     4.32 &      0.18 &   2.14 &    0.13 &   3.03 &    0.19 &   0.032 &    0.125 &   0.007 &    0.108 \\
  201717274 &  May-05-2015 &     2.28 &      0.32 &     1.06 &      0.32 &     1.42 &      0.32 &   1.52 &    0.23 &   2.10 &    0.26 &  -0.257 &    0.154 &  -0.188 &    0.132 \\
  201855371 &  Apr-16-2015 &     8.15 &      0.25 &     5.33 &      0.26 &     4.00 &      0.21 &   1.42 &    0.15 &   2.41 &    0.19 &    $\cdots$ & $\cdots$ &  $\cdots$ & $\cdots$ \\
  205924614 &  Sep-24-2015 &     8.30 &      0.22 &     5.81 &      0.20 &     4.17 &      0.17 &   1.39 &    0.13 &   2.17 &    0.16 &   0.246 &    0.125 &   0.170 &    0.108 \\
  205924614 &  Nov-26-2015 &     7.94 &      0.12 &     5.50 &      0.12 &     3.80 &      0.10 &   1.36 &    0.08 &   2.21 &    0.11 &   0.376 &    0.095 &   0.168 &    0.089 \\
  206011691 &  Aug-07-2015 &     8.13 &      0.08 &     5.60 &      0.08 &     4.42 &      0.07 &   1.66 &    0.06 &   2.33 &    0.09 &  -0.121 &    0.088 &  -0.122 &    0.085 \\
  206011691 &  Sep-24-2015 &     7.85 &      0.08 &     5.78 &      0.10 &     4.47 &      0.08 &   1.73 &    0.06 &   2.25 &    0.09 &  -0.034 &    0.090 &  -0.057 &    0.086 \\
  206119924 &  Sep-24-2015 &     8.34 &      0.07 &     5.68 &      0.08 &     3.88 &      0.06 &   1.52 &    0.06 &   2.25 &    0.08 &   0.337 &    0.086 &   0.204 &    0.084 \\
  206209135 &  Sep-24-2015 &     2.54 &      0.12 &     1.65 &      0.11 &     2.30 &      0.10 &   1.36 &    0.07 &   1.54 &    0.10 &  -0.271 &    0.093 &  -0.278 &    0.089 \\
  206312951 &  Sep-24-2015 &     4.95 &      0.11 &     3.28 &      0.11 &     3.10 &      0.09 &   1.64 &    0.07 &   2.39 &    0.08 &   0.097 &    0.092 &   0.066 &    0.087 \\
  206318379 &  Sep-24-2015 &     2.33 &      0.13 &     1.41 &      0.12 &     1.96 &      0.11 &   1.26 &    0.08 &   1.64 &    0.10 &   0.332 &    0.096 &   0.208 &    0.090 \\
  210448987 &  Nov-27-2015 &     7.41 &      0.10 &     4.88 &      0.10 &     3.14 &      0.09 &   1.39 &    0.07 &   1.59 &    0.10 &    $\cdots$ & $\cdots$ &  $\cdots$ & $\cdots$ \\
  210489231 &  Sep-24-2015 &     6.32 &      0.12 &     3.63 &      0.12 &     3.16 &      0.11 &   1.24 &    0.09 &   1.99 &    0.13 &   0.524 &    0.098 &   0.349 &    0.091 \\
  210508766 &  Sep-24-2015 &     5.82 &      0.08 &     4.28 &      0.09 &     3.95 &      0.08 &   1.72 &    0.07 &   2.21 &    0.09 &  -0.107 &    0.089 &  -0.060 &    0.085 \\
  210558622 &  Oct-14-2015 &     8.16 &      0.10 &     5.45 &      0.12 &     3.81 &      0.10 &   1.42 &    0.09 &   2.37 &    0.11 &   0.025 &    0.096 &   0.012 &    0.089 \\
  210558622 &  Nov-26-2015 &     8.21 &      0.11 &     5.58 &      0.11 &     3.76 &      0.10 &   1.22 &    0.08 &   2.14 &    0.11 &   0.094 &    0.094 &   0.050 &    0.090 \\
  210564155 &  Nov-27-2015 &     2.00 &      0.11 &     1.39 &      0.11 &     1.53 &      0.10 &   1.20 &    0.08 &   1.44 &    0.10 &  -0.149 &    0.092 &  -0.124 &    0.088 \\
  210707130 &  Sep-24-2015 &     8.48 &      0.07 &     5.71 &      0.07 &     3.84 &      0.06 &   1.57 &    0.06 &   2.22 &    0.09 &     $\cdots$ & $\cdots$ &  $\cdots$ & $\cdots$ \\
  210750726 &  Sep-24-2015 &     3.67 &      0.08 &     2.87 &      0.08 &     2.64 &      0.07 &   1.32 &    0.07 &   1.80 &    0.10 &   0.100 &    0.088 &   0.034 &    0.085 \\
  210838726 &  Sep-24-2015 &     5.28 &      0.06 &     3.55 &      0.08 &     3.51 &      0.07 &   1.63 &    0.05 &   2.21 &    0.07 &   0.180 &    0.085 &   0.111 &    0.083 \\
  210968143 &  Sep-24-2015 &     7.93 &      0.07 &     5.39 &      0.08 &     4.03 &      0.06 &   1.59 &    0.06 &   2.02 &    0.08 &    $\cdots$ & $\cdots$ &  $\cdots$ & $\cdots$ \\
  211077024 &  Nov-26-2015 &     2.96 &      0.08 &     1.73 &      0.08 &     1.79 &      0.08 &   1.22 &    0.05 &   1.62 &    0.07 &   0.170 &    0.087 &   0.062 &    0.085 \\
  211305568 &  Nov-27-2015 &     3.99 &      0.09 &     2.89 &      0.09 &     2.79 &      0.08 &   1.23 &    0.07 &   1.96 &    0.10 &  -0.175 &    0.090 &  -0.105 &    0.087 \\
  211331236 &  Nov-26-2015 &     4.68 &      0.10 &     2.97 &      0.10 &     3.05 &      0.09 &   1.59 &    0.07 &   2.02 &    0.10 &   0.037 &    0.091 &   0.083 &    0.088 \\
  211331236 &  Apr-18-2016 &     5.02 &      0.11 &     3.23 &      0.09 &     3.19 &      0.07 &   1.93 &    0.07 &   2.28 &    0.10 &   0.106 &    0.088 &  -0.001 &    0.085 \\
  211336288 &  Mar-27-2016 &     6.42 &      0.08 &     4.76 &      0.06 &     4.05 &      0.05 &   1.81 &    0.05 &   2.33 &    0.08 &  -0.075 &    0.084 &  -0.123 &    0.084 \\
  211357309 &  Nov-27-2015 &     4.49 &      0.10 &     3.23 &      0.10 &     2.88 &      0.09 &   1.65 &    0.08 &   2.03 &    0.11 &  -0.175 &    0.092 &  -0.085 &    0.088 \\
  211428897 &  Nov-26-2015 &     3.19 &      0.10 &     1.59 &      0.10 &     1.87 &      0.09 &   1.13 &    0.07 &   1.46 &    0.09 &  -0.131 &    0.087 &  -0.154 &    0.085 \\
  211509553 &  Mar-27-2016 &     5.77 &      0.17 &     3.64 &      0.11 &     4.03 &      0.09 &   2.17 &    0.07 &   2.83 &    0.11 &   0.044 &    0.096 &  -0.177 &    0.092 \\
  211680698 &  Mar-28-2016 &     7.44 &      0.14 &     4.87 &      0.09 &     2.77 &      0.07 &   1.15 &    0.07 &   1.56 &    0.09 &    $\cdots$ & $\cdots$ &  $\cdots$ & $\cdots$\\
 211694226a &   Mar-8-2016 &     4.13 &      0.18 &     2.98 &      0.17 &     2.83 &      0.17 &   1.67 &    0.12 &   2.74 &    0.13 &   0.043 &    0.108 &   0.053 &    0.101 \\
 211694226b &   Mar-8-2016 &     3.54 &      0.24 &     2.75 &      0.22 &     2.39 &      0.24 &   1.71 &    0.15 &   2.42 &    0.16 &   0.261 &    0.131 &   0.117 &    0.110 \\
  211762841 &   Mar-4-2016 &     7.63 &      0.09 &     5.21 &      0.09 &     4.06 &      0.09 &   1.62 &    0.07 &   2.36 &    0.10 &   0.218 &    0.089 &   0.241 &    0.086 \\
  211770795 &  Apr-18-2016 &     7.38 &      0.17 &     5.35 &      0.12 &     3.27 &      0.09 &   1.33 &    0.07 &   1.54 &    0.10 &    $\cdots$ & $\cdots$ &  $\cdots$ & $\cdots$ \\
  211791178 &  Mar-27-2016 &     7.30 &      0.15 &     4.79 &      0.11 &     3.24 &      0.09 &   1.26 &    0.06 &   1.79 &    0.07 &  -0.399 &    0.096 &  -0.095 &    0.092 \\
  211799258 &   Mar-8-2016 &     3.58 &      0.39 &     2.18 &      0.32 &     1.15 &      0.35 &   0.73 &    0.23 &   1.07 &    0.25 &   0.120 &    0.167 &   0.181 &    0.145 \\
  211817229 &   Mar-4-2016 &     1.23 &      0.12 &     0.90 &      0.11 &     0.95 &      0.11 &   0.63 &    0.08 &   0.62 &    0.11 &  -0.401 &    0.090 &  -0.327 &    0.088 \\
  211818569 &  Feb-19-2016 &     7.66 &      0.10 &     5.30 &      0.08 &     3.22 &      0.06 &   1.12 &    0.06 &   1.73 &    0.09 &  $\cdots$ & $\cdots$ &  $\cdots$ & $\cdots$ \\
  211822797 &  Mar-27-2016 &     6.41 &      0.08 &     4.64 &      0.07 &     3.94 &      0.06 &   1.90 &    0.05 &   2.08 &    0.07 &   0.322 &    0.084 &   0.179 &    0.083 \\
  211826814 &  Feb-19-2016 &     2.60 &      0.35 &     0.97 &      0.27 &     1.33 &      0.21 &   0.63 &    0.15 &   1.06 &    0.19 &  -0.254 &    0.130 &  -0.317 &    0.123 \\
  211831378 &  Apr-18-2016 &     5.47 &      0.40 &     3.91 &      0.23 &     3.76 &      0.19 &   1.88 &    0.13 &   2.60 &    0.16 &   0.257 &    0.138 &   0.111 &    0.128 \\
  211839798 &   Mar-4-2016 &     1.69 &      0.12 &     1.18 &      0.12 &     1.48 &      0.12 &   0.92 &    0.08 &   0.98 &    0.11 &  -0.078 &    0.095 &  -0.010 &    0.089 \\
  211924657 &   Mar-8-2016 &     2.42 &      0.13 &     1.69 &      0.13 &     1.72 &      0.13 &   1.01 &    0.09 &   1.32 &    0.11 &  -0.004 &    0.096 &  -0.006 &    0.091 \\
  211965883 &  Mar-27-2016 &     7.40 &      0.08 &     5.10 &      0.07 &     4.16 &      0.05 &   1.86 &    0.04 &   2.33 &    0.06 &  -0.196 &    0.084 &   0.024 &    0.083 \\
  211969807 &   Mar-8-2016 &     3.87 &      0.25 &     3.47 &      0.21 &     3.26 &      0.23 &   1.73 &    0.15 &   2.67 &    0.18 &   0.179 &    0.125 &   0.200 &    0.116 \\
  211970234 &  Apr-18-2016 &     1.46 &      0.28 &     1.08 &      0.16 &     1.22 &      0.13 &   1.05 &    0.09 &   0.83 &    0.12 &  -0.177 &    0.109 &  -0.087 &    0.102 \\
  211988320 &  Mar-27-2016 &     7.20 &      0.08 &     5.13 &      0.06 &     4.20 &      0.04 &   1.50 &    0.04 &   1.97 &    0.06 &  -0.369 &    0.084 &  -0.157 &    0.083 \\
  212006344 &  Nov-26-2015 &     7.30 &      0.07 &     4.95 &      0.07 &     4.35 &      0.06 &   1.99 &    0.06 &   2.86 &    0.08 &   0.444 &    0.085 &   0.341 &    0.083 \\
  212006344 &  Feb-19-2016 &     7.25 &      0.10 &     5.38 &      0.09 &     4.12 &      0.06 &   2.37 &    0.06 &   3.15 &    0.08 &   0.521 &    0.086 &   0.309 &    0.085 \\
  212069861 &  Nov-26-2015 &     7.08 &      0.08 &     4.64 &      0.08 &     3.90 &      0.07 &   1.75 &    0.06 &   2.38 &    0.09 &   0.324 &    0.088 &   0.195 &    0.085 \\
  212154564 &  Mar-27-2016 &     3.35 &      0.11 &     2.00 &      0.10 &     2.63 &      0.07 &   1.14 &    0.05 &   1.64 &    0.07 &  -0.093 &    0.088 &  -0.238 &    0.086 \\
  212354731 &  Mar-28-2016 &     3.36 &      0.30 &     2.61 &      0.17 &     2.13 &      0.15 &   1.39 &    0.11 &   1.79 &    0.11 &  -0.009 &    0.124 &   0.018 &    0.107 \\
  212398486 &   Mar-4-2016 &     4.09 &      0.16 &     2.58 &      0.15 &     2.50 &      0.17 &   1.58 &    0.10 &   1.61 &    0.12 &  -0.278 &    0.103 &  -0.197 &    0.096 \\
  212443973 &  Mar-27-2016 &     2.31 &      0.08 &     1.99 &      0.06 &     2.44 &      0.05 &   1.12 &    0.05 &   1.32 &    0.08 &   0.201 &    0.084 &  -0.054 &    0.083 \\
  212460519 &   Mar-8-2016 &     7.57 &      0.11 &     4.77 &      0.12 &     3.68 &      0.11 &   1.42 &    0.10 &   1.71 &    0.13 &  -0.116 &    0.095 &  -0.140 &    0.091 \\
  212554013 &  Apr-18-2016 &     6.85 &      0.22 &     4.79 &      0.15 &     2.86 &      0.12 &   1.36 &    0.09 &   1.84 &    0.11 &  $\cdots$ & $\cdots$ &  $\cdots$ & $\cdots$ \\
  212565386 &  Mar-10-2016 &     6.09 &      0.14 &     4.23 &      0.14 &     2.97 &      0.14 &   1.30 &    0.10 &   1.88 &    0.13 &   0.020 &    0.103 &  -0.002 &    0.095 \\
  212572452 &  Mar-10-2016 &     8.64 &      0.18 &     6.09 &      0.17 &     4.44 &      0.16 &   1.55 &    0.13 &   2.37 &    0.18 &   0.206 &    0.112 &   0.184 &    0.102 \\
  212572452 &  Mar-27-2016 &     8.77 &      0.31 &     5.78 &      0.38 &     4.16 &      0.22 &   1.52 &    0.19 &   1.77 &    0.24 &   0.249 &    0.131 &   0.222 &    0.124 \\
  212628098 &  Apr-18-2016 &     6.89 &      0.15 &     4.63 &      0.12 &     3.13 &      0.08 &   1.83 &    0.07 &   2.75 &    0.10 &  -0.008 &    0.093 &   0.015 &    0.088 \\
  212634172 &   Mar-4-2016 &     2.47 &      0.13 &     1.79 &      0.13 &     1.67 &      0.12 &   1.07 &    0.09 &   1.24 &    0.11 &   0.405 &    0.096 &   0.299 &    0.092 \\
  212679181 &   Mar-4-2016 &     3.72 &      0.11 &     2.89 &      0.12 &     2.70 &      0.10 &   1.29 &    0.08 &   1.70 &    0.11 &   0.084 &    0.092 &   0.027 &    0.089 \\
  212679798 &  Apr-18-2016 &     6.08 &      0.22 &     4.44 &      0.16 &     3.08 &      0.12 &   1.82 &    0.10 &   2.77 &    0.15 &   0.402 &    0.104 &   0.296 &    0.097 \\
  212686205 &   Mar-8-2016 &     7.51 &      0.10 &     4.97 &      0.11 &     3.22 &      0.10 &   0.95 &    0.10 &   1.54 &    0.15 & $\cdots$ & $\cdots$ &  $\cdots$ & $\cdots$ \\
  212690867 &   Mar-8-2016 &     3.56 &      0.18 &     2.62 &      0.18 &     3.07 &      0.18 &   1.59 &    0.15 &   2.11 &    0.20 &  -0.188 &    0.114 &  -0.187 &    0.104 \\
  212773272 &  Apr-18-2016 &     2.79 &      0.18 &     2.32 &      0.12 &     2.08 &      0.11 &   1.68 &    0.07 &   2.36 &    0.10 &   0.329 &    0.098 &   0.250 &    0.092 \\
  212773309 &  Mar-28-2016 &     6.87 &      0.09 &     4.72 &      0.08 &     3.28 &      0.05 &   1.53 &    0.05 &   2.10 &    0.08 &   0.288 &    0.085 &   0.123 &    0.084 \\
  212773309B &  Mar-28-2016 &     3.37 &      0.33 &     2.07 &      0.20 &     2.62 &      0.17 &   1.61 &    0.11 &   2.20 &    0.14 &   0.595 &    0.144 &   0.251 &    0.115 \\
  213951550 &   May-6-2016 &     4.84 &      0.17 &     3.23 &      0.15 &     3.15 &      0.15 &   1.58 &    0.11 &   2.66 &    0.13 &   0.153 &    0.110 &   0.099 &    0.097 \\
  214254518 &   May-5-2016 &     7.57 &      0.08 &     5.09 &      0.09 &     3.75 &      0.08 &   1.38 &    0.07 &   1.95 &    0.09 &  -0.074 &    0.090 &  -0.055 &    0.086 \\
  214254518 &  Oct-26-2016 &     8.00 &      0.08 &     5.53 &      0.09 &     3.73 &      0.07 &   1.30 &    0.07 &   1.73 &    0.10 &  -0.130 &    0.089 &  -0.058 &    0.086 \\
  214522613 &   May-5-2016 &     4.35 &      0.21 &     2.90 &      0.20 &     2.93 &      0.19 &   1.73 &    0.13 &   2.45 &    0.17 &   0.407 &    0.118 &   0.125 &    0.106 \\
  214787262 &   May-5-2016 &     2.70 &      0.09 &     1.96 &      0.09 &     1.89 &      0.08 &   0.99 &    0.06 &   1.32 &    0.08 &   0.006 &    0.089 &   0.023 &    0.086 \\
  216892056 &   May-5-2016 &     2.94 &      0.10 &     2.31 &      0.11 &     2.50 &      0.10 &   1.52 &    0.08 &   1.85 &    0.11 &  -0.111 &    0.093 &  -0.116 &    0.089 \\
  217941732 &   May-5-2016 &     7.26 &      0.24 &     5.05 &      0.22 &     3.86 &      0.23 &   1.08 &    0.17 &   1.69 &    0.20 &  $\cdots$ & $\cdots$ &  $\cdots$ & $\cdots$ \\
  217941732 &  Oct-26-2016 &     7.03 &      0.23 &     5.08 &      0.23 &     4.21 &      0.21 &   1.37 &    0.16 &   1.95 &    0.20 &  $\cdots$ & $\cdots$ &  $\cdots$ & $\cdots$ \\
\enddata
\tablenotetext{1}{Estimated by taking the average of the $H$-band and $K$-band estimates determined using the spectral indices introduced by \citet{mann_et_al2013a}. We do not report [Fe/H] and [M/H] for K3 -- K5 dwarfs because the \citet{mann_et_al2013a} relations are not valid for those stars.}
\end{deluxetable*}

\appendix
\section{Reduced Stellar Spectra}
\label{sec:appendix}
\begin{figure*}[tbhp]
\centering
\includegraphics[width=0.49\textwidth]{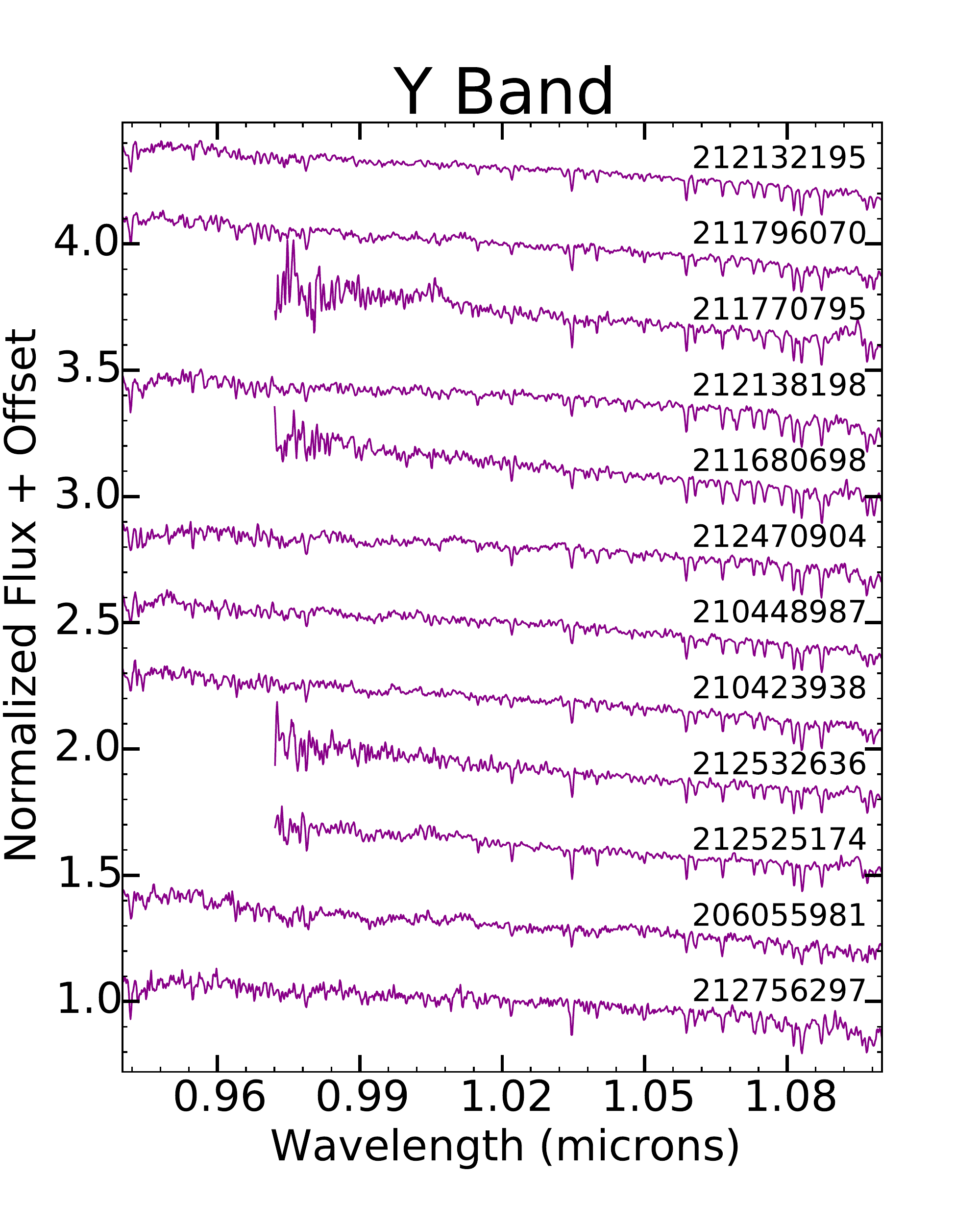}
\includegraphics[width=0.49\textwidth]{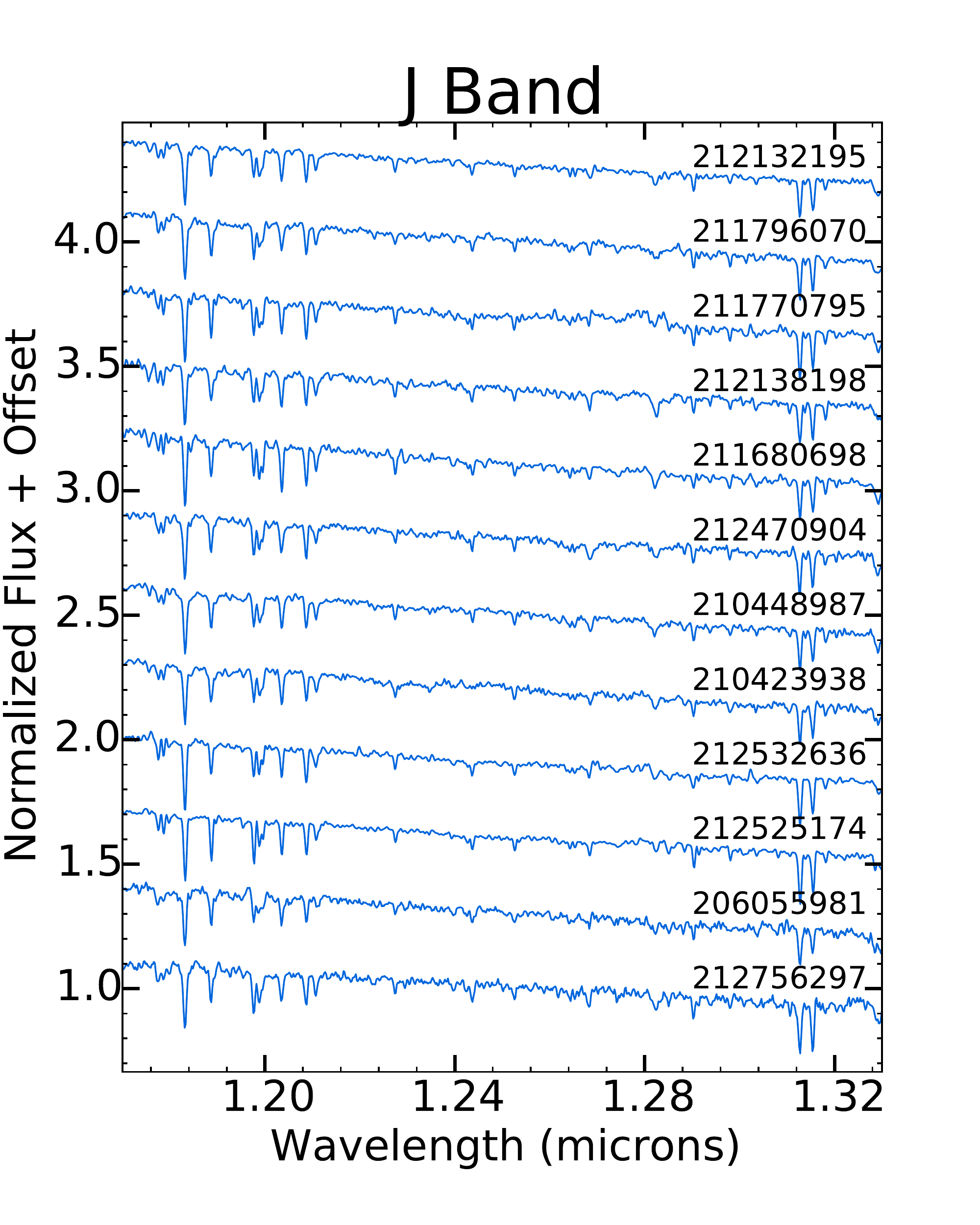}\\
\includegraphics[width=0.49\textwidth]{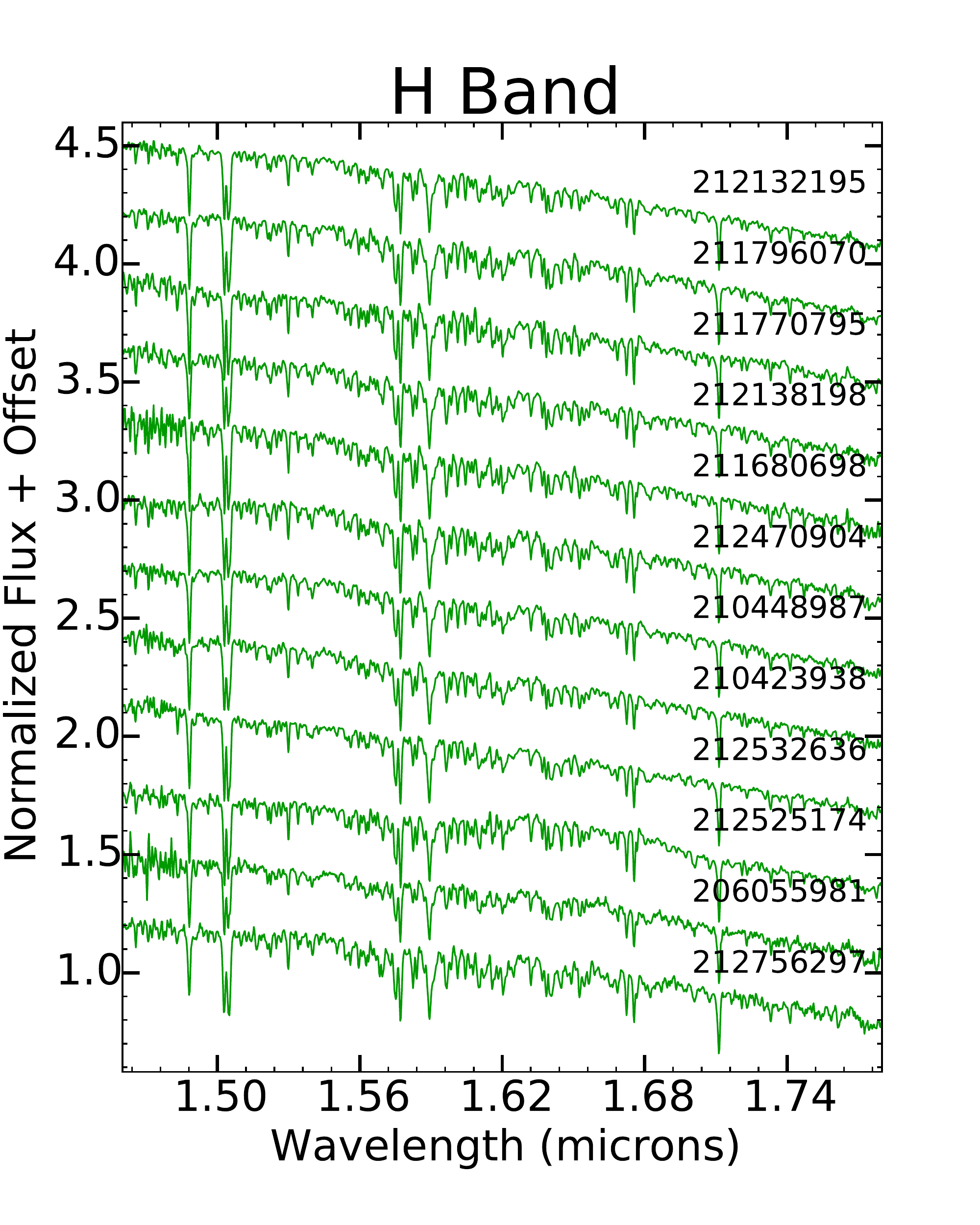}
\includegraphics[width=0.49\textwidth]{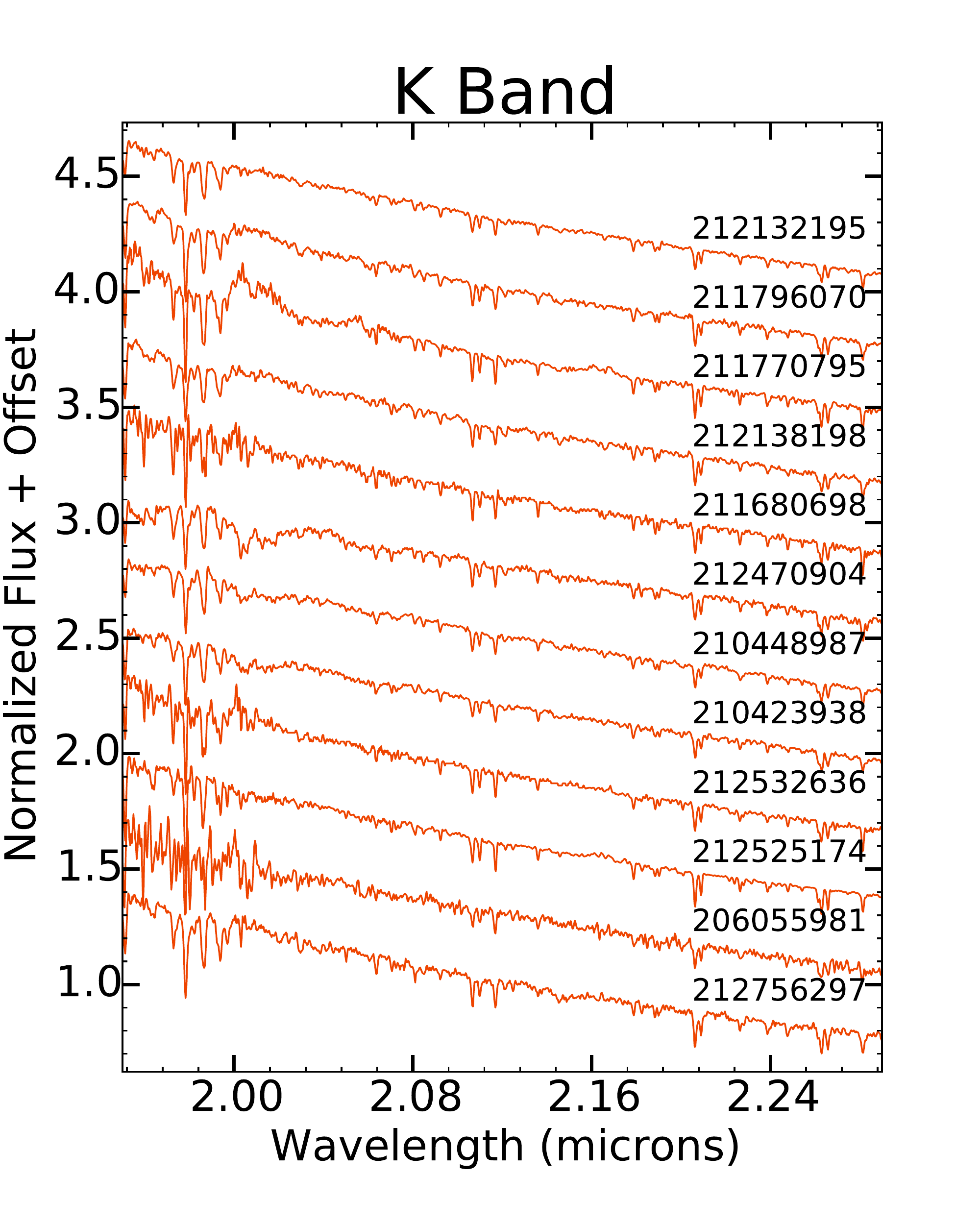}\\
\caption{$Y$-band (top left), $J$-band (top right), $H$-band (bottom left), and $K$-band (bottom right) spectra of cool dwarfs with effective temperatures between 4800K and 4480K. The hottest stars are shown at the top of the plots. Stars with truncated $Y$-band coverage were observed at the Palomar 200'' Hale Telescope using TripleSpec; the other stars were observed at the IRTF using SpeX. \label{fig:dwarfspec5}}
\end{figure*}

\begin{figure*}[tbhp]
\centering
\includegraphics[width=0.49\textwidth]{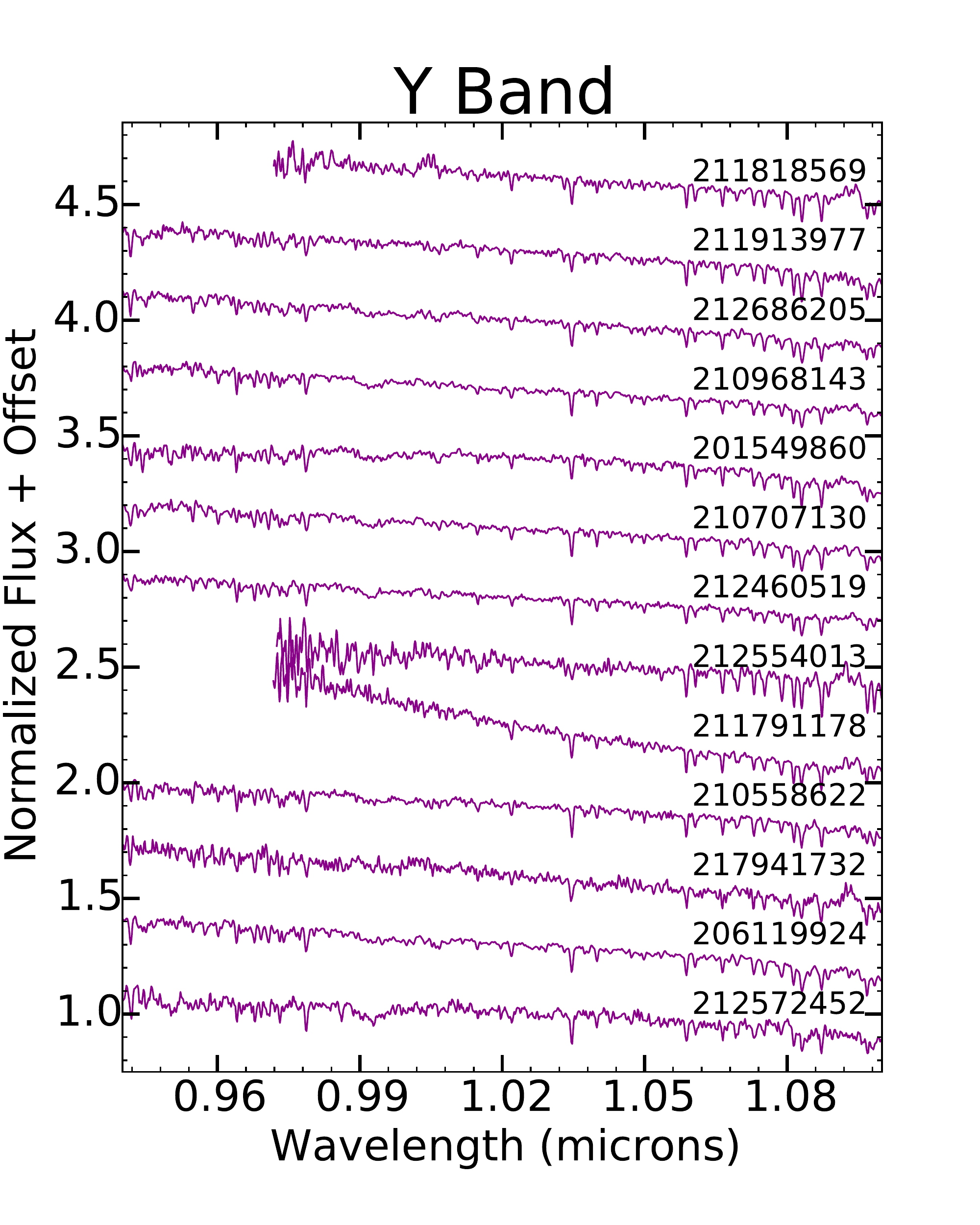}
\includegraphics[width=0.49\textwidth]{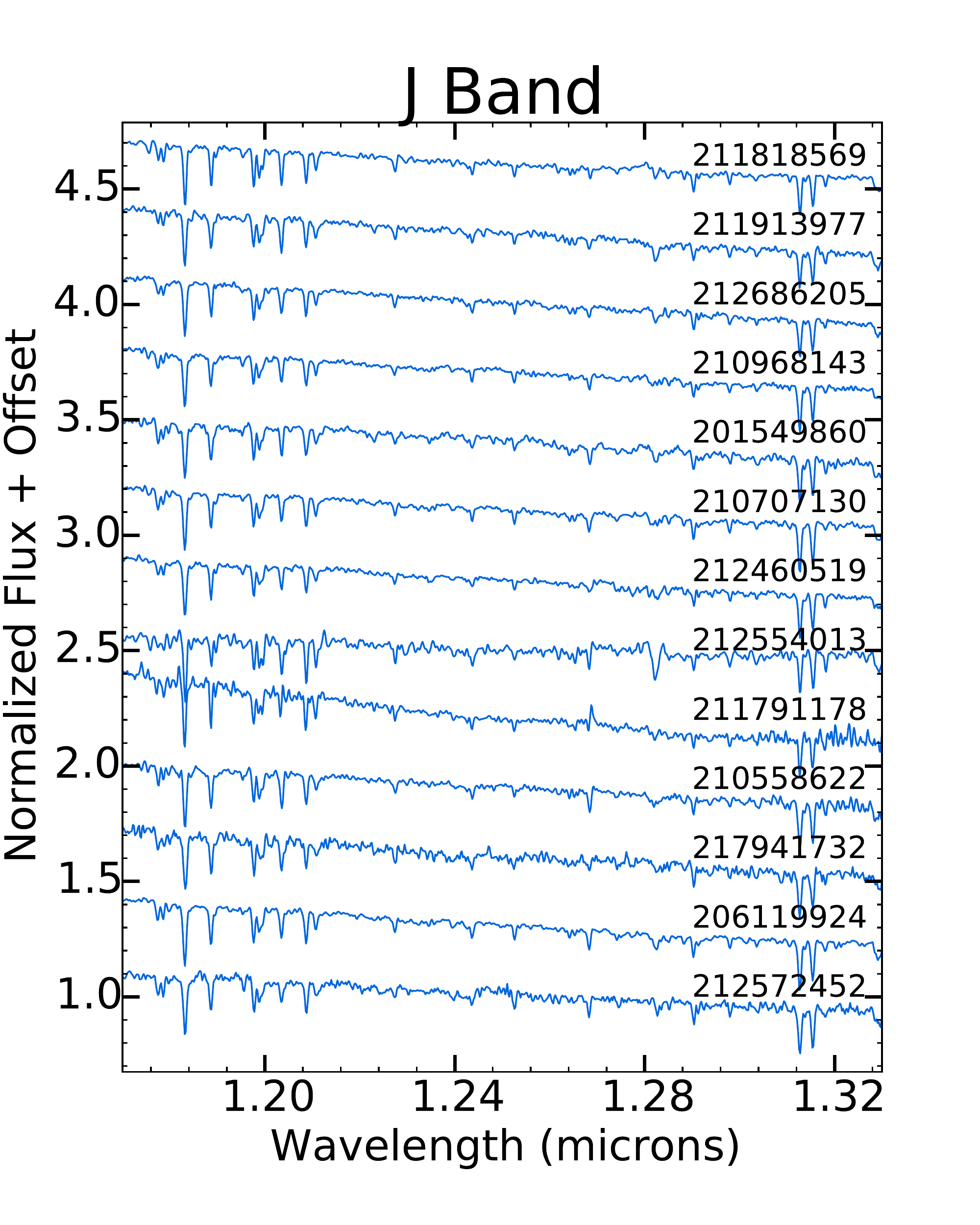}\\
\includegraphics[width=0.49\textwidth]{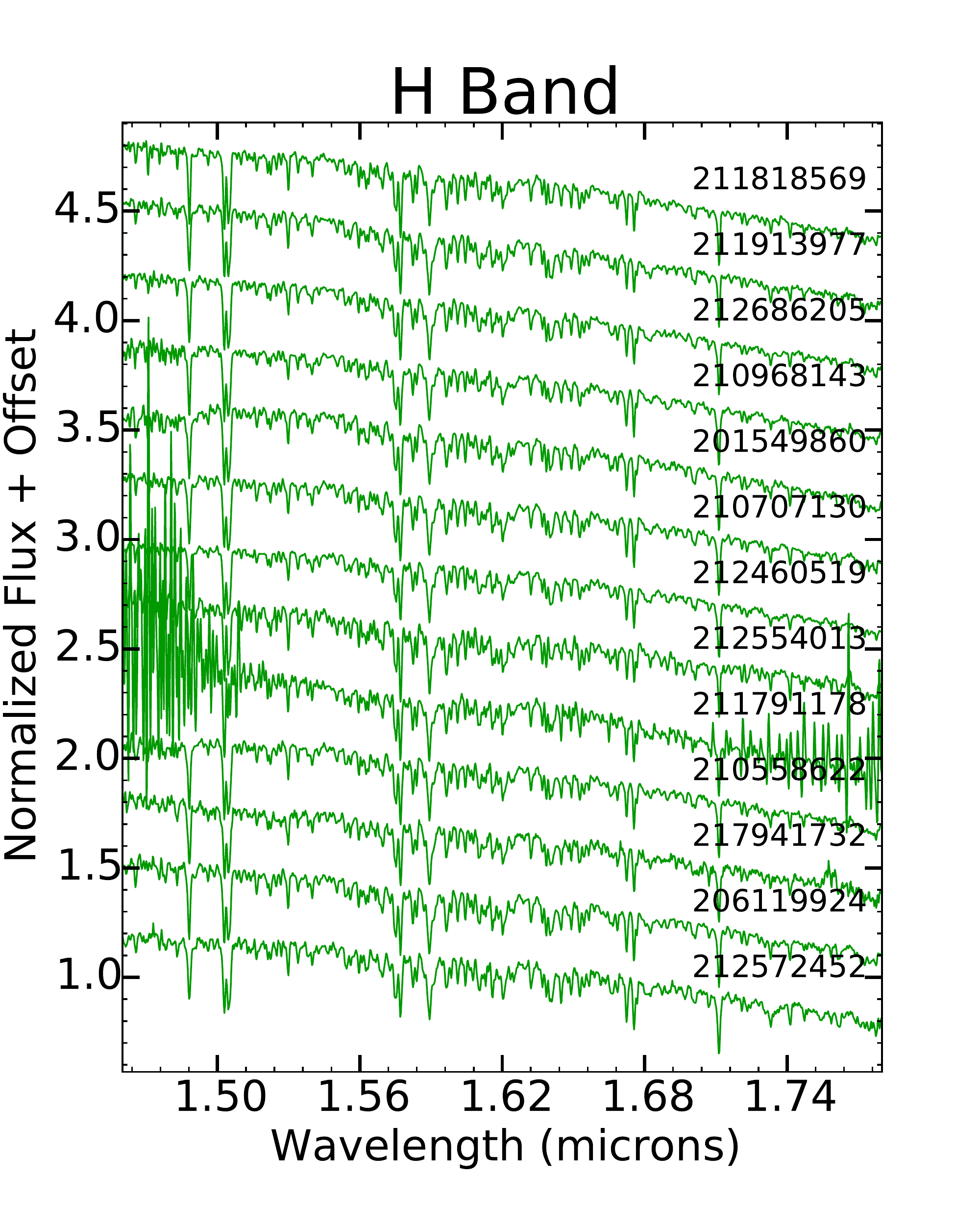}
\includegraphics[width=0.49\textwidth]{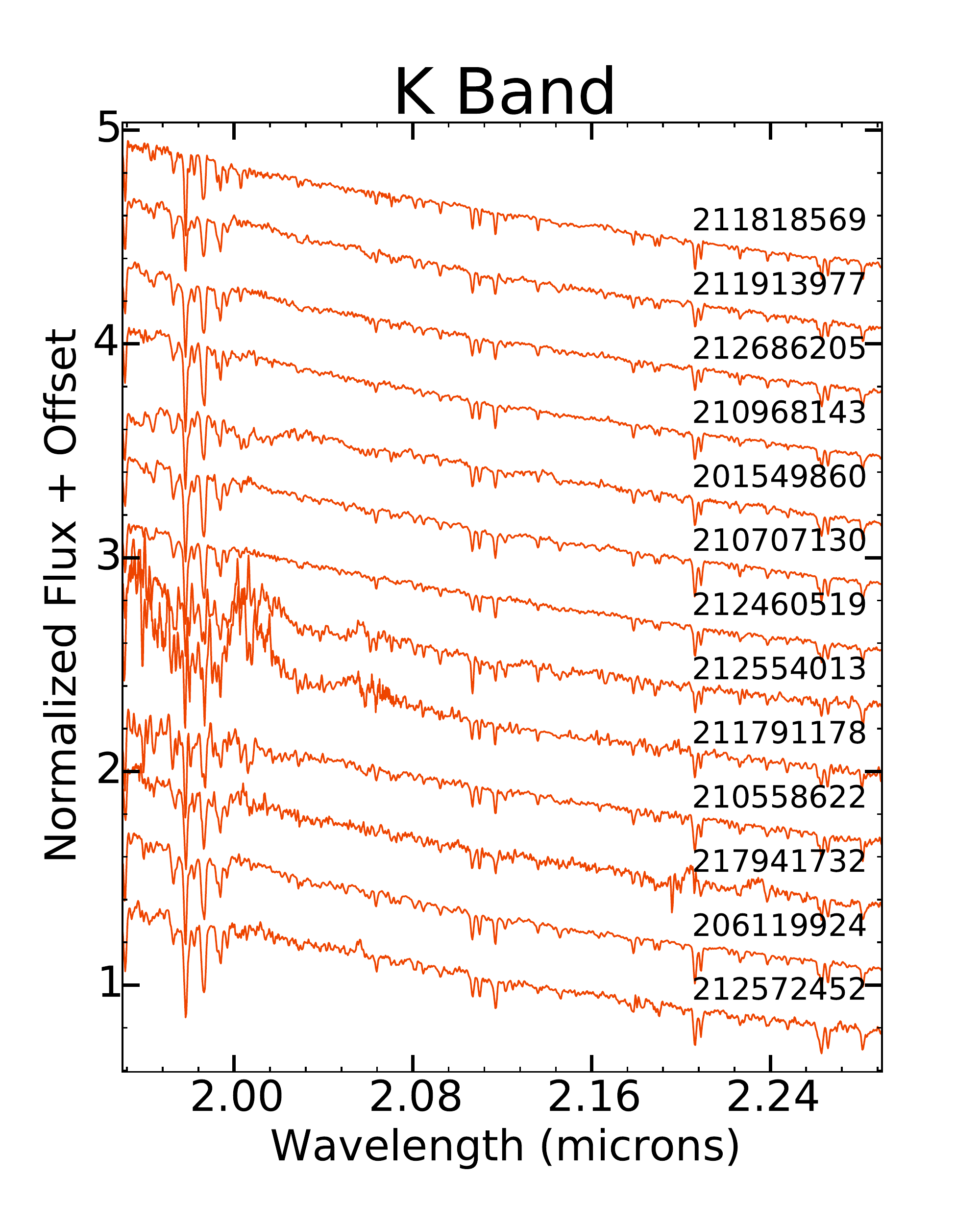}\\
\caption{Same as Figure~\ref{fig:dwarfspec5} for cool dwarfs with effective temperatures between 4480K and 4333K. \label{fig:dwarfspec4}}
\end{figure*}

\begin{figure*}[tbhp]
\centering
\includegraphics[width=0.49\textwidth]{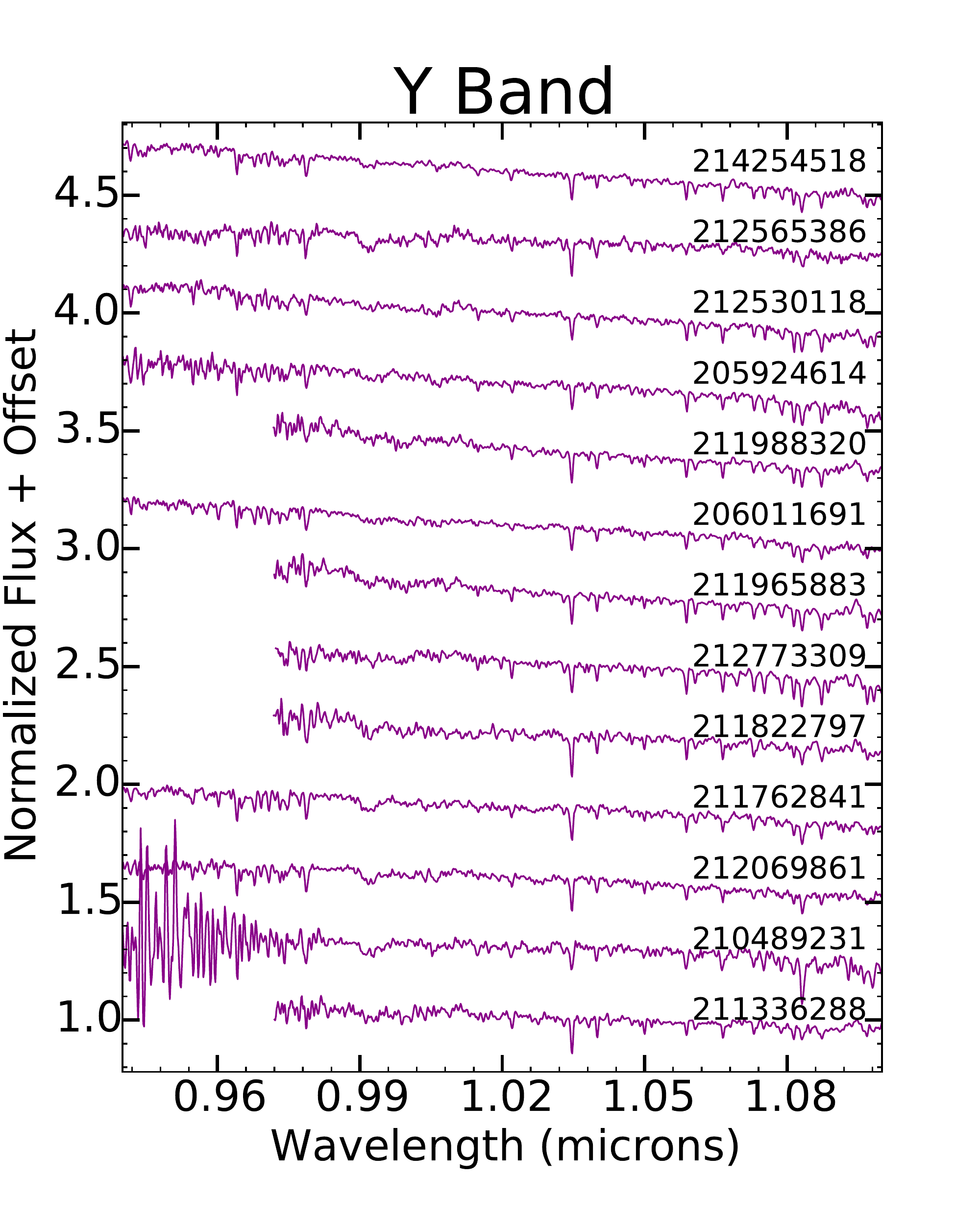}
\includegraphics[width=0.49\textwidth]{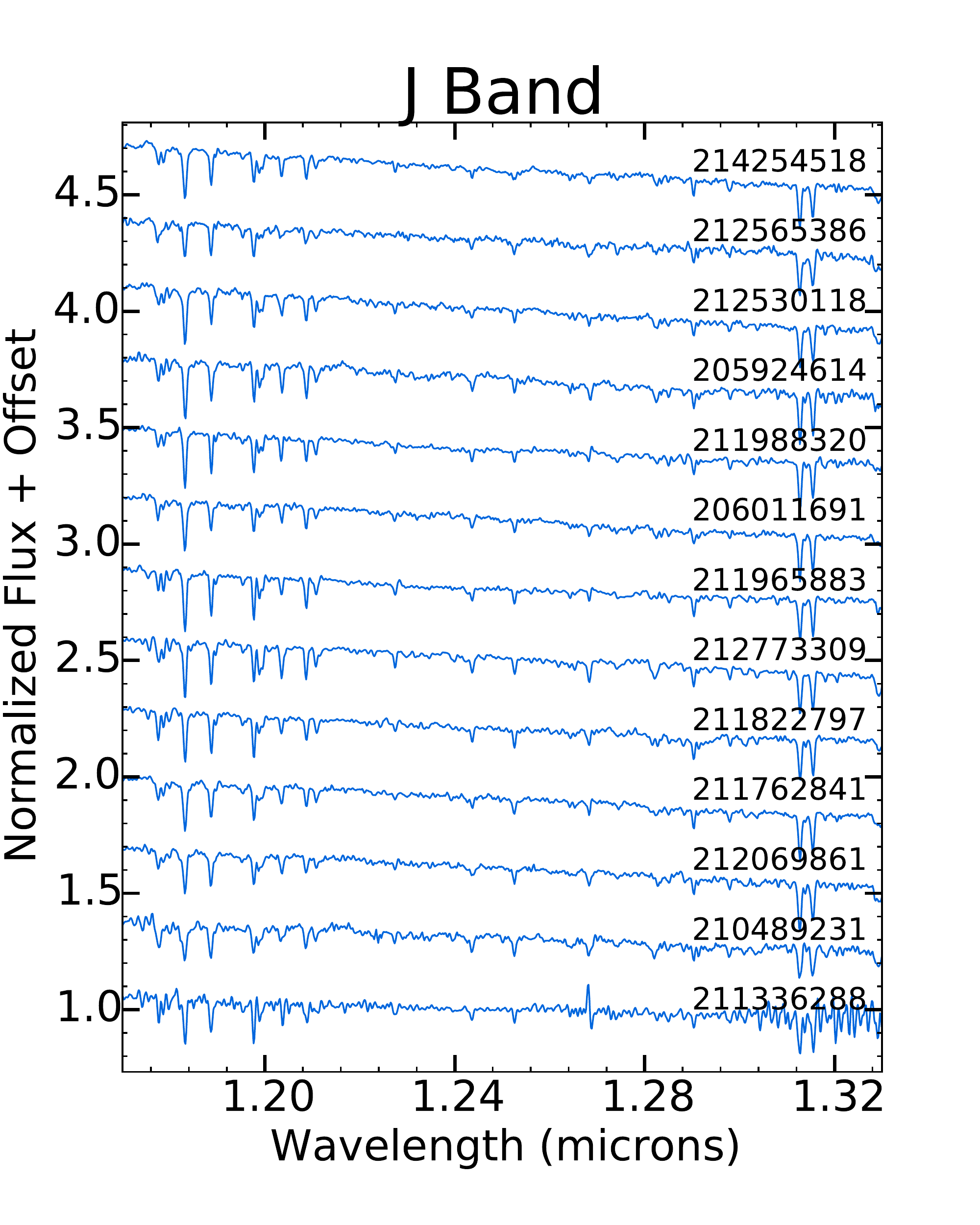}\\
\includegraphics[width=0.49\textwidth]{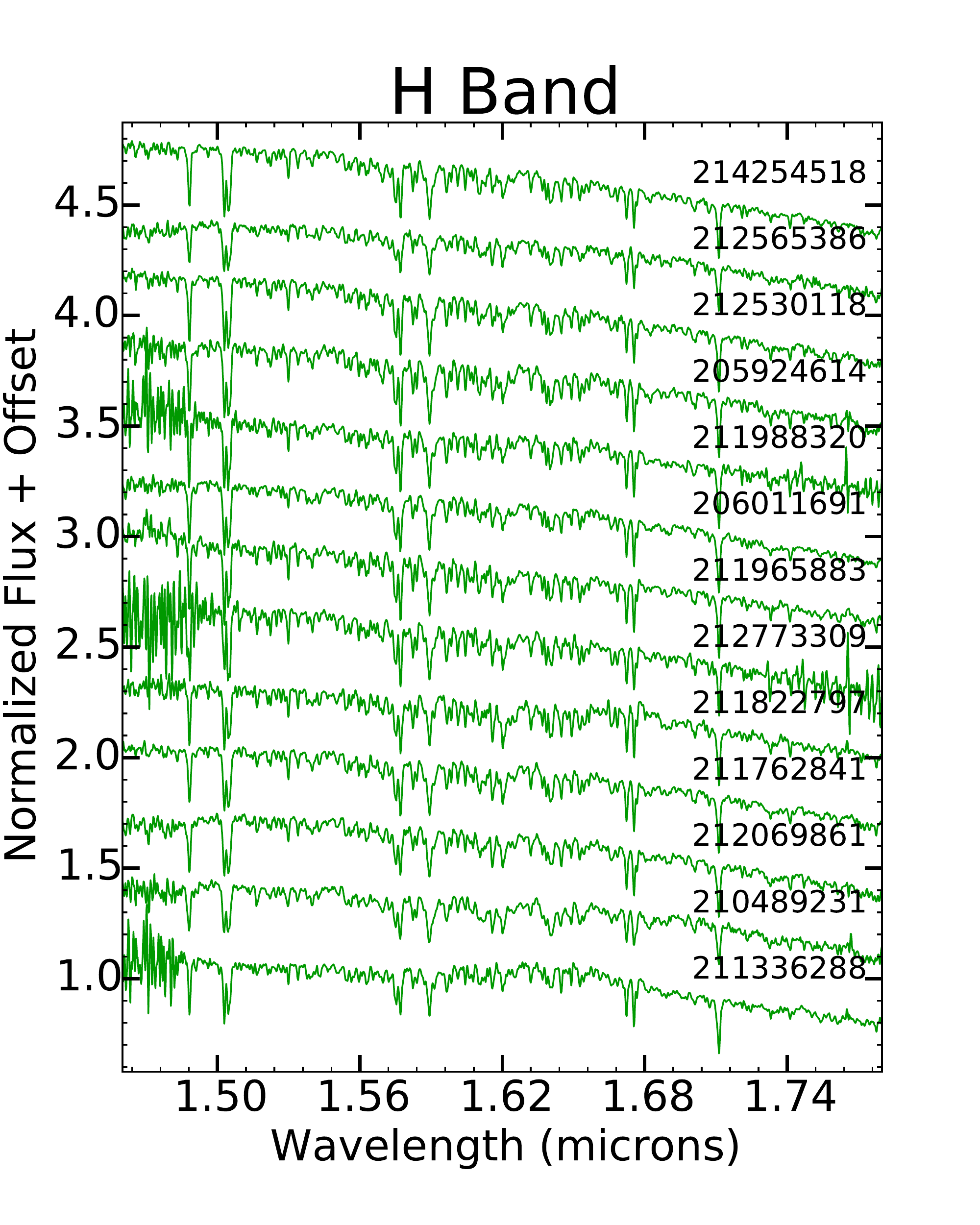}
\includegraphics[width=0.49\textwidth]{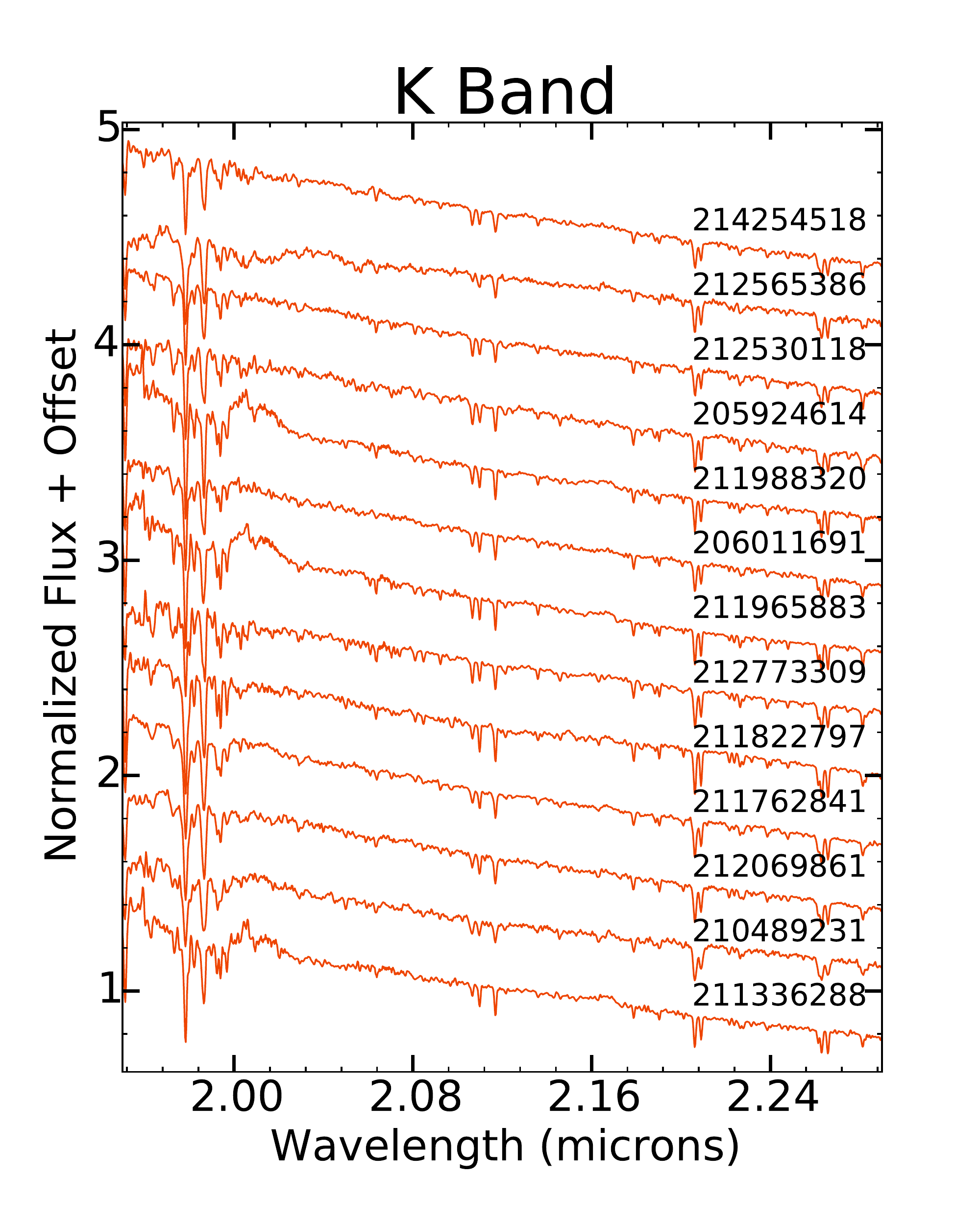}\\
\caption{Same as Figure~\ref{fig:dwarfspec5} for cool dwarfs with effective temperatures between 4333K and 3995K. \label{fig:dwarfspec3}}
\end{figure*}

\begin{figure*}[tbhp]
\centering
\includegraphics[width=0.49\textwidth]{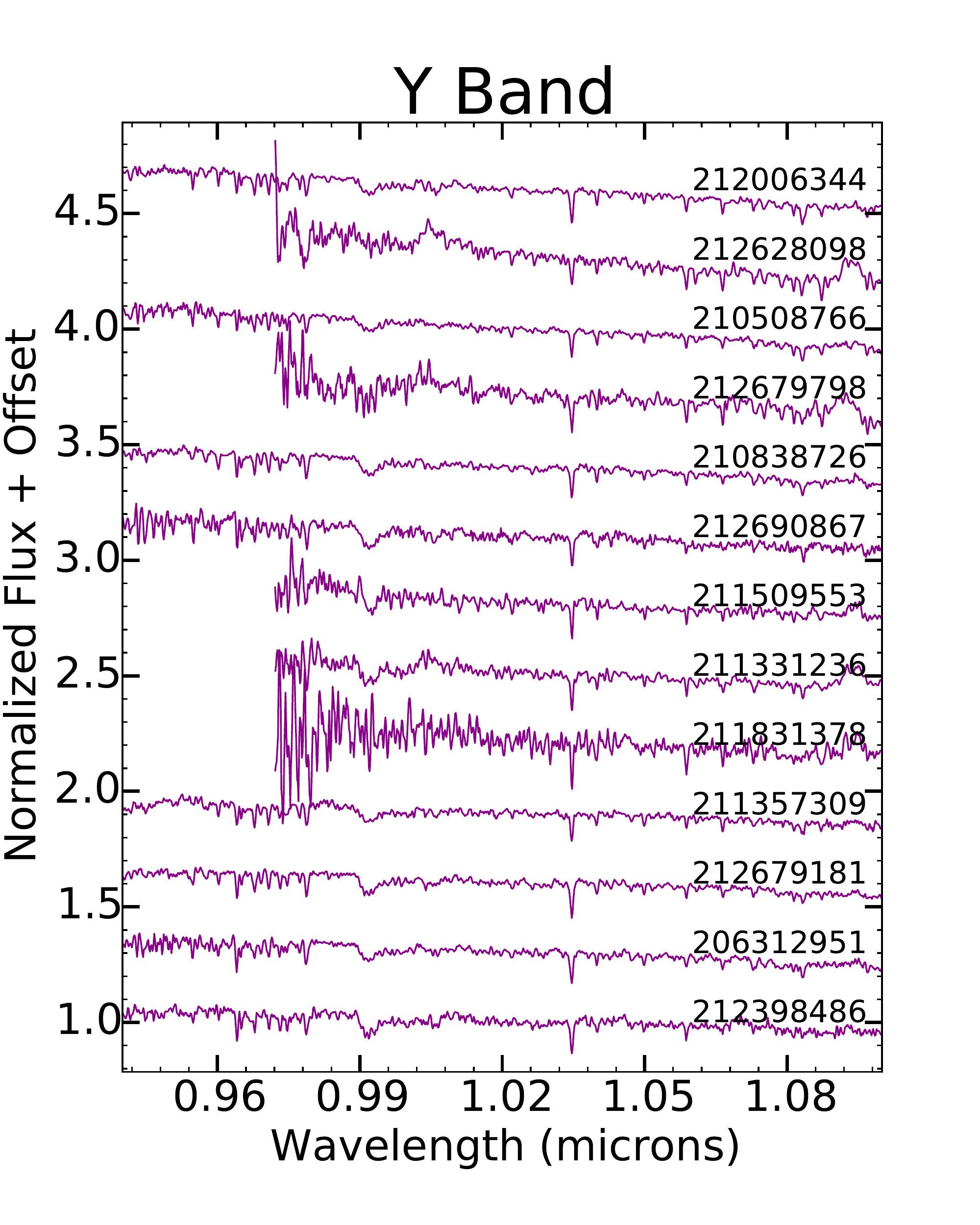}
\includegraphics[width=0.49\textwidth]{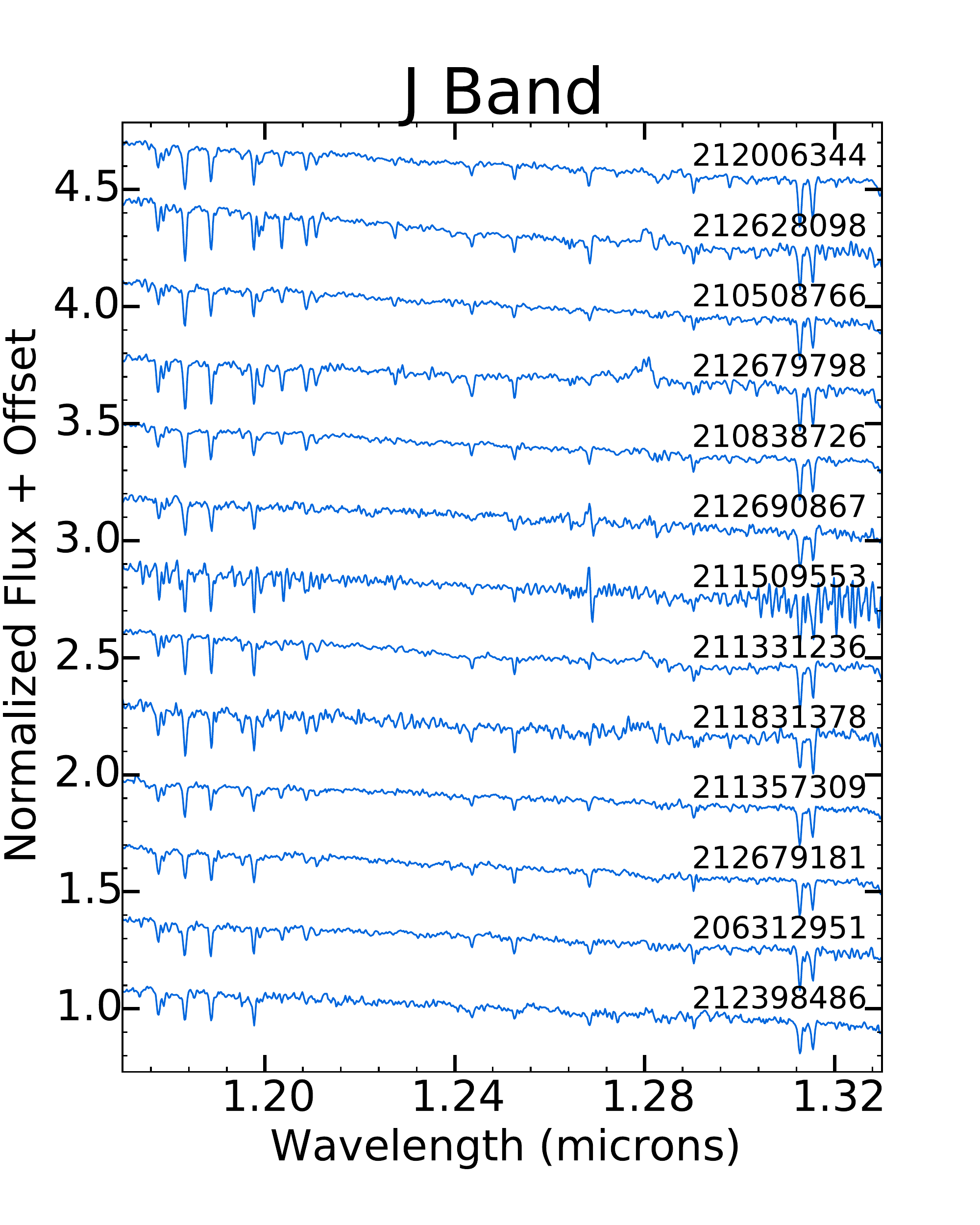}\\
\includegraphics[width=0.49\textwidth]{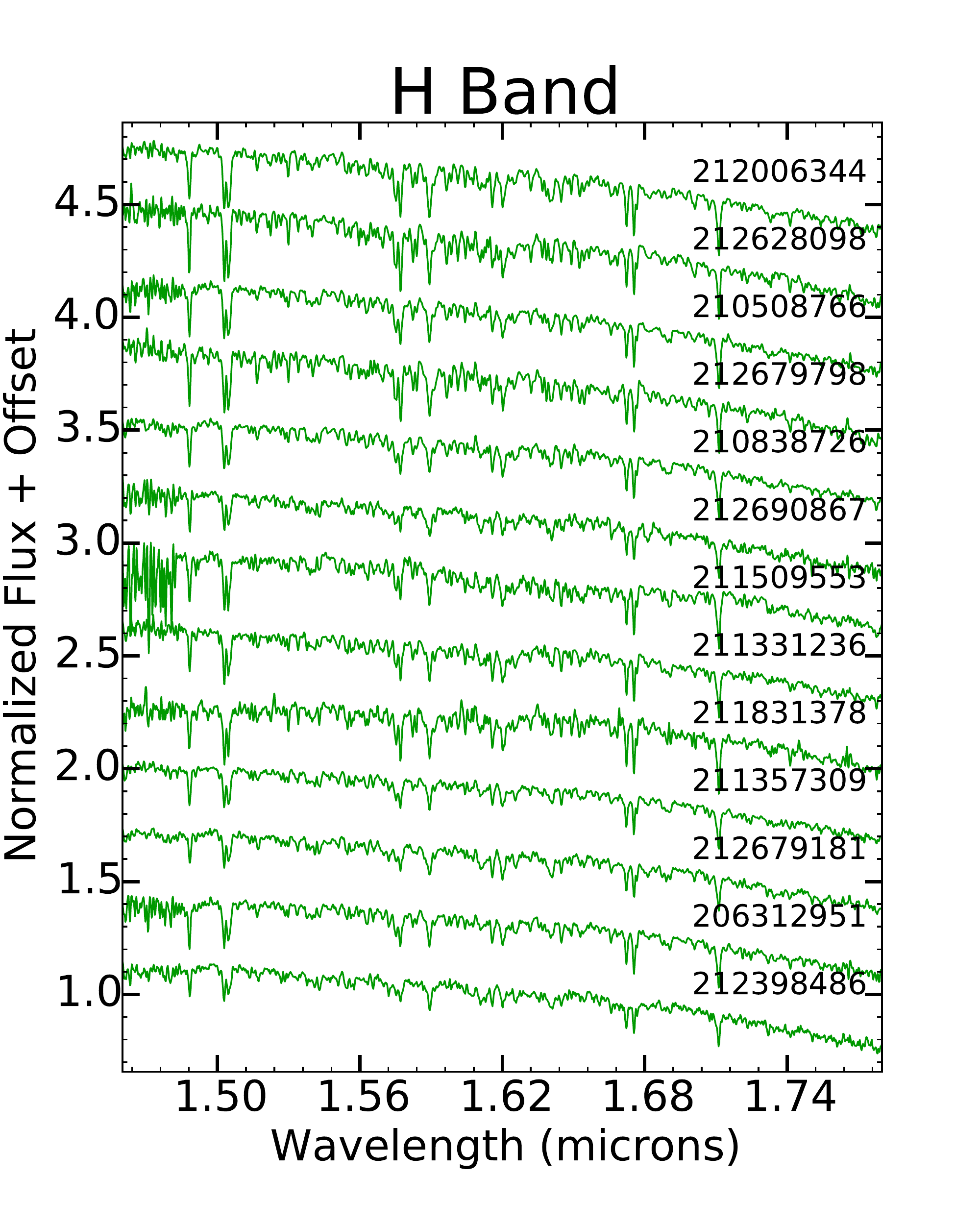}
\includegraphics[width=0.49\textwidth]{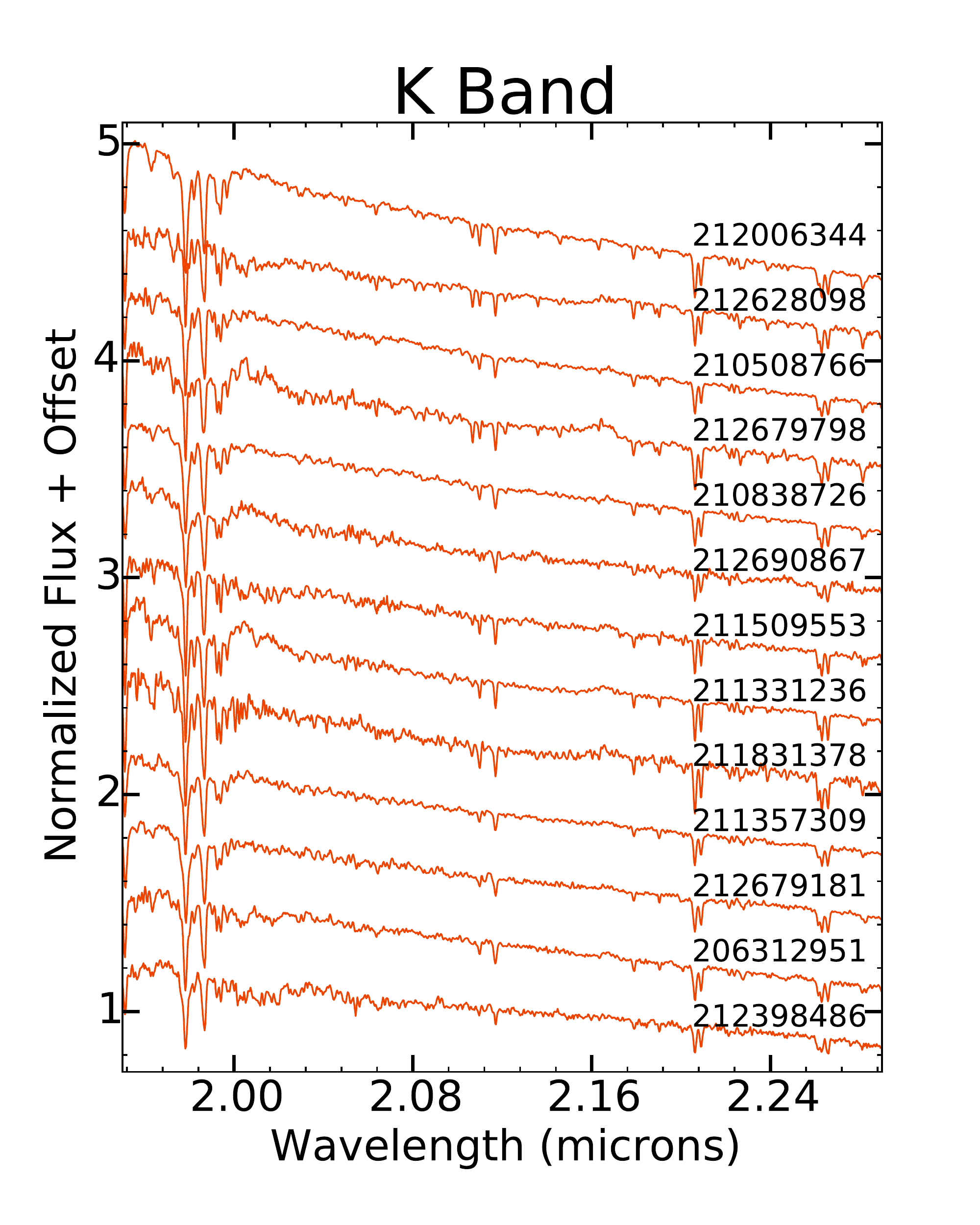}\\
\caption{Same as Figure~\ref{fig:dwarfspec5} for cool dwarfs with effective temperatures between 3995K and 3650K. \label{fig:dwarfspec2}}
\end{figure*}

\begin{figure*}[tbhp]
\centering
\includegraphics[width=0.49\textwidth]{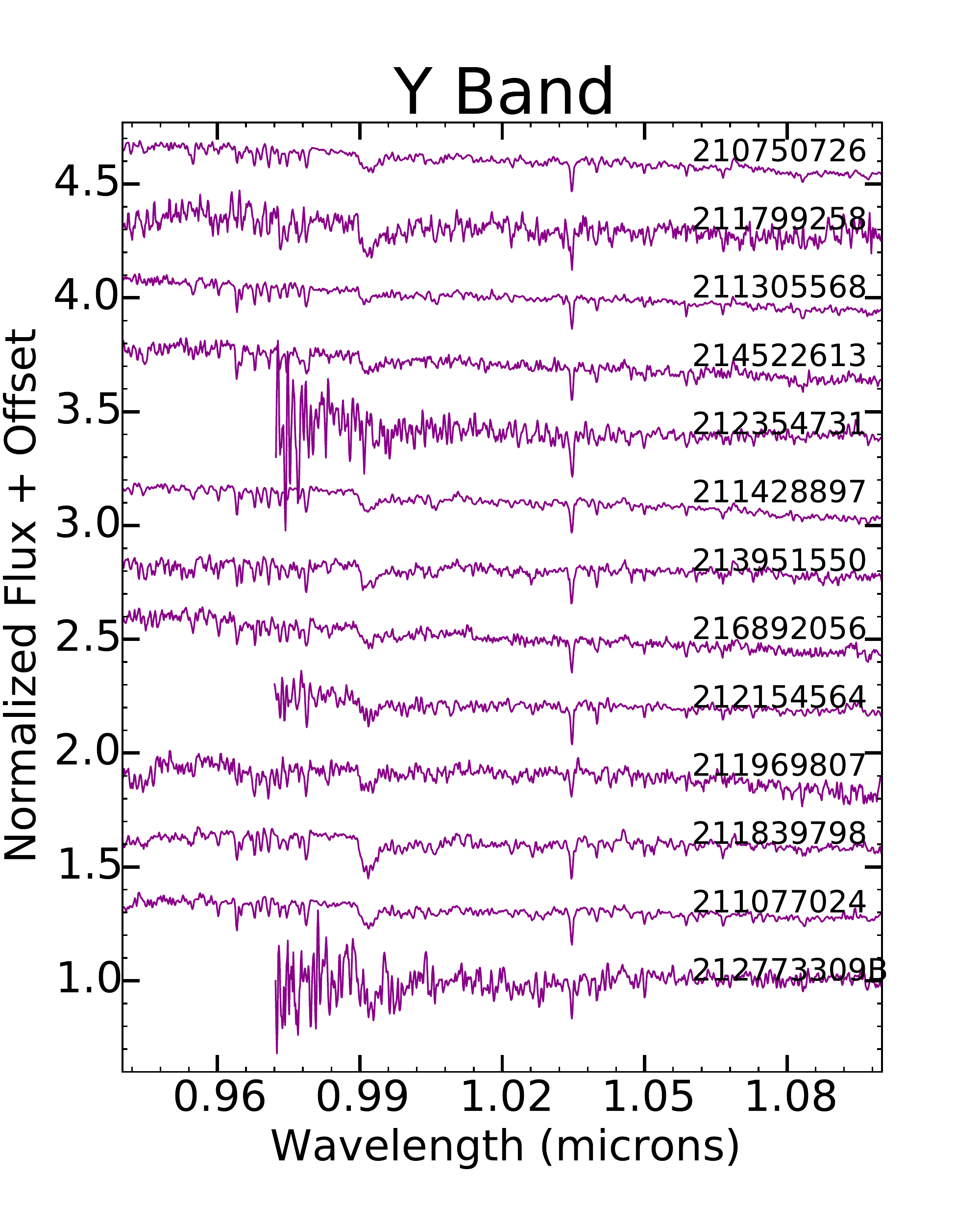}
\includegraphics[width=0.49\textwidth]{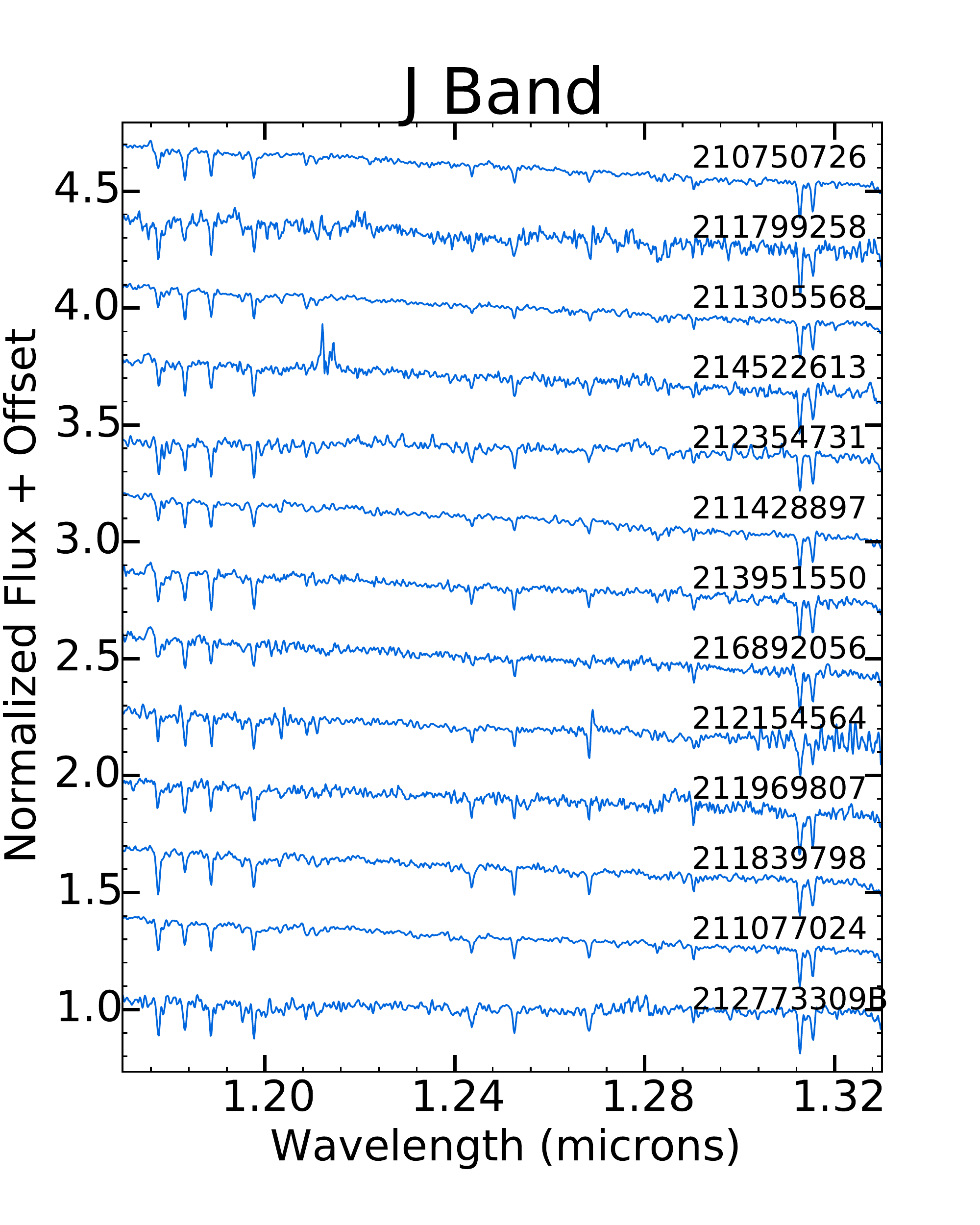}\\
\includegraphics[width=0.49\textwidth]{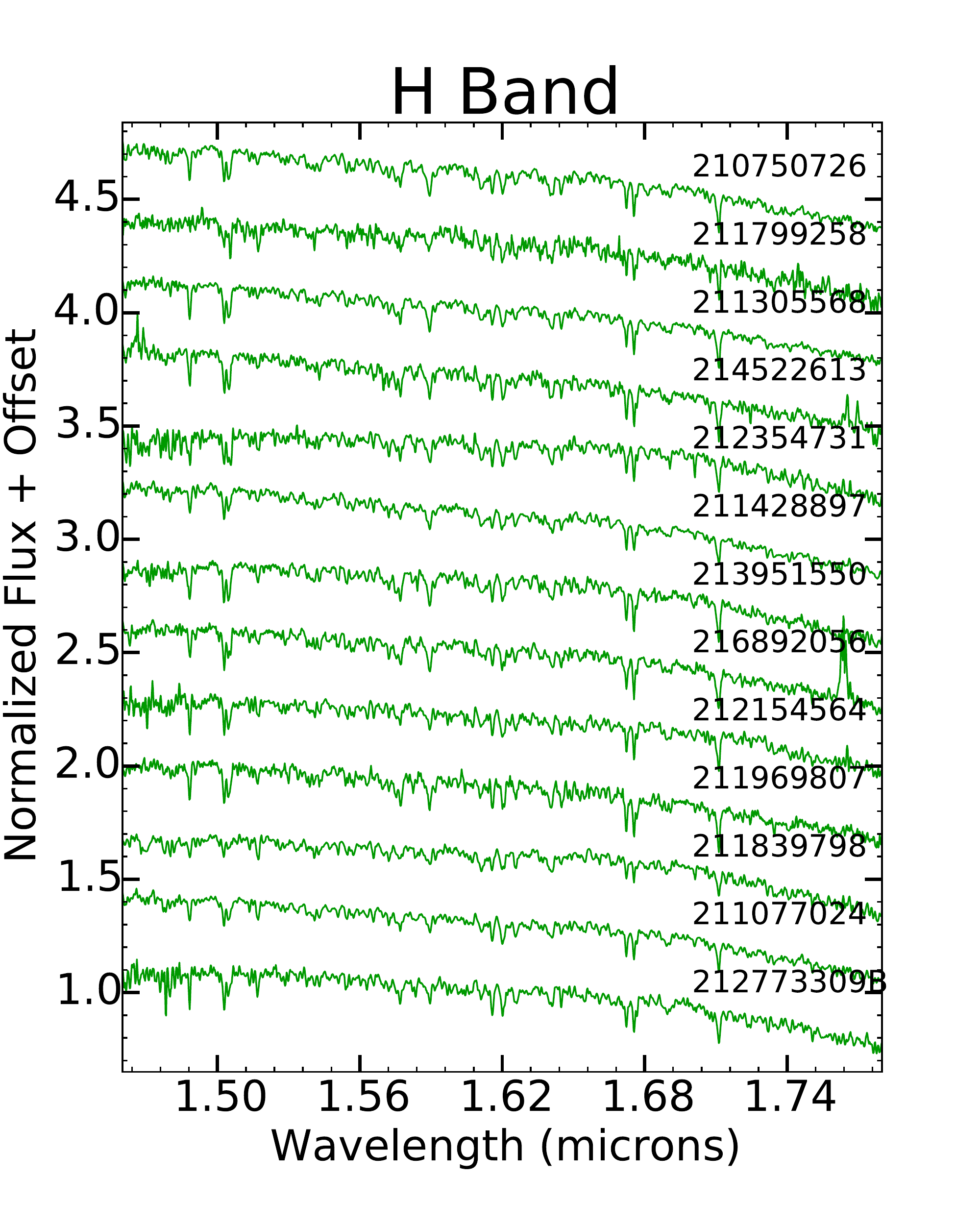}
\includegraphics[width=0.49\textwidth]{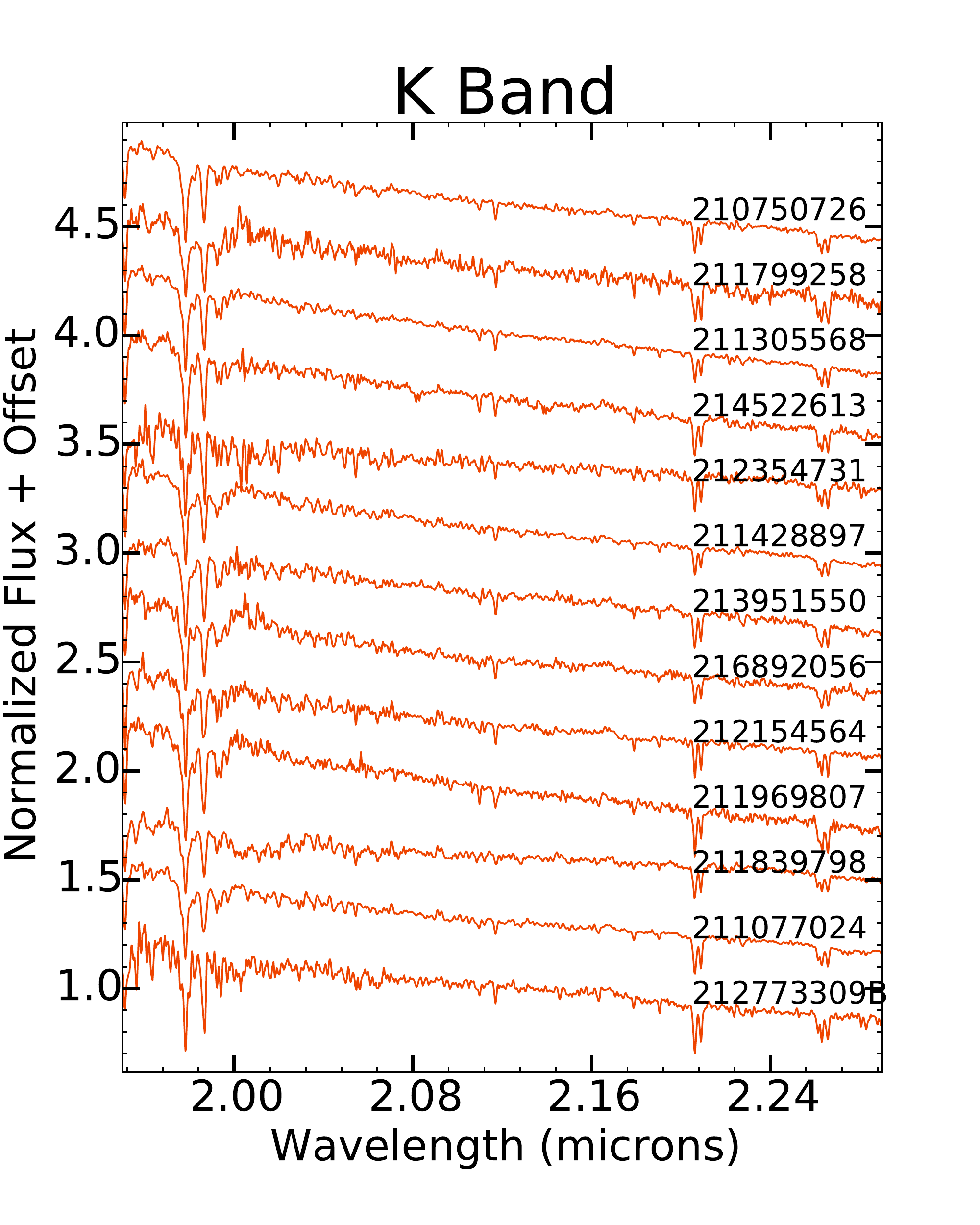}\\
\caption{Same as Figure~\ref{fig:dwarfspec5} for cool dwarfs with effective temperatures between 3650K and 3465K. \label{fig:dwarfspec1}}
\end{figure*}

\begin{figure*}[tbhp]
\centering
\includegraphics[width=0.49\textwidth]{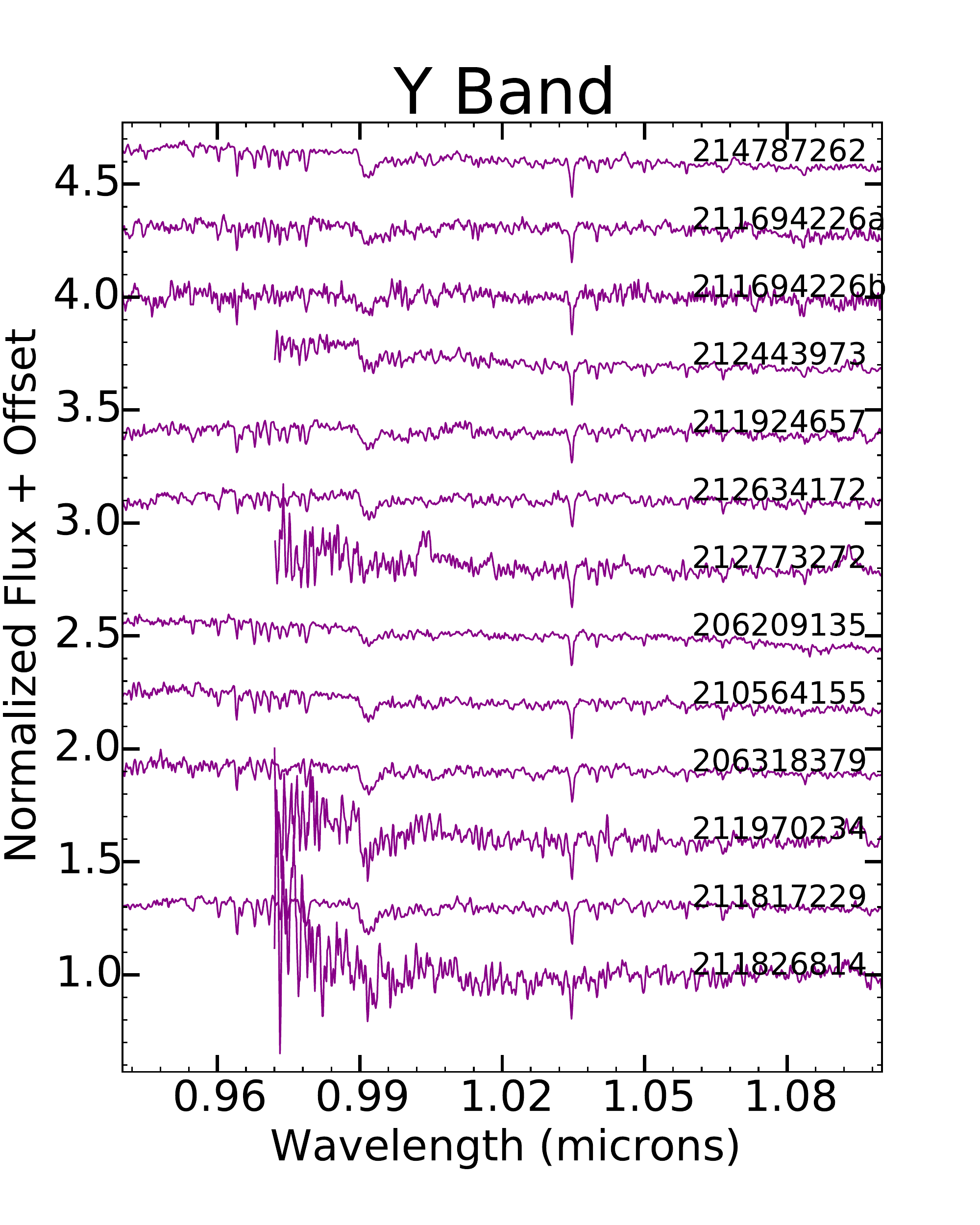}
\includegraphics[width=0.49\textwidth]{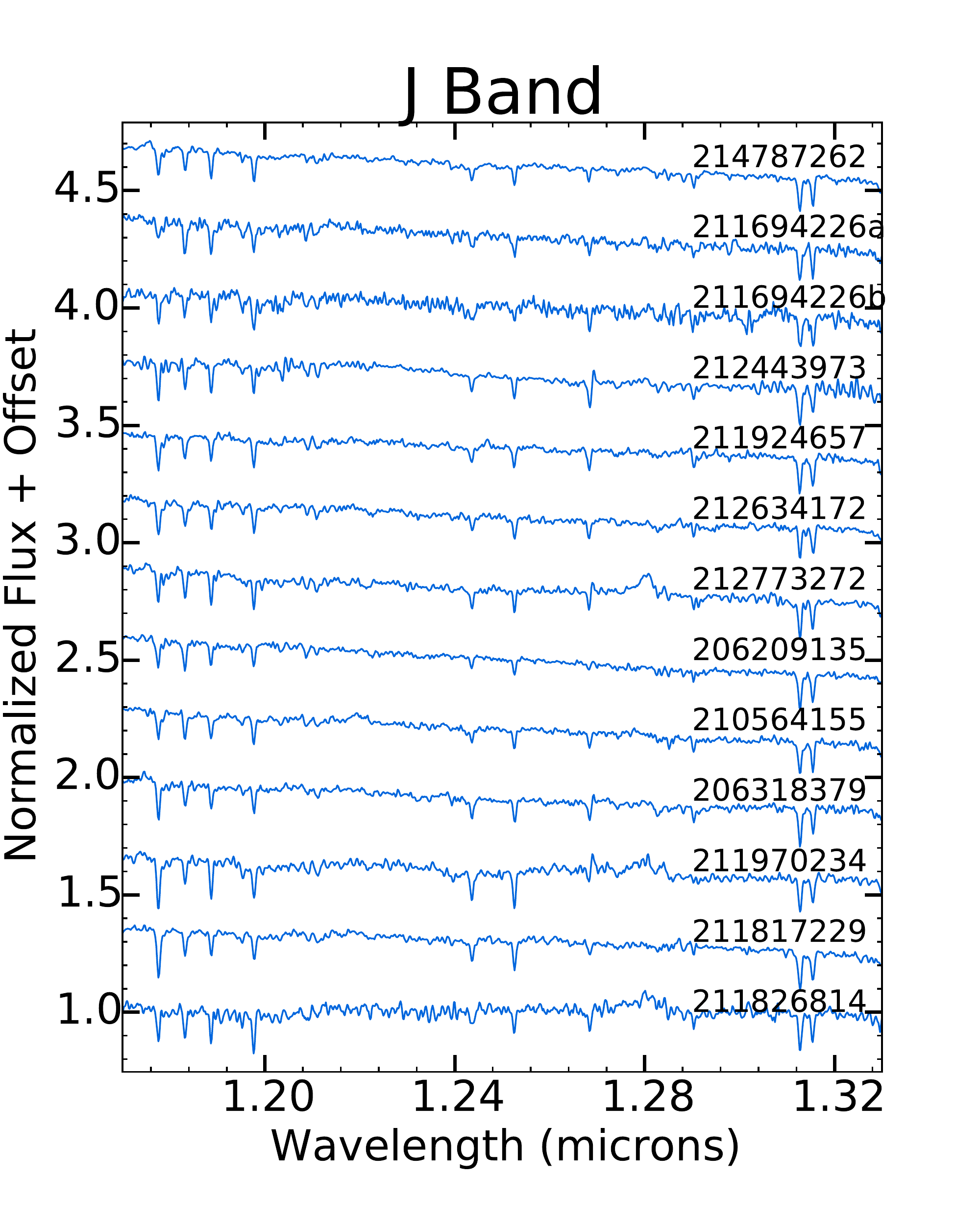}\\
\includegraphics[width=0.49\textwidth]{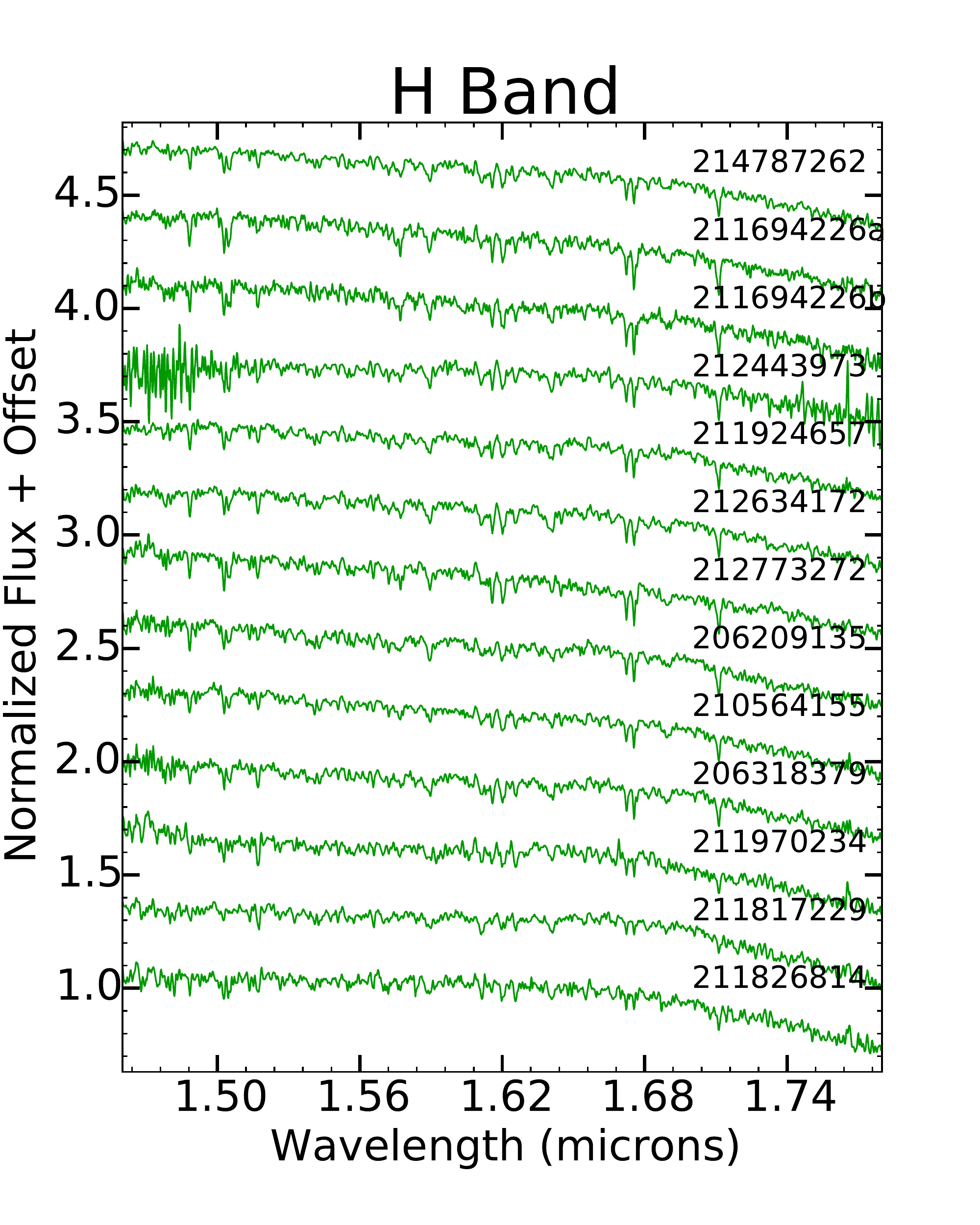}
\includegraphics[width=0.49\textwidth]{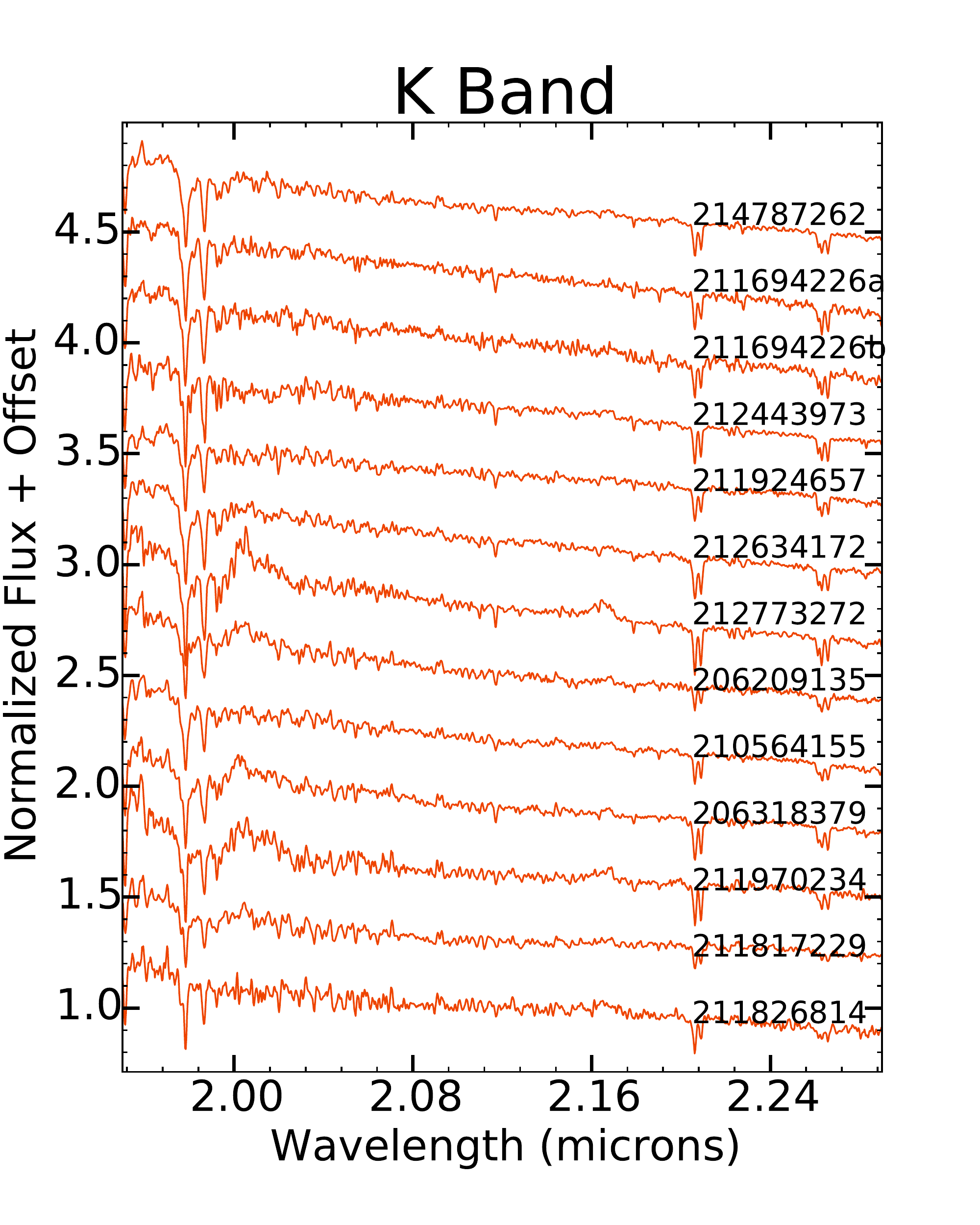}\\
\caption{Same as Figure~\ref{fig:dwarfspec5} for cool dwarfs with effective temperatures between 3465K and 3220K.  \label{fig:dwarfspec0}}
\end{figure*}

\begin{figure*}[tbhp]
\centering
\includegraphics[width=0.49\textwidth]{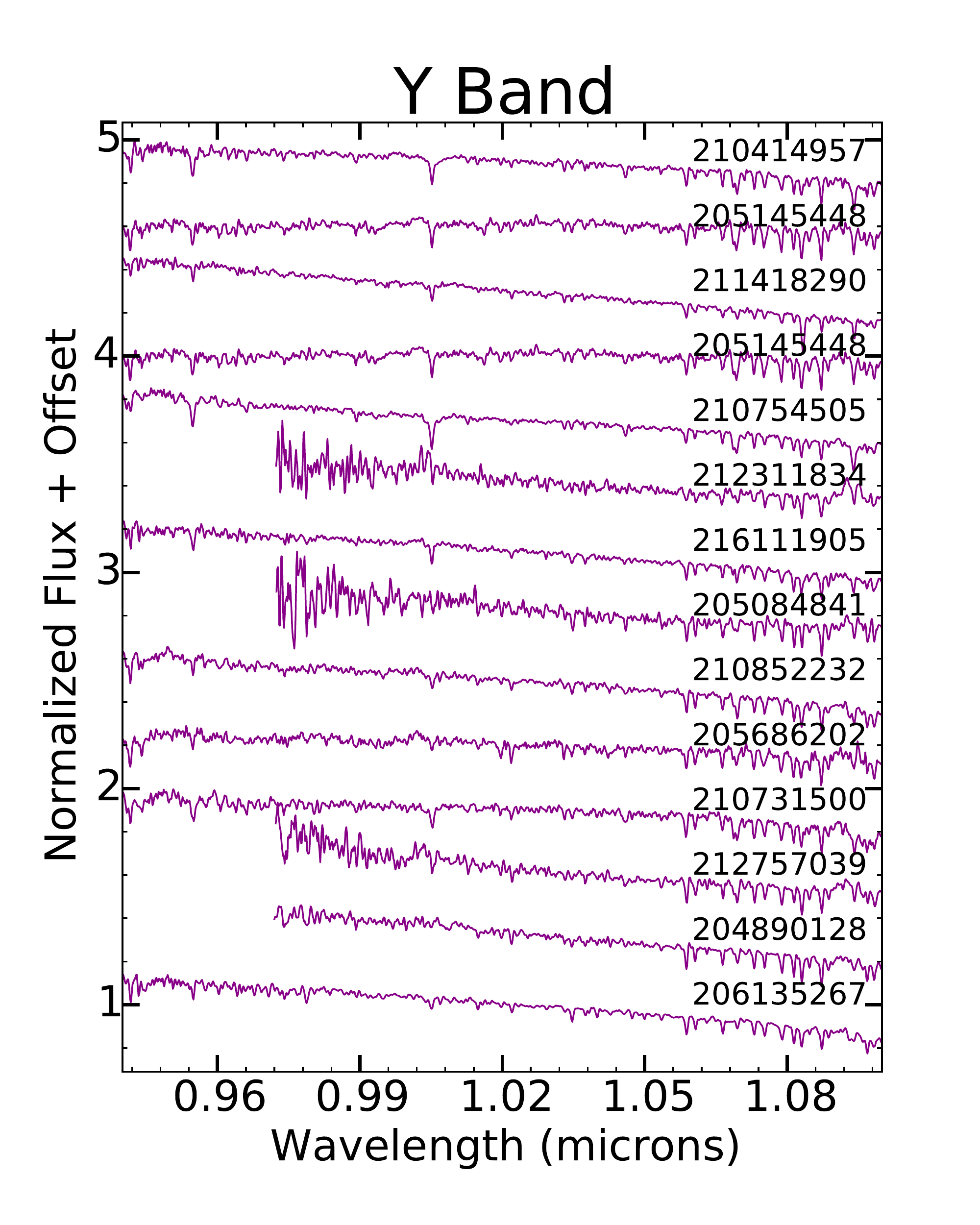}
\includegraphics[width=0.49\textwidth]{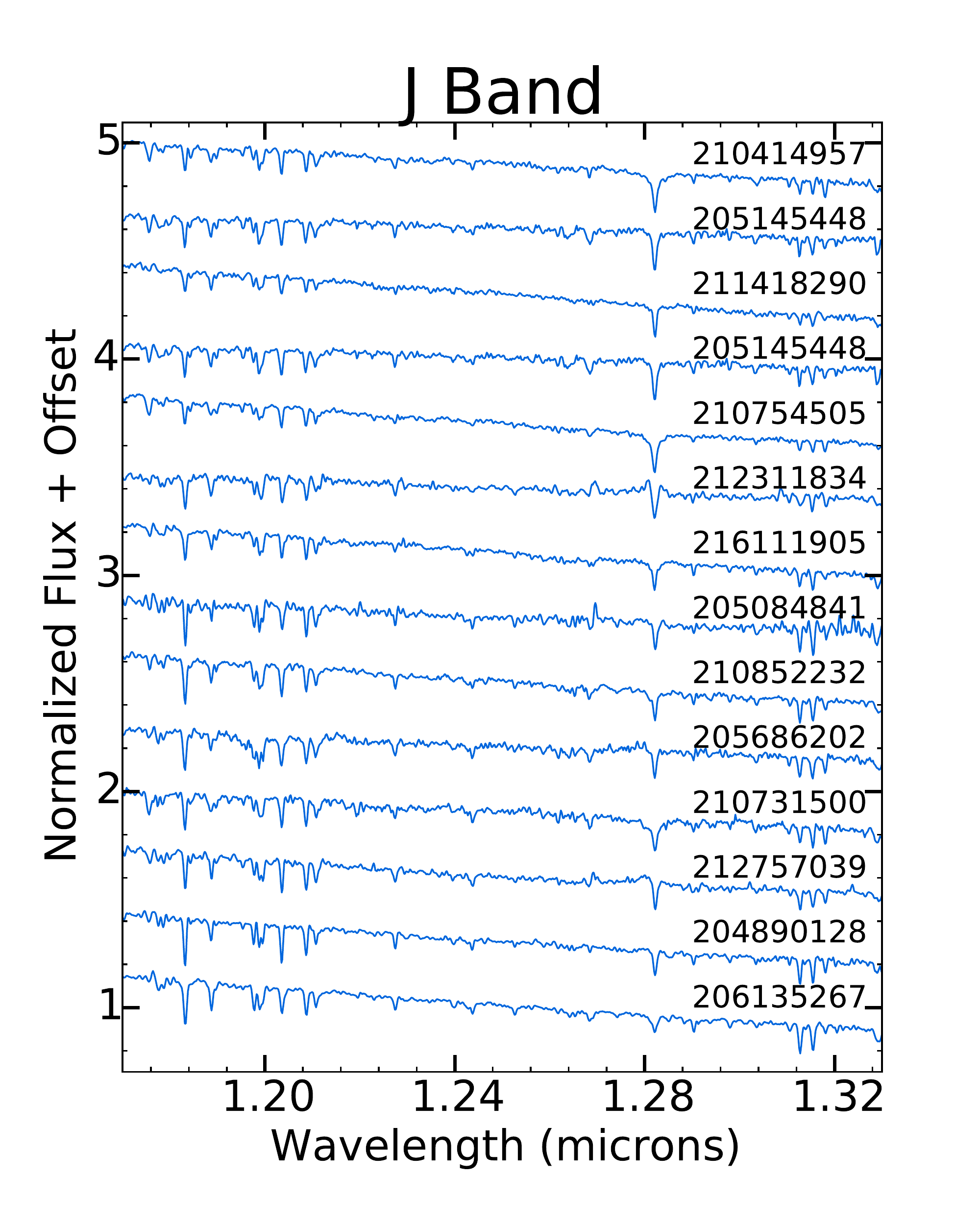}\\
\includegraphics[width=0.49\textwidth]{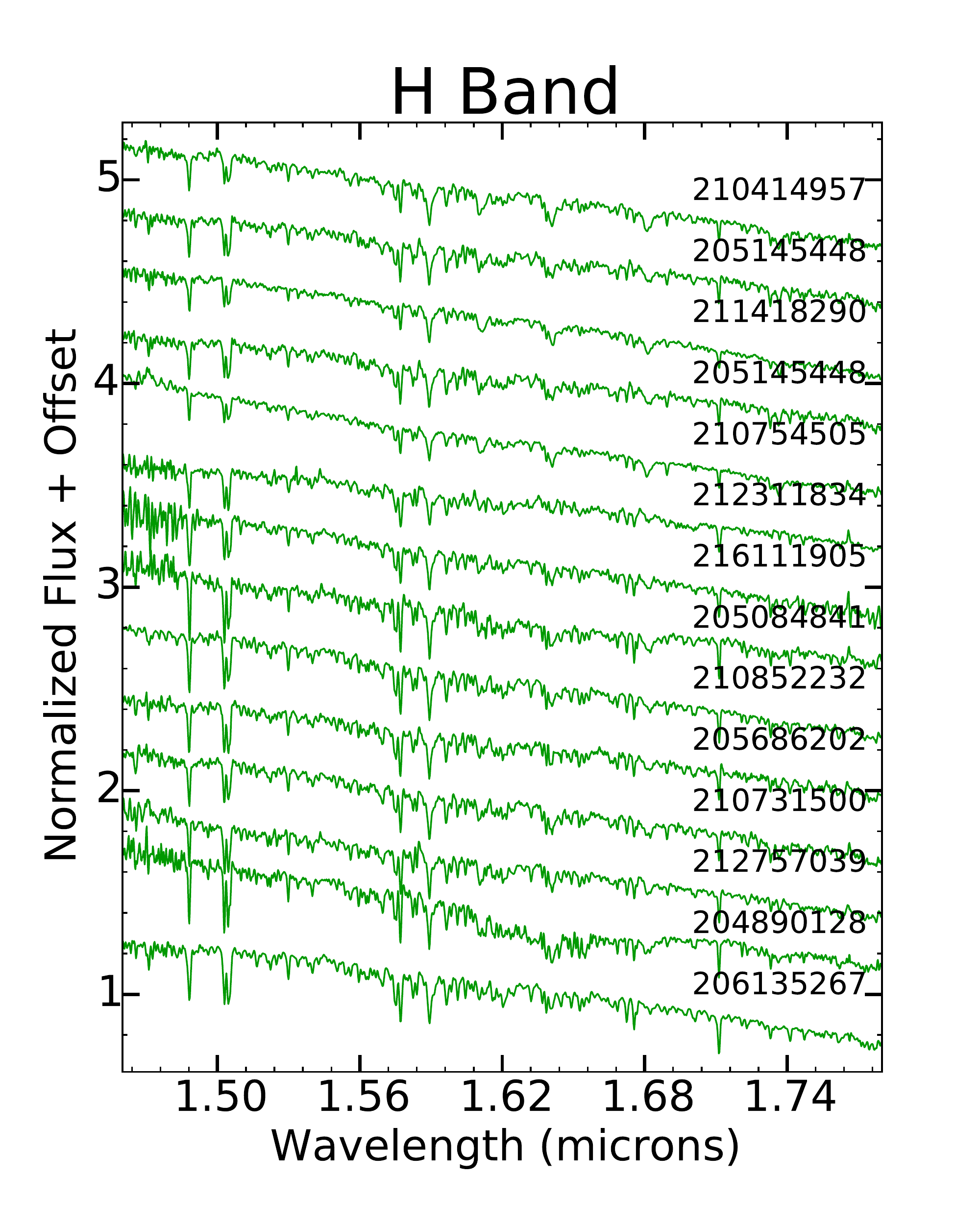}
\includegraphics[width=0.49\textwidth]{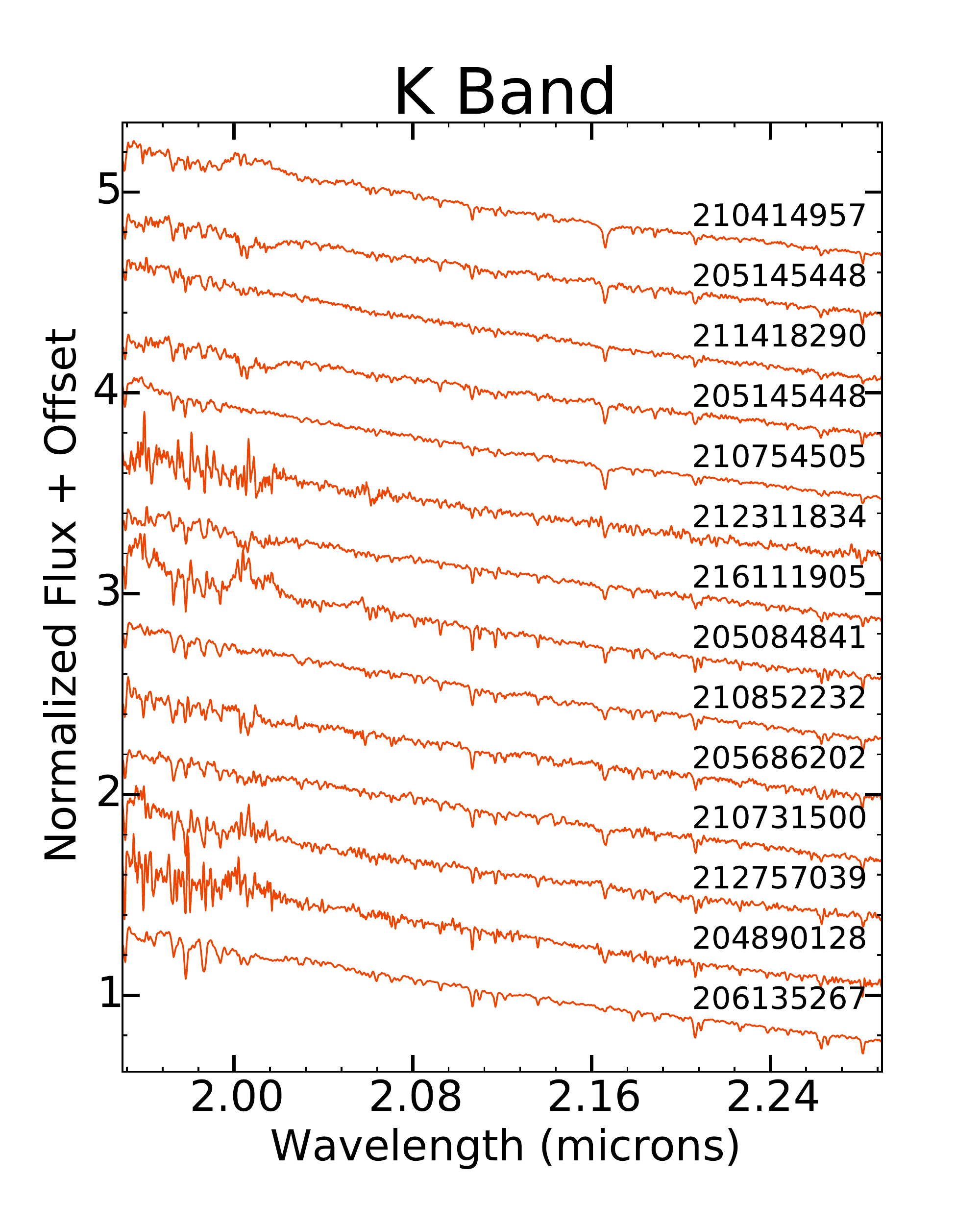}\\
\caption{$Y$-band (top left), $J$-band (top right), $H$-band (bottom left), and $K$-band (bottom right) spectra of hotter dwarfs with spectral types between K2 and G2. The hottest stars are shown at the top of the plots. Stars with truncated $Y$-band coverage were observed at the Palomar 200'' Hale Telescope using TripleSpec; the other stars were observed at the IRTF using SpeX. \label{fig:hotspec2}}
\end{figure*}

\begin{figure*}[tbhp]
\centering
\includegraphics[width=0.49\textwidth]{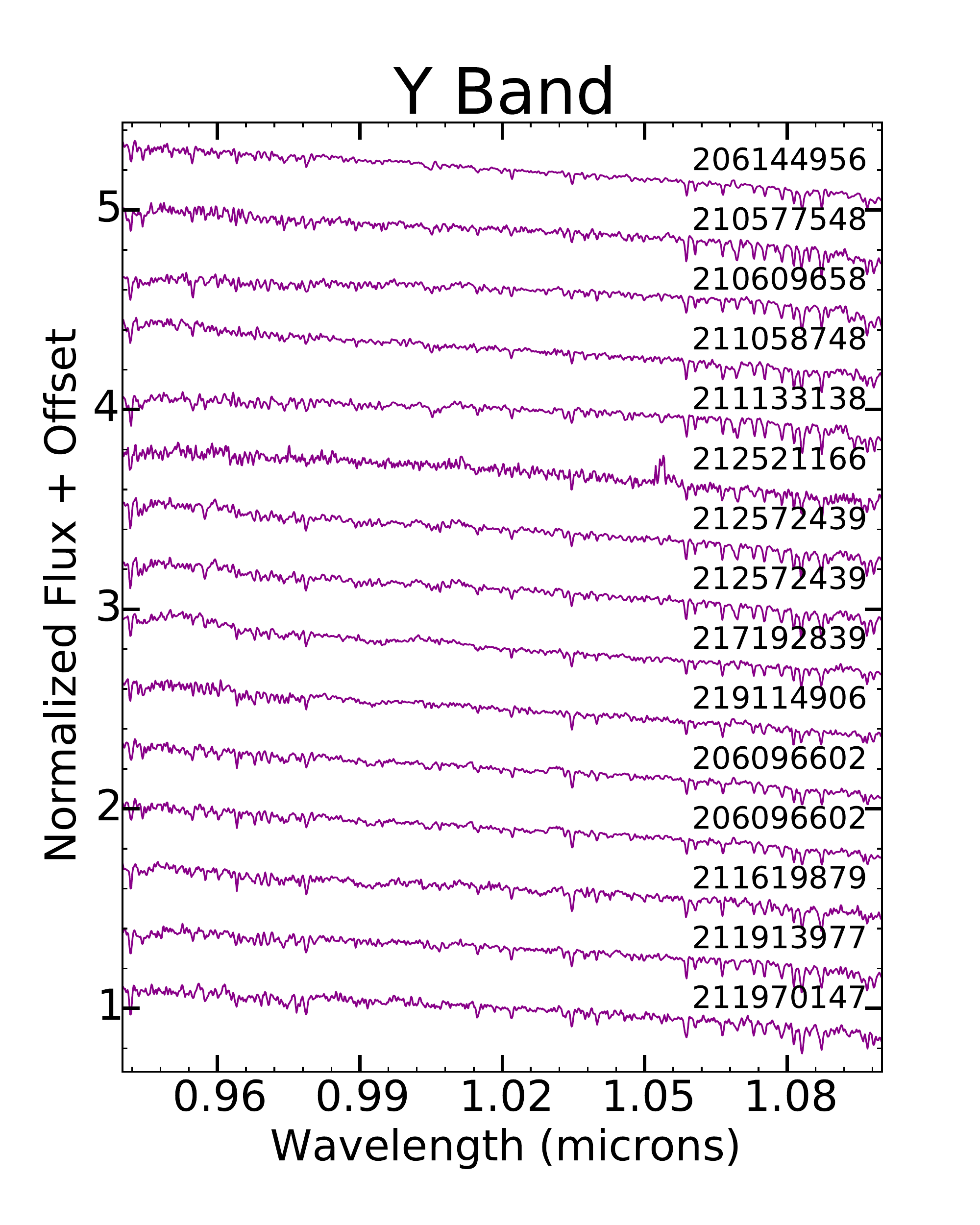}
\includegraphics[width=0.49\textwidth]{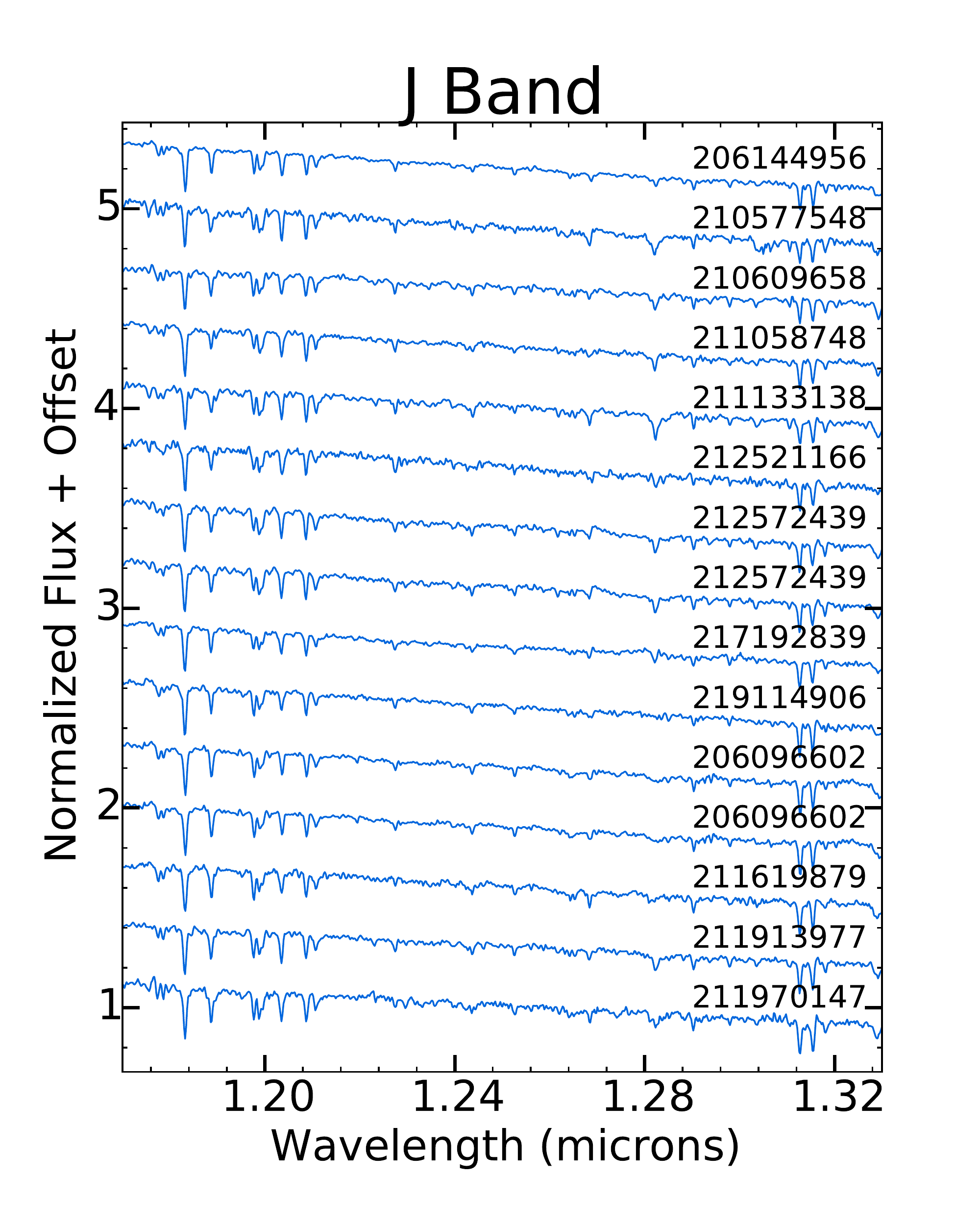}\\
\includegraphics[width=0.49\textwidth]{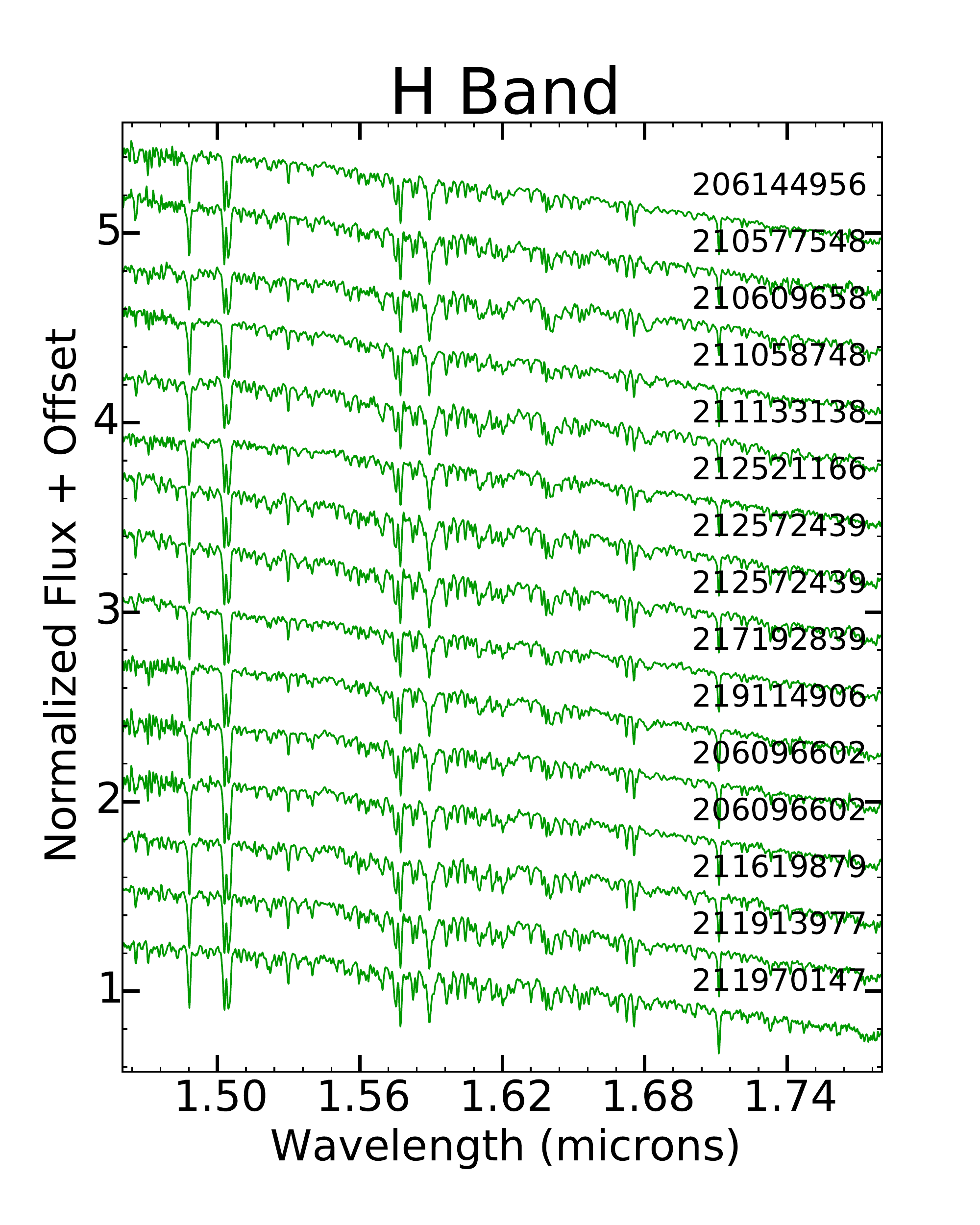}
\includegraphics[width=0.49\textwidth]{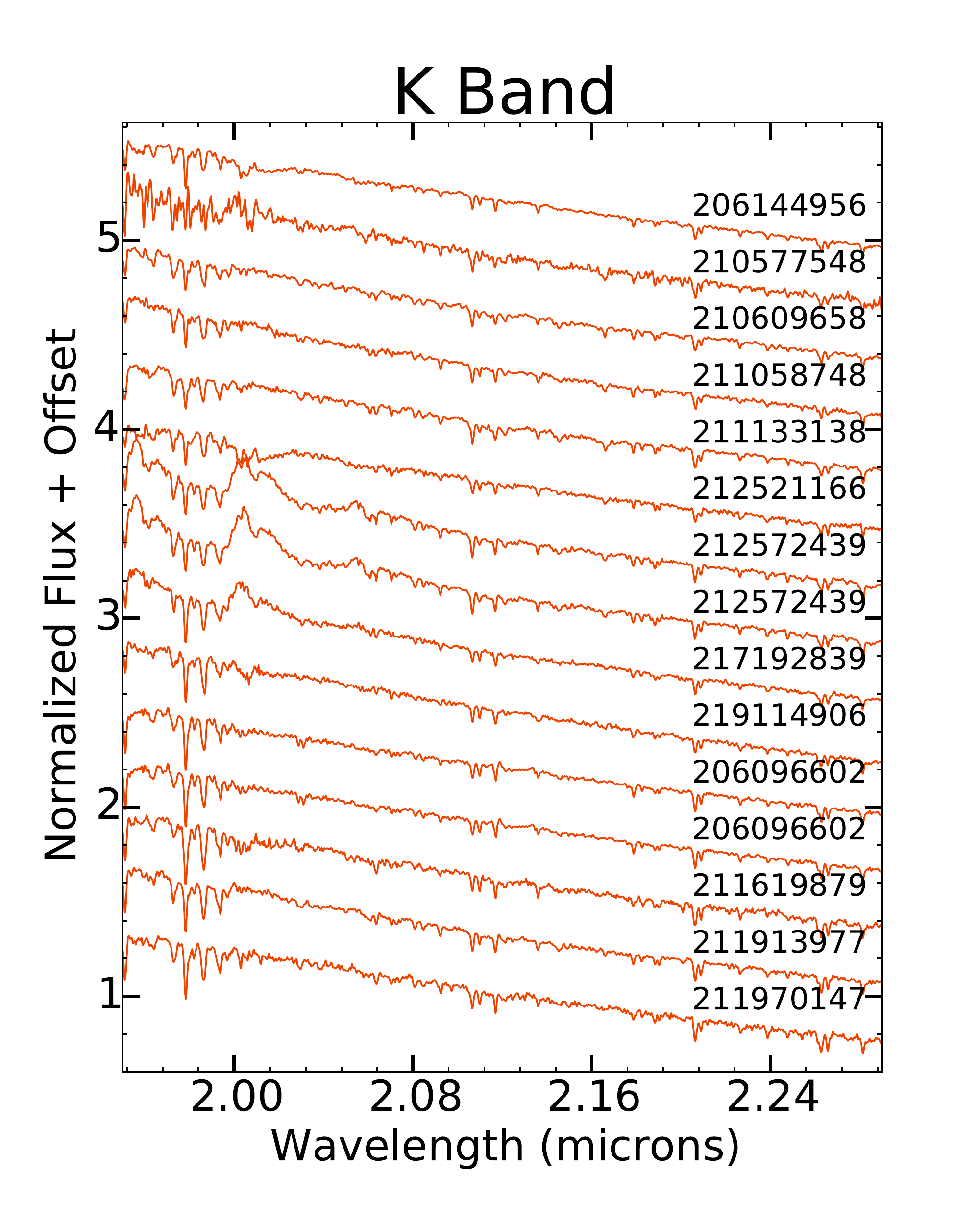}\\
\caption{Same as Figure~\ref{fig:hotspec2} for hotter dwarfs with spectral types between K3 and K2.  \label{fig:hotspec1}}
\end{figure*}

\begin{figure*}[tbhp]
\centering
\includegraphics[width=0.49\textwidth]{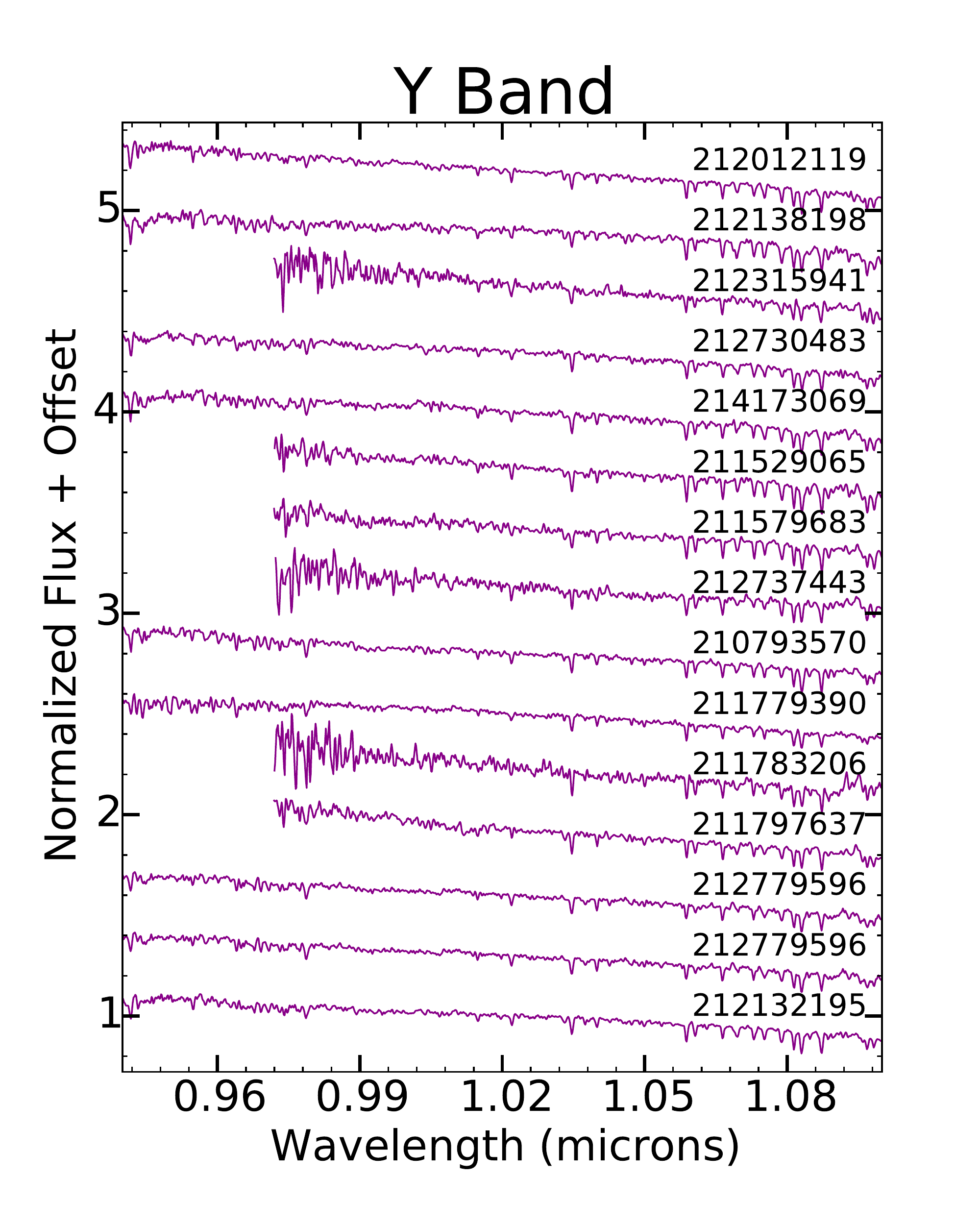}
\includegraphics[width=0.49\textwidth]{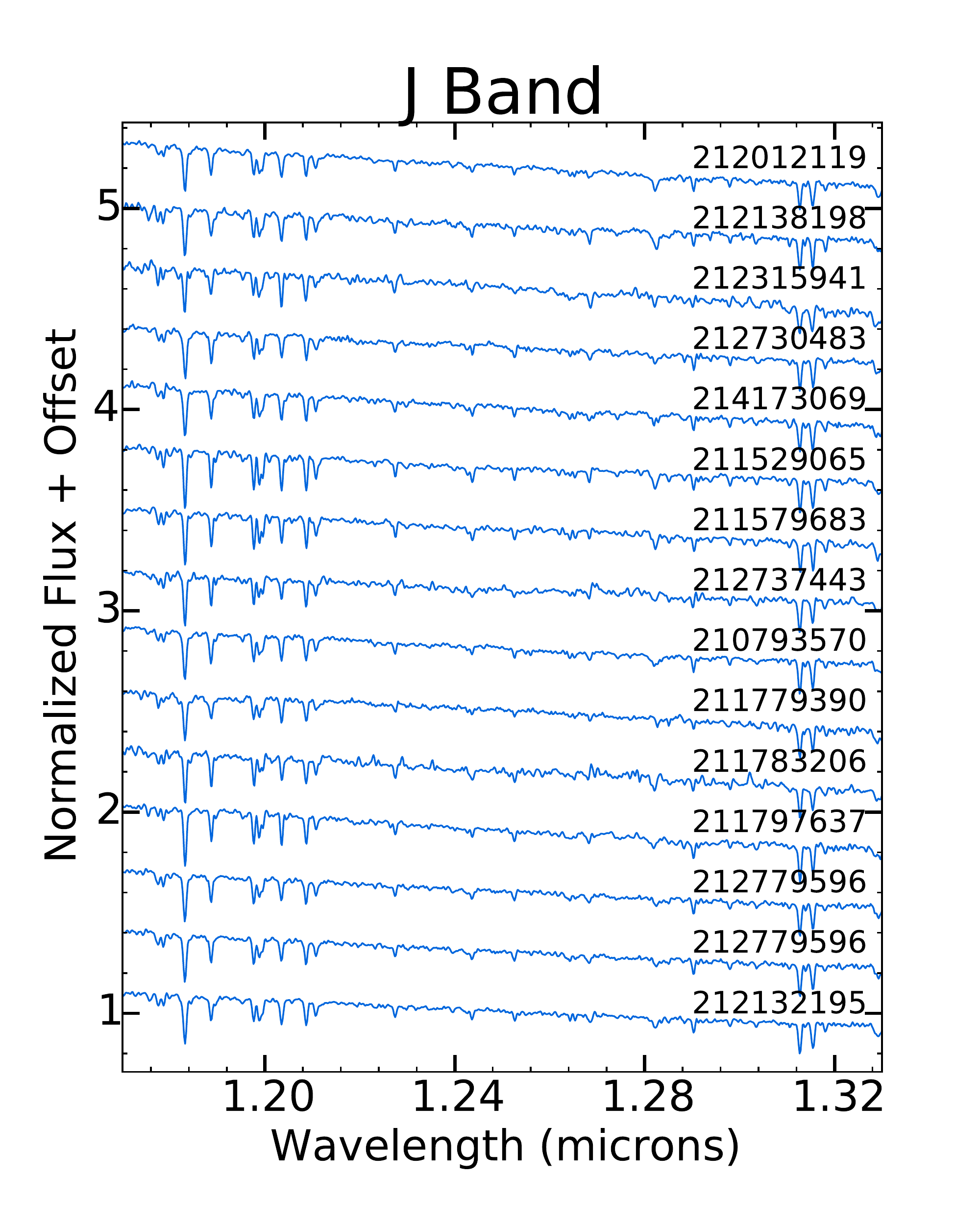}\\
\includegraphics[width=0.49\textwidth]{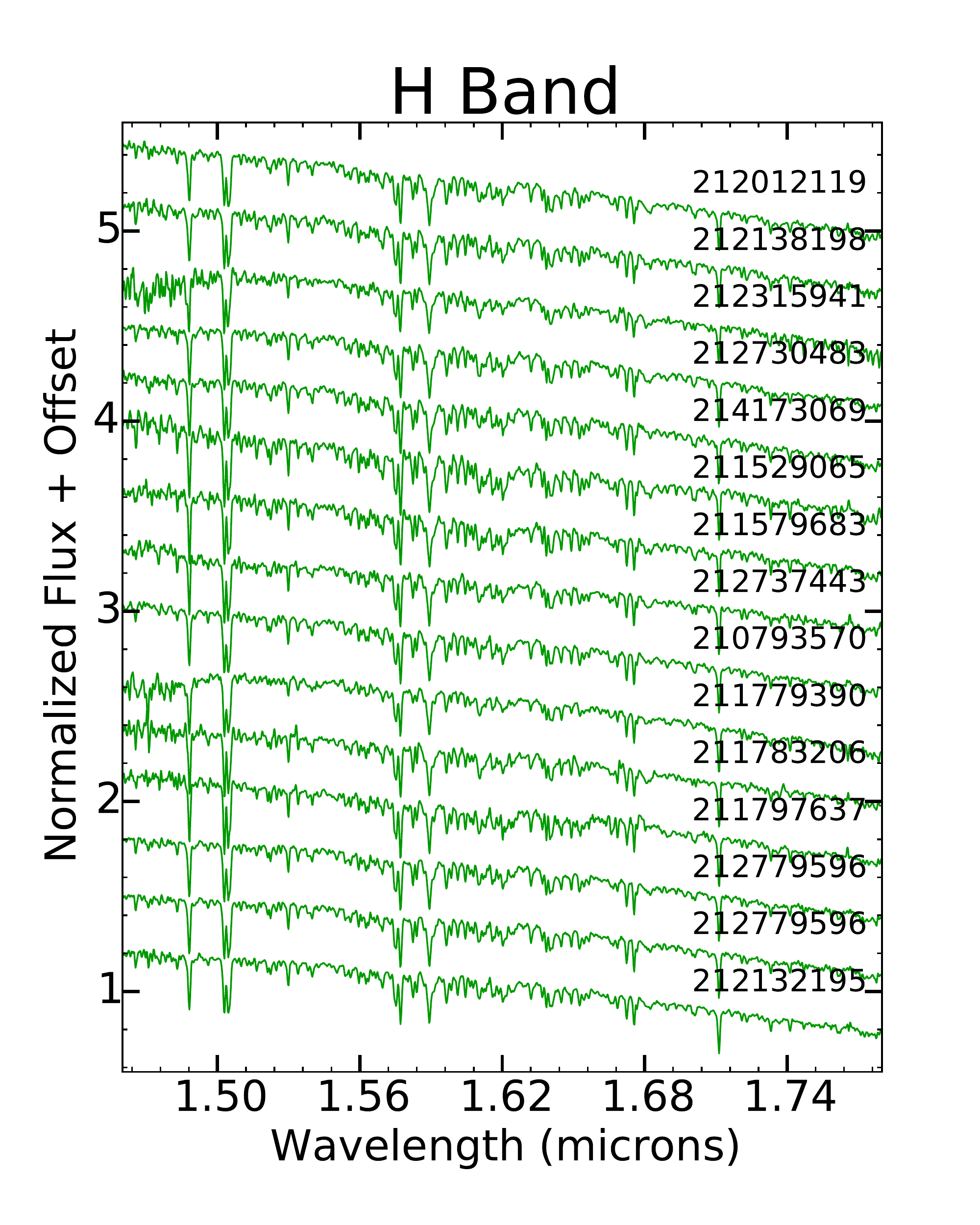}
\includegraphics[width=0.49\textwidth]{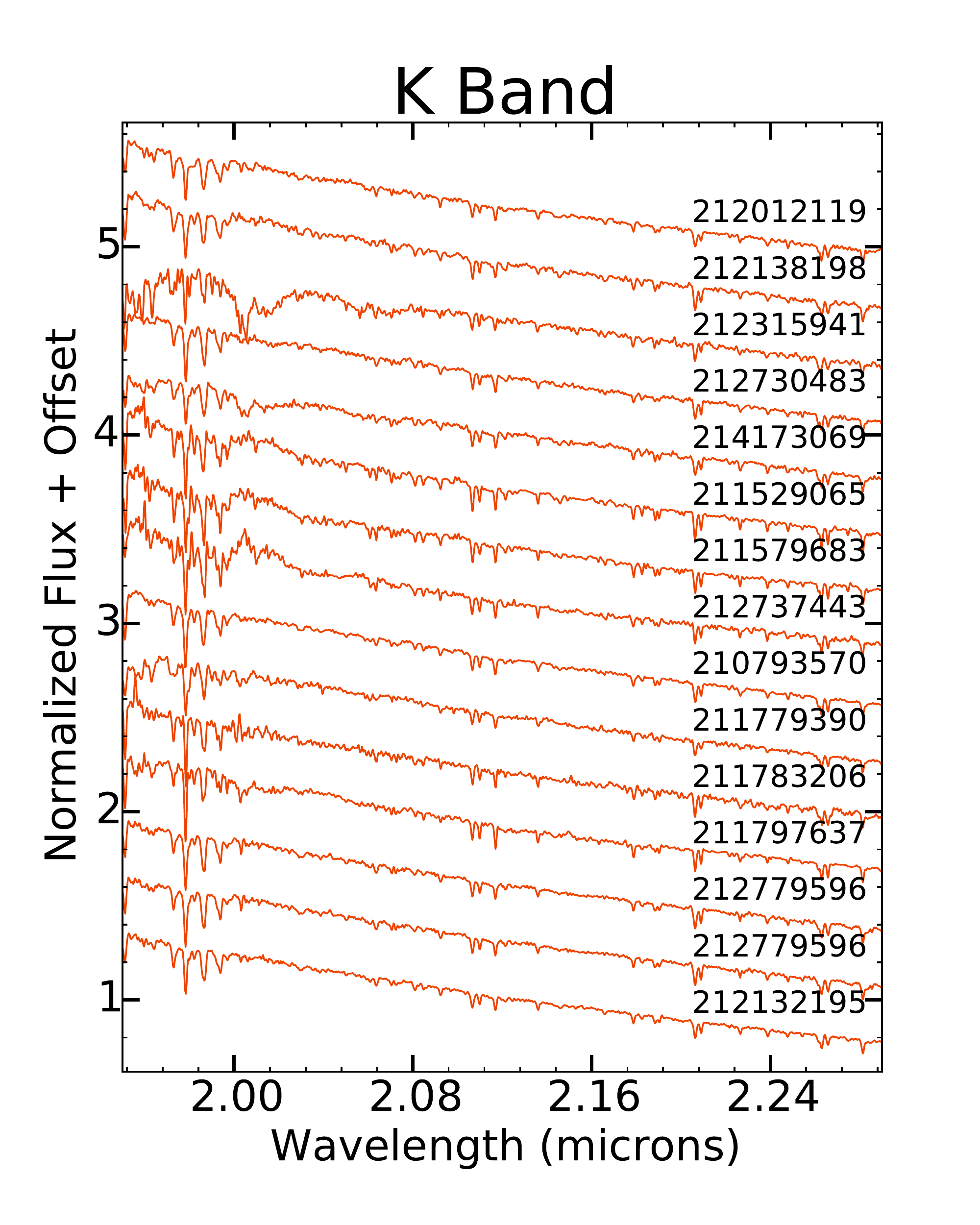}\\
\caption{Same as Figure~\ref{fig:hotspec2} for hotter dwarfs with spectral types between K5 and K3. Although some of these stars were expected to be cool enough for the \citet{newton_et_al2015} relations, they were assigned temperatures hotter than 4800~K and therefore excluded from the cool dwarf analysis.  \label{fig:hotspec0}}
\end{figure*}

\begin{figure*}[tbhp]
\centering
\includegraphics[width=0.49\textwidth]{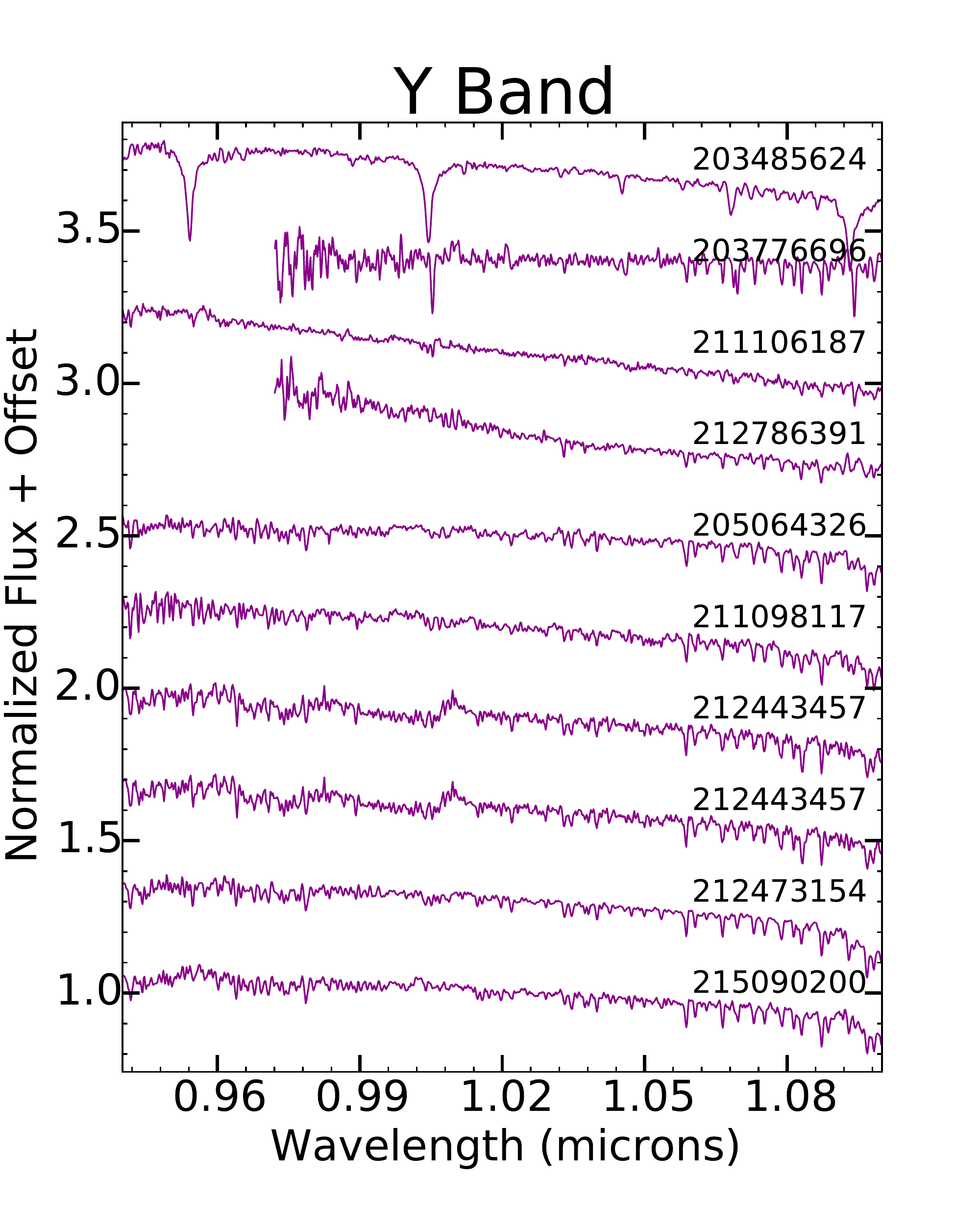}
\includegraphics[width=0.49\textwidth]{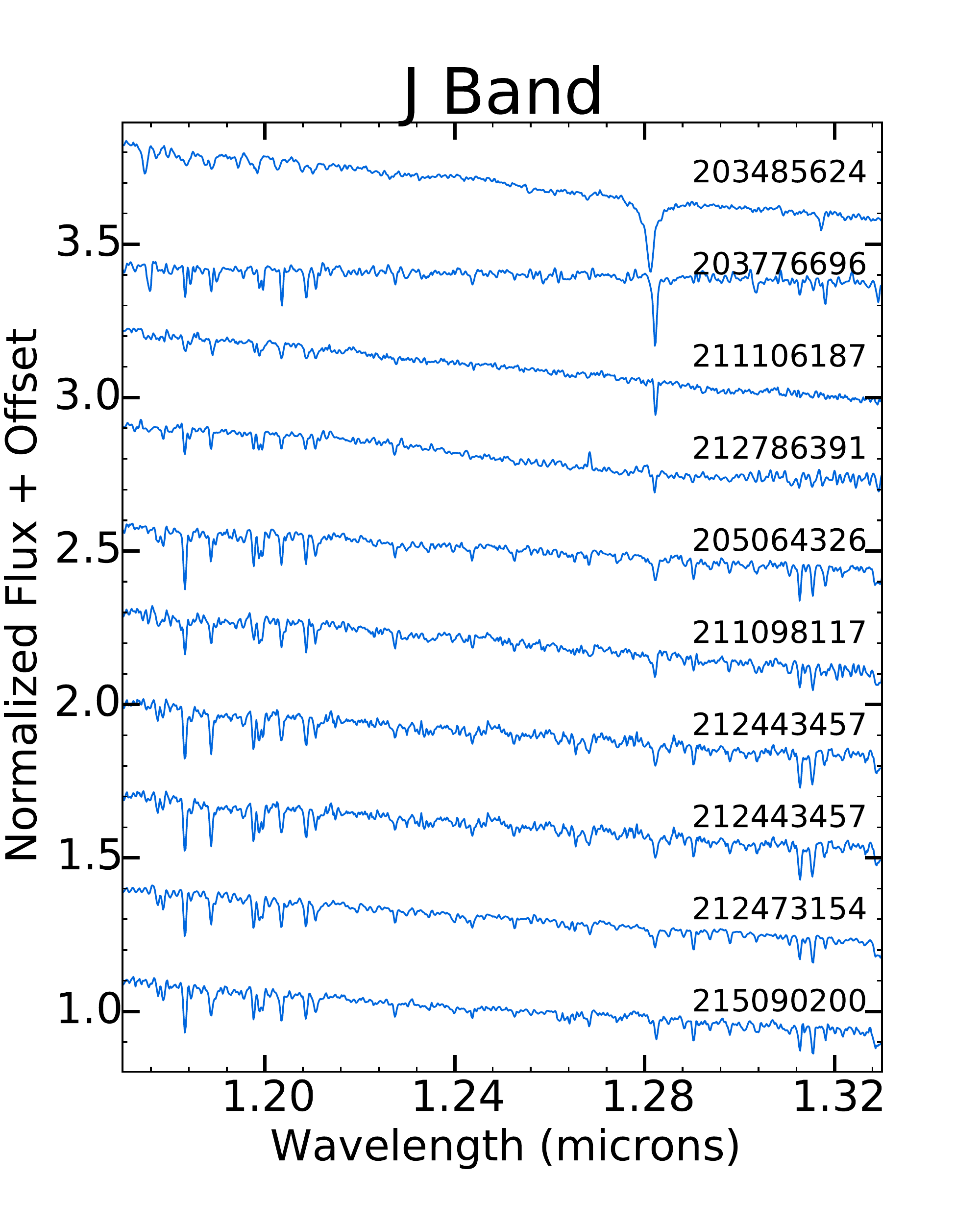}\\
\includegraphics[width=0.49\textwidth]{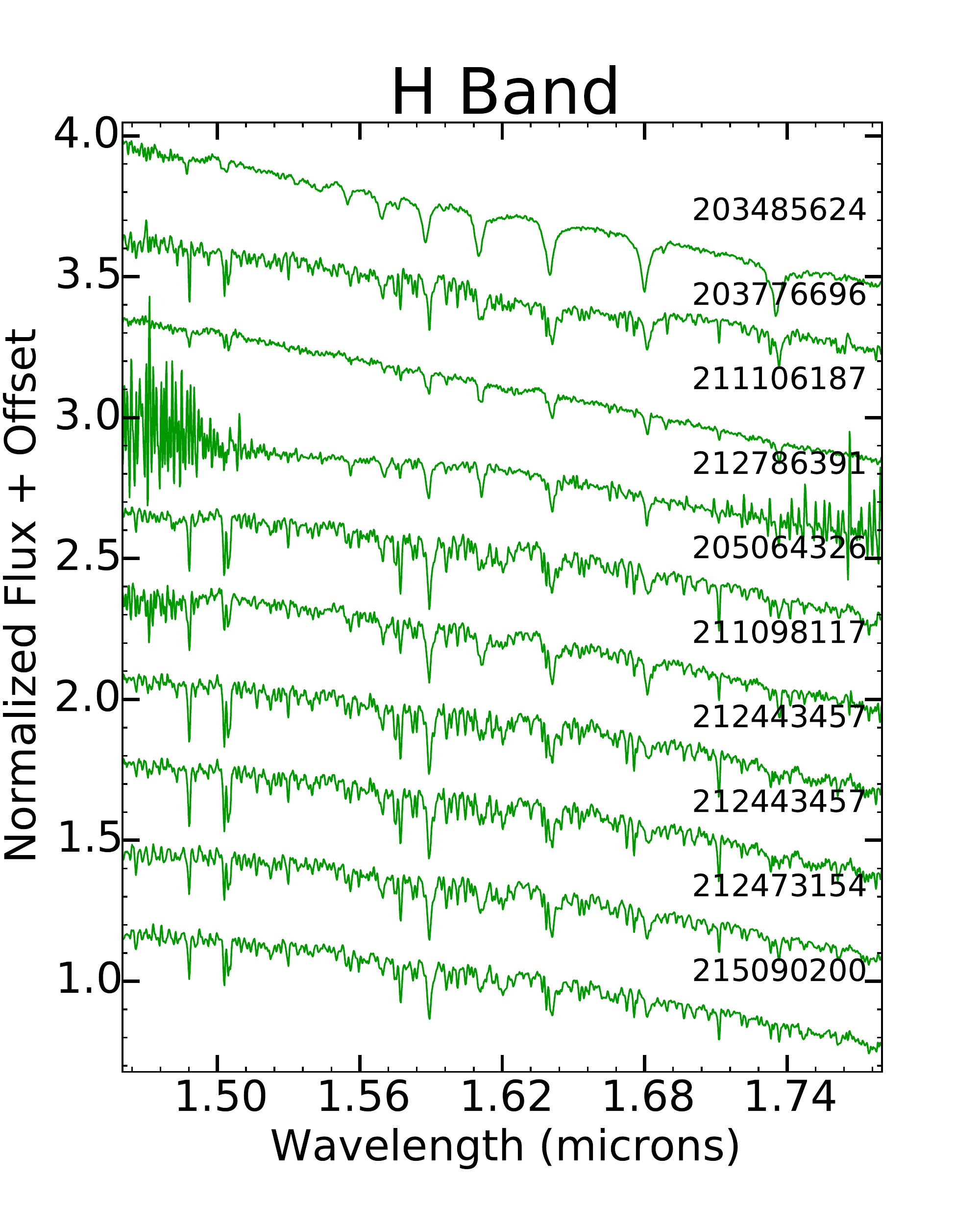}
\includegraphics[width=0.49\textwidth]{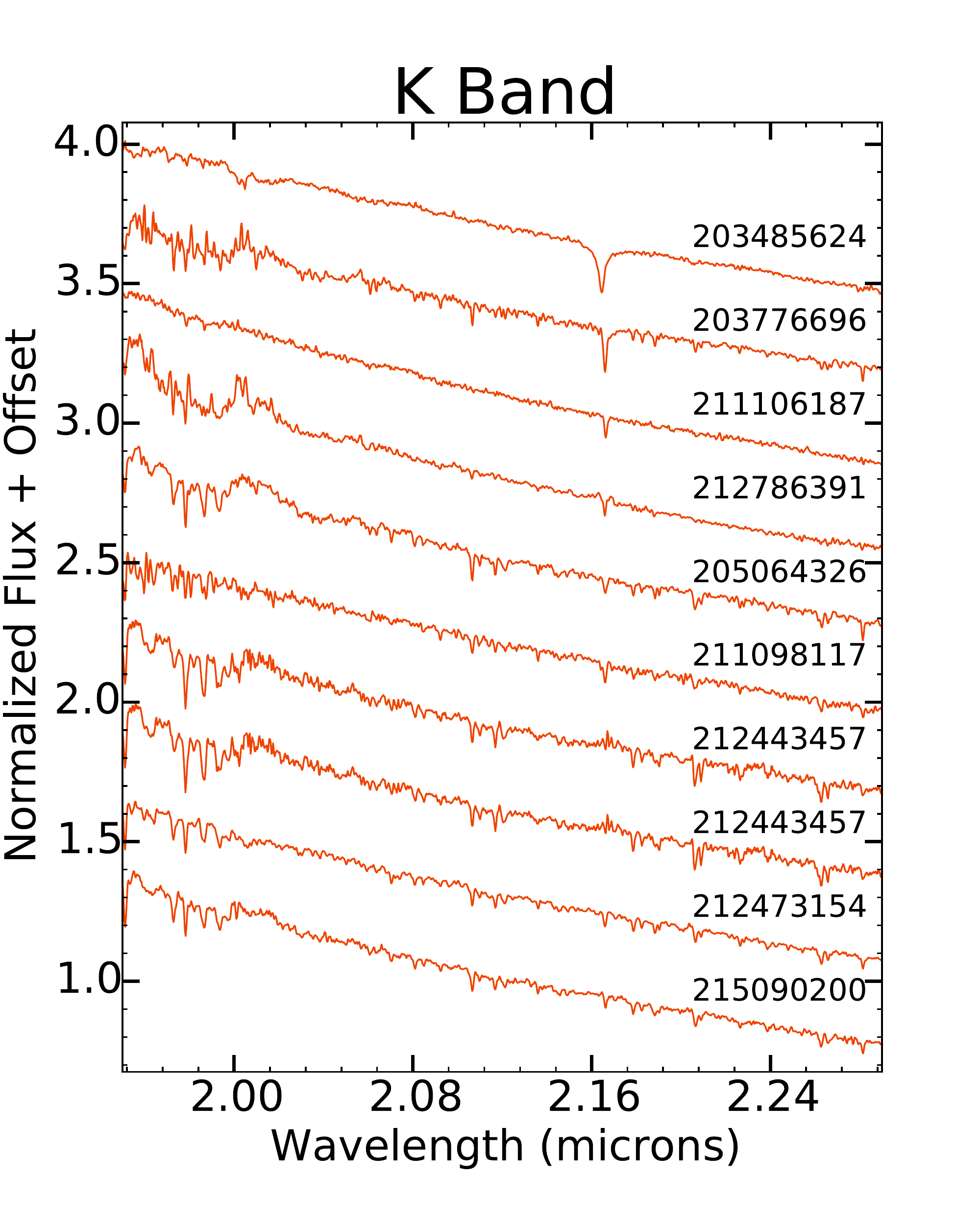}\\
\caption{$Y$-band (top left), $J$-band (top right), $H$-band (bottom left), and $K$-band (bottom right) spectra of giant stars with spectral types between F2 and K0. The hottest stars are shown at the top of the plots. Stars with truncated $Y$-band coverage were observed at the Palomar 200'' Hale Telescope using TripleSpec; the other stars were observed at the IRTF using SpeX. \label{fig:giantspec1}}
\end{figure*}

\begin{figure*}[tbhp]
\centering
\includegraphics[width=0.49\textwidth]{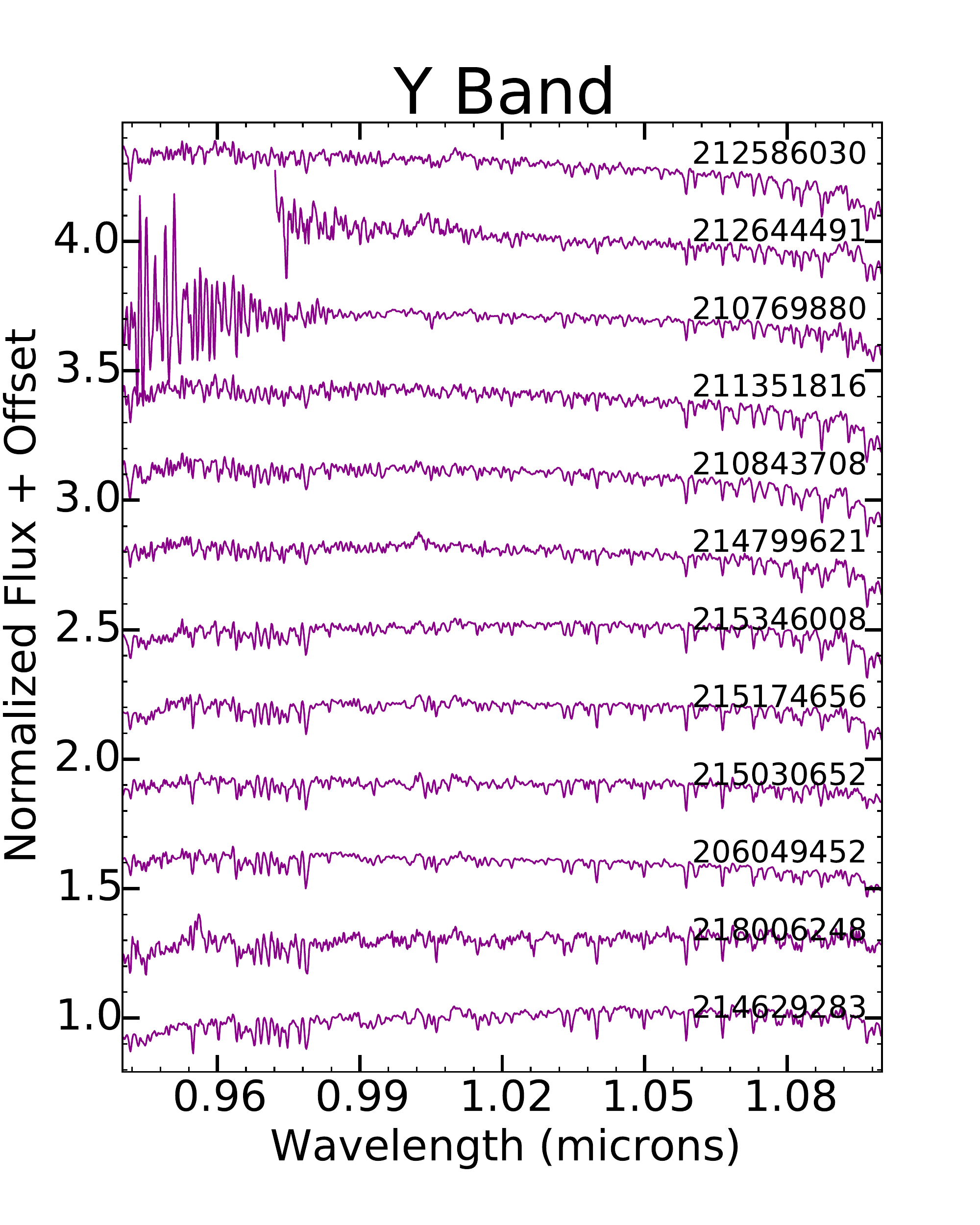}
\includegraphics[width=0.49\textwidth]{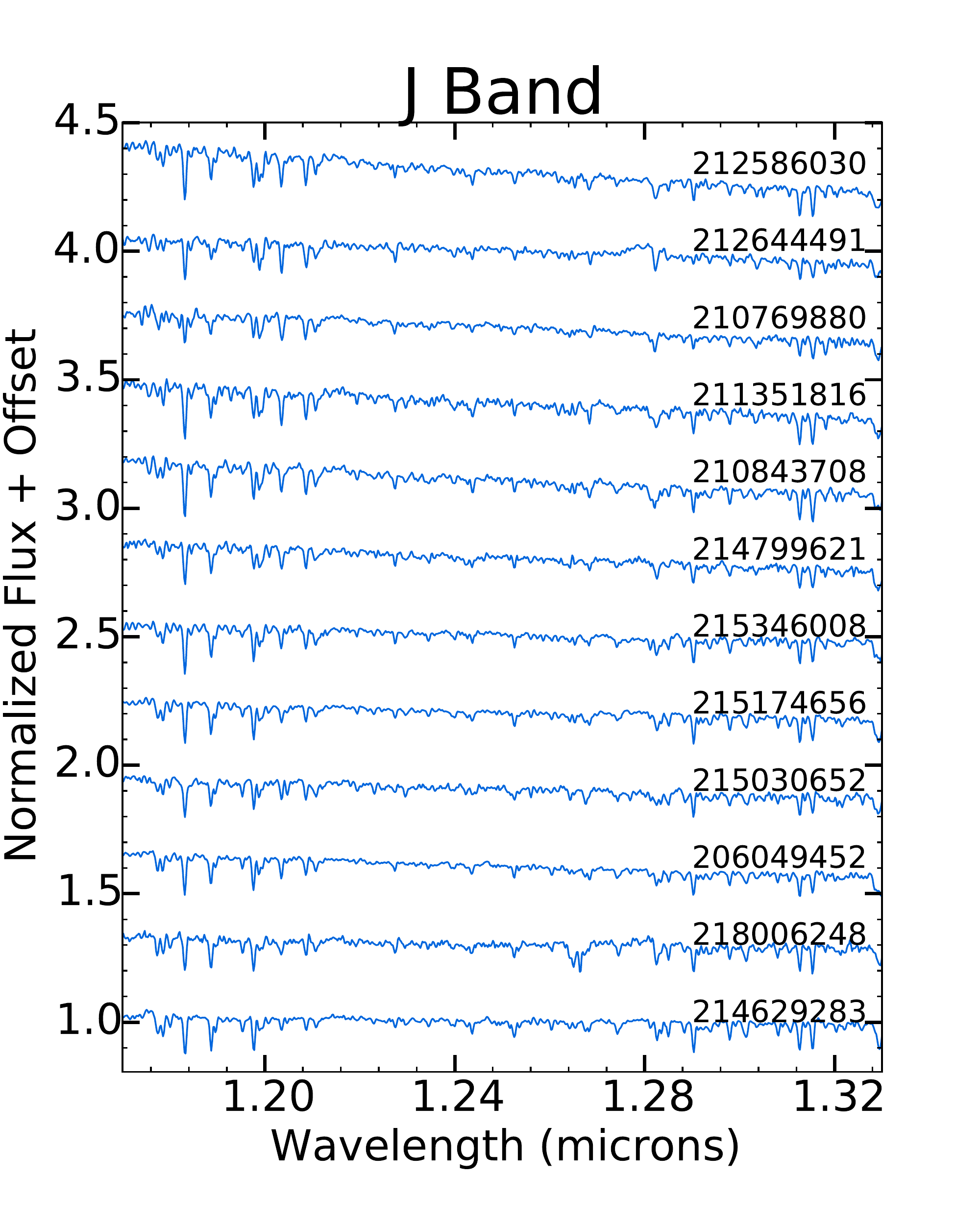}\\
\includegraphics[width=0.49\textwidth]{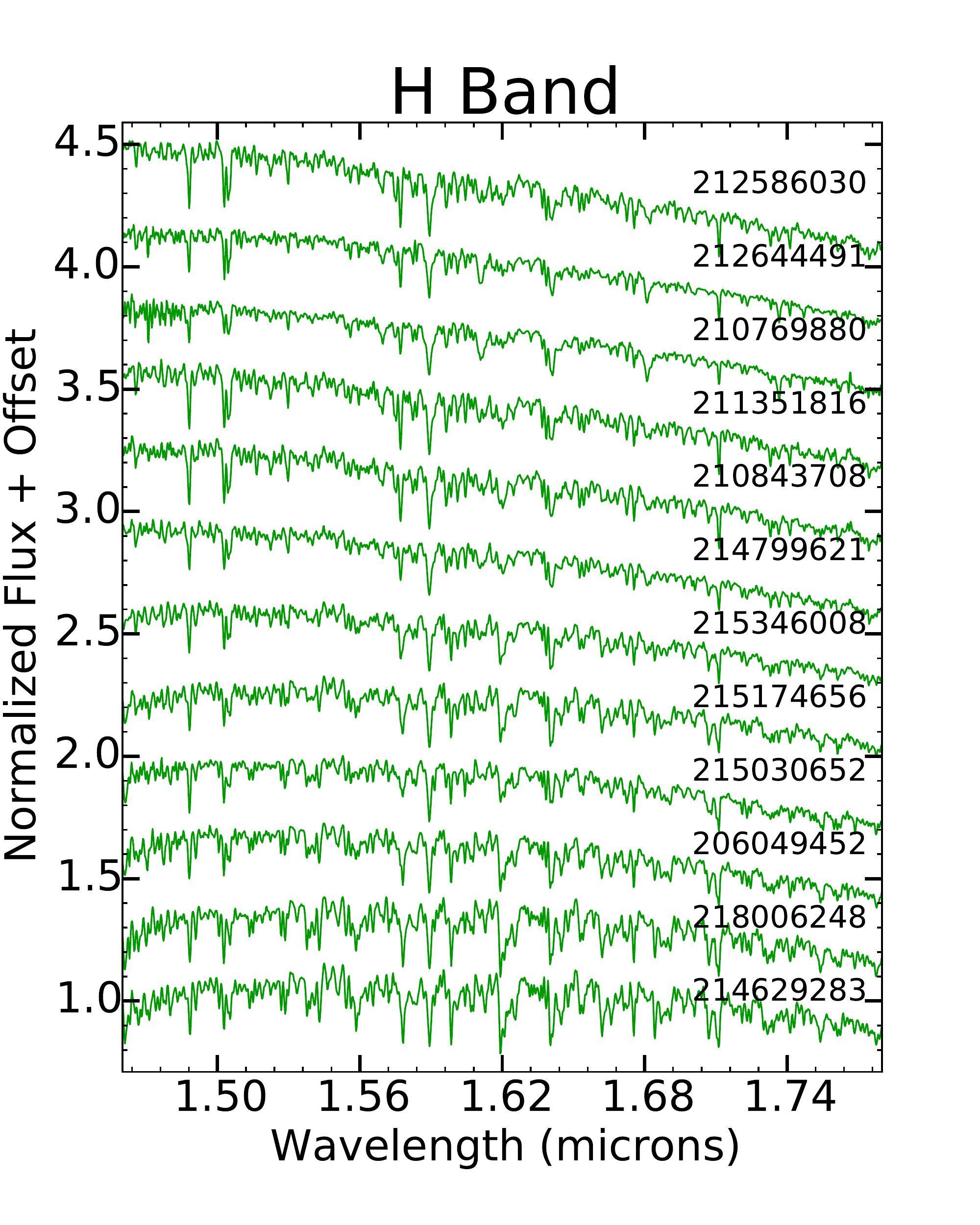}
\includegraphics[width=0.49\textwidth]{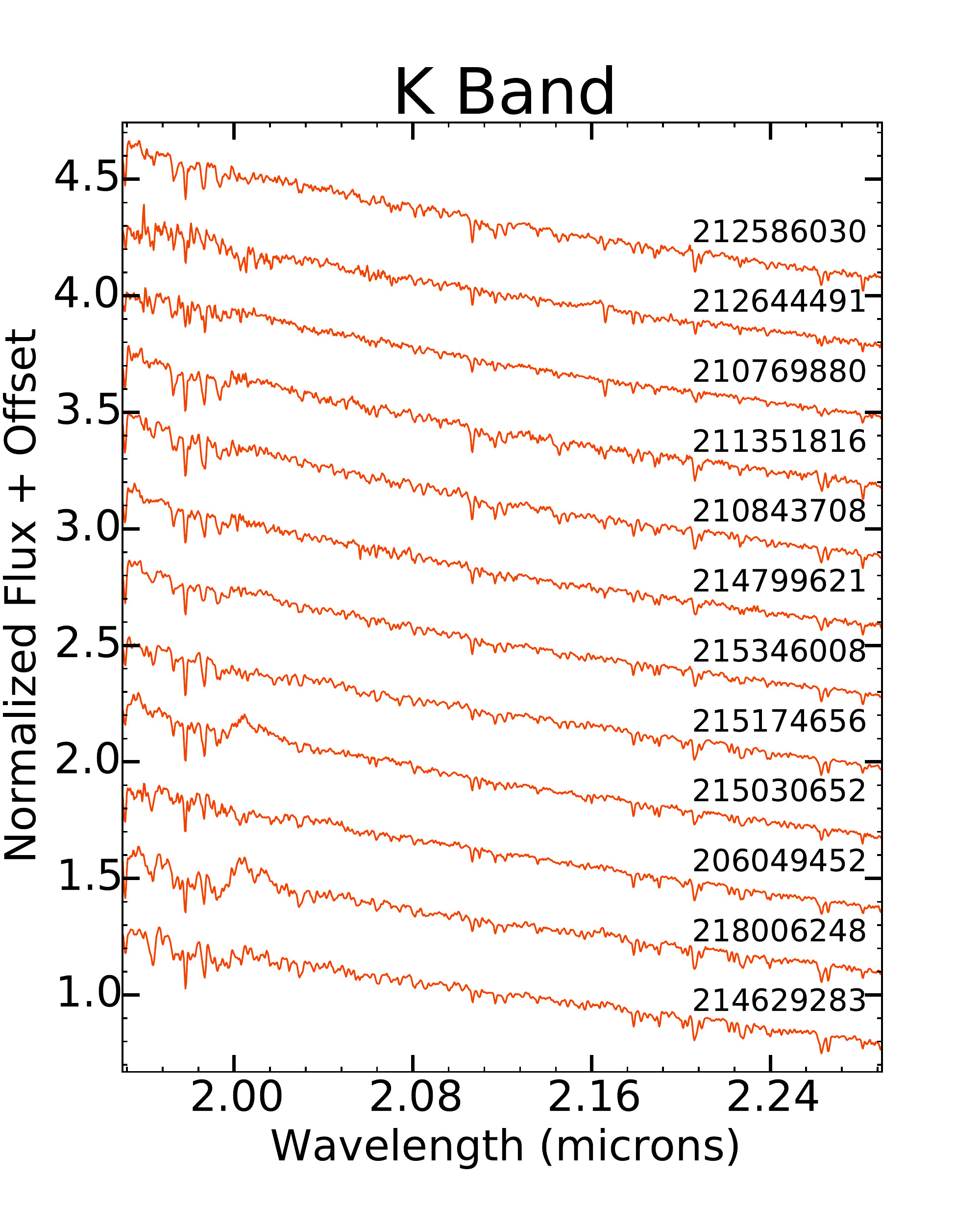}\\
\caption{Same as Figure~\ref{fig:giantspec1} for giant stars with spectral types between K1 and M2. \label{fig:giantspec0}}
\end{figure*}

As mentioned in Section~\ref{ssec:initial_class}, all of our reduced spectra are posted on the ExoFOP website. We also display the spectra in Figures~\ref{fig:dwarfspec5}-\ref{fig:dwarfspec0} for cool dwarfs, Figures~\ref{fig:hotspec2}-\ref{fig:hotspec0} for hotter dwarfs, and Figures~\ref{fig:giantspec1}-\ref{fig:giantspec0} for giants.

\end{document}